\documentclass[a4paper,11pt]{article}

\usepackage{jheppub} 

\usepackage[T1]{fontenc} 
\usepackage{besphysics}
\usepackage{lineno}
\usepackage{float}
\usepackage{bm}
\usepackage{cleveref}
\crefname{equation}{Equation}{Equations}
\crefname{figure}{Figure}{Figures}
\crefname{table}{Table}{Tables}

\title{\boldmath Study of $e^+e^- \to \gamma\phi\jpsi$ from $\sqrt{s}=4.600$ to $4.951\gev$}
\author{
\noindent
M.~Ablikim$^{1}$, M.~N.~Achasov$^{12,b}$, P.~Adlarson$^{72}$, M.~Albrecht$^{4}$, R.~Aliberti$^{33}$, A.~Amoroso$^{71A,71C}$, M.~R.~An$^{37}$, Q.~An$^{68,55}$, Y.~Bai$^{54}$, O.~Bakina$^{34}$
% ...
\\
\vspace{0.2cm}
(BESIII Collaboration)\\
\vspace{0.2cm} {\it
$^{1}$ Institute of High Energy Physics, Beijing 100049, People's Republic of China\\
% ...
$^{4}$ Bochum  Ruhr-University, D-44780 Bochum, Germany\\
% ...
$^{12}$ G.I. Budker Institute of Nuclear Physics SB RAS (BINP), Novosibirsk 630090, Russia\\
% ...
$^{33}$ Johannes Gutenberg University of Mainz, Johann-Joachim-Becher-Weg 45, D-55099 Mainz, Germany\\
$^{34}$ Joint Institute for Nuclear Research, 141980 Dubna, Moscow region, Russia\\
% ...
$^{37}$ Liaoning Normal University, Dalian 116029, People's Republic of China\\
% ...
$^{54}$ Southeast University, Nanjing 211100, People's Republic of China\\
$^{55}$ State Key Laboratory of Particle Detection and Electronics, Beijing 100049, Hefei 230026, People's Republic of China\\
% ...
$^{68}$ University of Science and Technology of China, Hefei 230026, People's Republic of China\\
% ...
$^{71}$ University of Turin and INFN, (A)University of Turin, I-10125, Turin, Italy; (B)University of Eastern Piedmont, I-15121, Alessandria, Italy; (C)INFN, I-10125, Turin, Italy\\
$^{72}$ Uppsala University, Box 516, SE-75120 Uppsala, Sweden\\
% ...
$^{b}$ Also at the Novosibirsk State University, Novosibirsk, 630090, Russia\\
% ...
}
}
\collaboration{The BESIII Collaboration}
\emailAdd{besiii-publications@ihep.ac.cn}

\collaborationImg{\includegraphics[height=1.68cm]{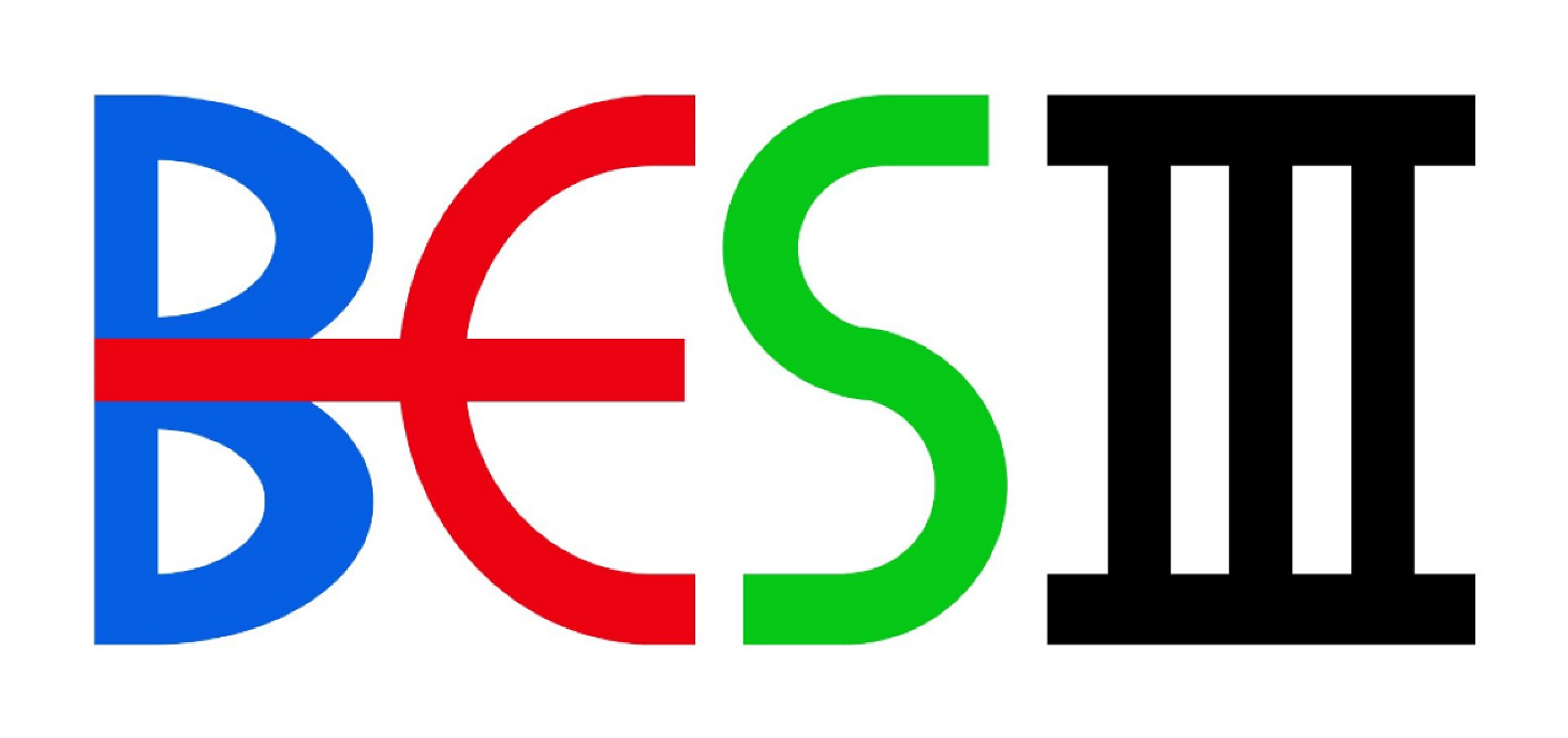}}

\abstract{Using data samples with an integrated luminosity of
  $6.4$~fb$^{-1}$ collected by the BESIII detector operating at the
  BEPCII storage ring, the process of $e^+e^-\to\gamma\phi J/\psi$ is
  studied. The processes of $e^+e^-\to\phi\chi_{c1,c2}$,
  $\chi_{c1,c2}\to\gamma J/\psi$ are observed with a significance of
  more than $10\sigma$. The $\sqrt{s}$-dependent cross section of $e^+e^-
  \to \phi\chi_{c1,c2}$ is measured between 4.600 and 4.951~GeV, and
  evidence of a resonance structure is found for the first time in the
  $\phi\chi_{c2}$ process. We also search for the processes of
  $e^+e^-\to\gamma X(4140)$, $\gamma X(4274)$ and $\gamma X(4500)$ via
  the $\gamma\phi J/\psi$ final state, but no obvious structures are
  found. The upper limits on the production cross section times the
  branching fraction for these processes at the 90\% confidence level
  are reported.}

\begin{document} 
\maketitle
\flushbottom

\section{Introduction}
In the past decades, several charmonium-like states with $\jpc=1^{--}$
have been discovered, such as the
$Y(4260)$~\cite{babary4260,belley4260,cleoy4260},
$Y(4360)$~\cite{babary4360,belley4360} and
$Y(4660)$~\cite{belley4360,BaBar:2012hpr,Belle:2014wyt,BESIII:2021njb}. The
potential model predicts five vector charmonium states in the mass
region between 4.0 and 4.7~$\gevcc$, namely $3S$, $2D$, $4S$, $3D$,
and $5S$~\cite{potential-mode}. The first three states have been
identified with the $\psi(4040)$, $\psi(4160)$ and
$\psi(4415)$~\cite{Zyla:2020zbs}. Together with the three observed
$Y$-states, we have at least six $1^{--}$ states in this mass region.
In addition, the masses of the undiscovered $3D$ and $5S$ states are
expected to be higher than 4.4~$\gevcc$, which leaves no room for $\y$
and $Y(4360)$ in the charmonium spectrum.  Unlike the conventional
$1^{--}$ charmonium states which predominantly decay to open charm
final states ($D^{(*)}\bar{D}^{(*)}$), the $Y$-states are found to
usually couple with hidden-charm final
states~\cite{babary4260,belley4260,cleoy4260,babary4360,belley4360}.
Considering these unusual properties, the $Y$-states are widely
regarded as good candidates for unconventional hadron states, such as
hybrids, tetraquarks, or meson
molecules\cite{Chen:2016qju,Brambilla:2019esw}.

At present, the inner structure of these $Y$-states remains
unclear. Experimentally, the $\EE$ annihilation process is one of the
most effective ways to probe the nature of $Y$-states.
The $Y(4660)$ resonance was first observed by the Belle Collaboration in $\EE\to\pp\psip$ process 
via initial-state-radiation (ISR)~\cite{belley4360}, and subsequently confirmed by the BaBar~\cite{BaBar:2012hpr} 
and BESIII Collaborations~\cite{BESIII:2021njb} in the same process. In the $Y(4660)\to\pp\psip$ decay, 
the $\pp$ system is found to be dominated by a $f_0(980)$ which has a significant $s\bar{s}$ component. 
Recently, the Belle experiment reported the first $Y(4626)$ resonance coupling
to the $\dsp\dsonem +\mathrm{c.c.}$ meson pair with a significance of
5.9$\sigma$~\cite{Jia:2019gfe}. Belle also reported evidence
(3.4$\sigma$) for a resonance with mass and width consistent with
$Y(4626)$ in the $\dsp D_{s2}^{*-}(2573) +\mathrm{c.c.}$
process~\cite{Jia:2020epr}. 
It is not clear whether $Y(4660)$ and $Y(4626)$ correspond to the same resonance or not.
The observation of the $Y(4660)$ state coupling to $f_0(980)\psip$ and the $Y(4626)$ state
coupling to the charmed-antistrange and anticharmed-strange meson pair
may indicate that $Y(4660)/Y(4626)$ have $c\bar{c}s\bar{s}$
components~\cite{Karliner:2016ith,He:2019csk,Deng:2019dbg}.
In such a case, the $Y(4660)/Y(4626)$ may also decay to the final states of $\phi\chico$ or
$\phi\chict$. 
The BESIII experiment has measured the cross section of
$\EE\to\phi\chi_{c1,c2}$ at 4.600 GeV~\cite{BESIII:2017qtm}, and
significant $\chi_{c1,c2}$ signals were found. With the data taken at
center-of-mass (c.m.)~energies up to $\sqrt{s}=4.951\gev$ at BESIII, which
fully covers the $Y(4626)$ and $Y(4660)$ mass region, we are able to
measure the cross section line shape of $\EE \to \phi\chi_{c1,c2}$. The measurements may shed light on the inner structure of the
$Y(4660)/Y(4626)$ states and help us understand their nature.

In addition to the $Y$-states, the non-vector $X$-states in the
$\phi\jpsi$ system also attract much interest.  A narrow ($\Gamma =
11.7\mev$) near-threshold peak around $4143\mevcc$ in the $\phi\jpsi$
mass spectrum was first reported by the CDF Collaboration in the
$B^+\to \phi\jpsi\kp$ process with 3.8$\sigma$ evidence (labeled as
$X(4140)$)~\cite{cdfy4140}. From the potential model, charmonium
states within this mass region are expected to have much larger widths
due to the open charm decay channels~\cite{Brambilla:2019esw}. The
$X(4140)$ is therefore suggested to be a candidate of an exotic
state. An updated analysis by the CDF Collaboration in
2011~\cite{cdf-X2} not only confirmed the existence of $X(4140)$ with
a 5$\sigma$ observation, but also reported evidence (3.1$\sigma$) of a
new narrow peak near $4274\mevcc$ in the $\phi\jpsi$
spectrum. Subsequent measurements were also carried out by the
Belle~\cite{ChengPing:2009vu},
LHCb~\cite{lhcb-X1,lhcb-X2,lhcb-X3,LHCb:2021uow}, CMS~\cite{cms-X},
D0~\cite{d0-X1,d0-X2}, BaBar~\cite{babar-X} and BESIII
experiments~\cite{BESIII:2017qtm,BESIII:2014fob}. 
Belle~\cite{ChengPing:2009vu}, BaBar~\cite{babar-X} and
BESIII~\cite{BESIII:2017qtm,BESIII:2014fob} found no evidence for the
$X(4140)$.

The LHCb Collaboration studied the $B^+ \to\phi\jpsi\kp$ process with
0.37~$\invfb$ of data, and no evidence of resonance structures was
found in the $\phi\jpsi$ system~\cite{lhcb-X1}. Later, with the full
Run1 data (3~$\invfb$), an updated analysis was performed by the LHCb
Collaboration with an amplitude analysis~\cite{lhcb-X2,lhcb-X3}. A
near-threshold structure with mass $4146.5\pm4.5^{+4.6}_{-2.8}\mevcc$
and width $83\pm21^{+21}_{-14}\mev$ was reported. In addition, they
also reported the existence of $X(4274)$, $X(4500)$ and $X(4700)$ in
the $\phi\jpsi$ system with significance more than $5\sigma$. Most
recently, with the Run1 and Run2 datasets (9 $\mathrm{fb}^{-1}$), LHCb
improved their amplitude analysis of $B^+ \to\phi\jpsi\kp$ with a new
model, and a total of seven structures have been observed in the
$\phi\jpsi$ system~\cite{LHCb:2021uow}. The abundant structures
observed are candidates for exotic hadrons
containing
$c\bar{c}s\bar{s}$~\cite{Ebert:2008kb,Chen:2010ze,Lu:2016cwr,Wu:2016gas,Wang:2016gxp,Chen:2016oma,Deng:2017xlb,Stancu:2009ka,Wang:2018djr,Liu:2021xje},
and provide new insight to exotic hadron spectroscopy. At BESIII, it
is possible to search for the $\phi\jpsi$ structures, such as the
$X(4140)$, $X(4274)$ and $X(4500)$, through the $\EE\to\gamma
\phi\jpsi$ process.

In this article, we present a study of the $\EE\to\gamma\phi\jpsi$
process with $6.4$~fb$^{-1}$ of data taken at $\EE$ c.m.~energies from
4.600 to 4.951~GeV~\cite{lum-4600,BESIII:2022ulv,BESIII:2015zbz}. The
$\sqrt{s}$-dependent cross section of $\EE\to\phi\chi_{c1,c2}$ is
measured and possible vector resonances are investigated. The
$\chi_{c1,c2}$ resonances are reconstructed with the $\gamma\jpsi$ and
$\jpsi\to\LL$ ($\ell=e,~\mu$) decays. We also search for the possible
$X$-state in the $\EE\to\gamma X, X\to\phi\jpsi$ process.  To increase
the number of candidates, both $\phi\to\kk$ and $\phi\to\ks\kl$ modes
are used to reconstruct $\phi$.

\section{BESIII detector and MC sample}
The BESIII detector~\cite{Ablikim:2009aa} records symmetric $\EE$ collisions provided by the BEPCII storage ring~\cite{Yu:IPAC2016-TUYA01}, which operates with a peak luminosity of $1\times10^{33}$~cm$^{-2}$s$^{-1}$
at $\EE$ center-of-mass energy 3.77~\gev.
BESIII has collected large data samples between 2.0 and 4.951~\gev~\cite{Ablikim:2019hff}. The cylindrical core of the BESIII detector covers 93\% of the full solid angle and consists of a helium-based multilayer drift chamber~(MDC), a plastic scintillator time-of-flight system~(TOF), and a CsI(Tl) electromagnetic calorimeter~(EMC), which are all enclosed in a superconducting solenoidal magnet providing a 1.0~T magnetic field. The solenoid is supported by an octagonal flux-return yoke with resistive plate counter muon chamber (MUC) system interleaved with steel. 
The charged-particle momentum resolution at $1~{\rm GeV}/c$ is
$0.5\%$, and the $dE/dx$ resolution is $6\%$ for electrons
from Bhabha scattering. The EMC measures photon energies with a
resolution of $2.5\%$ ($5\%$) at $1$~GeV in the barrel (end cap)
region. The time resolution in the TOF barrel region is 68~ps, while
that in the end cap region was 110~ps. 
The end cap TOF system was upgraded in 2015 using multi-gap resistive plate chamber technology, providing a time resolution of 60~ps~\cite{etof}. 
All the data sets with $\sqrt{s}>4.600$~GeV are taken with the new end cap TOF system.

Simulated samples produced with a {\sc geant4}-based~\cite{geant4} Monte
Carlo (MC) software, which includes the geometric description of the
BESIII detector and the detector response, are used to determine
detection efficiencies and to estimate backgrounds. The signal MC
samples of $\EE\to\phi\chi_{c1,c2}\to\gamma\phi\jpsi$ and
$\EE\to\gamma X\to\gamma\phi\jpsi$ are simulated at each c.m.~energy
point corresponding to the luminosity of data, with $\phi\to\kk$,
$\ks\kl$ and $\jpsi\to\EE$, $\MM$ being simulated according to the
branching fractions taken from the Particle Data Group
(PDG)~\cite{Zyla:2020zbs}. The inclusive MC sample includes the
production of open charm processes, the ISR production of vector
charmonium(-like) states, and the continuum processes simulated with
{\sc kkmc}~\cite{ref:kkmc}. The simulation models the beam energy
spread and ISR in the $\EE$ annihilations
with the generator {\sc kkmc}~\cite{ref:kkmc}. The known decay modes
of charmed hadrons are modelled with {\sc evtgen}~\cite{ref:evtgen}
using branching fractions taken from the PDG~\cite{Zyla:2020zbs}, and
the remaining unknown decays are modelled with {\sc
  lundcharm}~\cite{ref:lundcharm}. Final state radiation~(FSR) from
charged final state particles is incorporated using {\sc
  photos}~\cite{photos}.

\section{\boldmath Study of \texorpdfstring{\bm{$\EE\to\phi\chi_{c1,c2}$} with \bm{ $\chi_{c1,c2}\to\gamma\jpsi$}}{ee->phi chic1,chic2}\label{sec:event-selection}}
\subsection{Event Selection}
\label{common}

For candidate events of interest, the $\phi$ meson is reconstructed
from $\kk/\ks\kl$, where the $\ks$ is reconstructed from $\pp$ and the $\kl$
is missing due to the low detection efficiency with the BESIII detector.
The $\chi_{c1,c2}$
is reconstructed from $\gamma\jpsi$, and the $\jpsi$ is reconstructed from
the lepton pairs $\EE$ or $\MM$.  The following event selection
criteria are applied to both data and MC samples.

Charged tracks detected in the MDC are required to be within a polar
angle ($\theta$) range of $|\rm{cos\theta}|<0.93$ (the coverage of the MDC), where $\theta$ is
defined with respect to the $z$-axis, which is the symmetry axis of
the MDC. For charged tracks not used for $K_S^0$ reconstruction, the
distance of closest approach to the interaction point (IP) must be
less than 10\,cm along the $z$-axis, $|V_{z}|<10\cm$, and less than
1\,cm in the transverse plane, $|V_{xy}|<1\cm$, while those for
$K_S^0$ reconstruction, only a loose requirement of $|V_{z}|<20\cm$ is
applied.

Photon candidates are identified using showers in the EMC.  The
deposited energy of each shower must be greater than 25~MeV in the
barrel region ($|\cos \theta|< 0.80$) and greater than 50~MeV in the
end cap region ($0.86 <|\cos \theta|< 0.92$). To exclude the showers
that originate from charged tracks, the angle between the position of
each shower in the EMC and the closest extrapolated charged track must
be greater than 10 degrees. To suppress the electronic noise and the showers
unrelated to the event, the difference between the EMC time and the
event start time is required to be within [0, 700]\,ns.

For each event, the lepton pair ($\LL$) from $\jpsi$ decays and the
kaons from $\phi$ decays can be effectively distinguished by their
momenta in the lab-frame. The tracks with momentum larger than 1$\gevc$
are assigned as leptons, and the amount of deposited energy in the EMC
is further used to separate the muons from electrons.  For both muon
candidates, the deposited energy in the EMC is required to be less
than 0.4$\gev$, while it is required to be greater than 1.0$\gev$ for
electrons. For the tracks with momentum less than 1$\gevc$, particle
identification~(PID), which combines measurements of the energy
deposited in the MDC~(d$E$/d$x$) and the flight time in the TOF to
form likelihoods $\mathcal{L}(h)~(h=K,\pi)$ for each hadron $h$
hypothesis, is used.  Tracks are identified as kaons when the kaon
hypothesis has a larger likelihood than the pion hypothesis
($\mathcal{L}(K)>\mathcal{L}(\pi)$ and $\mathcal{L}(K)>0$).

\subsubsection{\boldmath 3-track events with $\phi\to\kp\km$\label{subsec:3trk}}

For the $\phi\to\kk$ channel, one of the kaons could be missing due to
an inefficiency. Together with the lepton pair from the $\jpsi$ decay,
there are three charged particles remaining in each signal event
(referred to as the 3-track events). Two of the charged tracks are
assigned as the lepton pair and the third as a kaon. The PID
likelihood of the kaon is required to satisfy
$\mathcal{L}(K)>\mathcal{L}(\pi)$ and $\mathcal{L}(K)>0$, and at least
one photon candidate is also required.

To improve the resolution and suppress background, a one-constraint
(1C) kinematic fit is applied to the 3-track event by constraining the
mass of the missing particle to the kaon nominal mass
($M(K^\mp_\mathrm{miss})=\sqrt{(P_{\EE}-P_{\gamma K^\pm \LL})^2}$)
inferred from the four momentum conservation. For the events with multiple
photons in the final state, the combination of $\gamma K^\pm
K^\mp_\mathrm{miss}\LL$ with the smallest $\chi^2$ from the kinematic
fit is retained, and $\chi^2<20$ is required.

To reduce the $\pi$ misidentification background in the $\MM$ channel,
the MUC is used to identify muons. At least one of muon candidate
should have a hit depth $>30\cm$ in the MUC. To veto the radiative
Bhabha background in $J/\psi \to e^+e^-$ events, the polar angle of
$e^+$ is required to satisfy $\cos(\theta_{e^+})<0.85$.

\begin{figure}[H]
	\centering
	\includegraphics[width=0.48\linewidth]{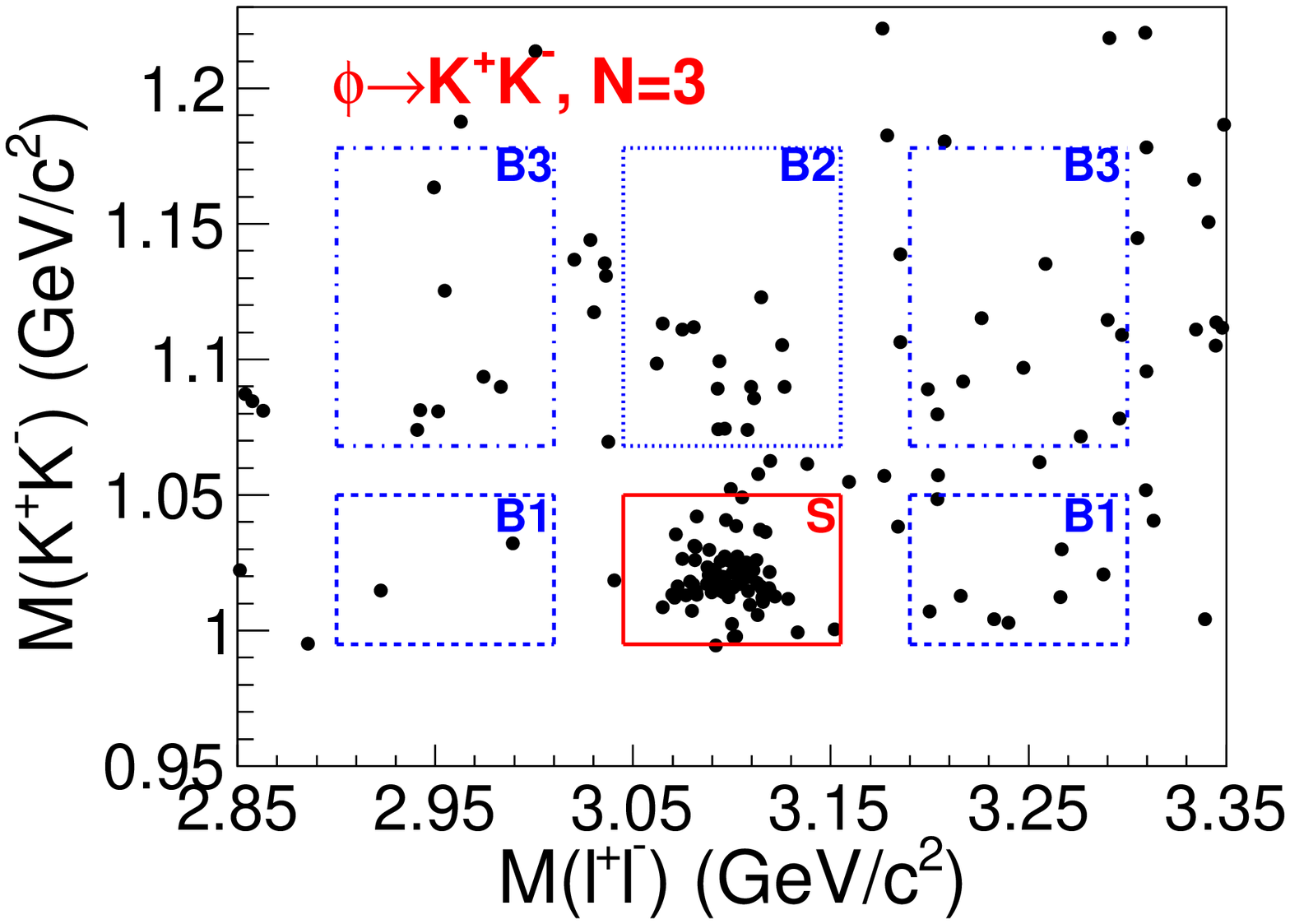}
	\includegraphics[width=0.48\linewidth]{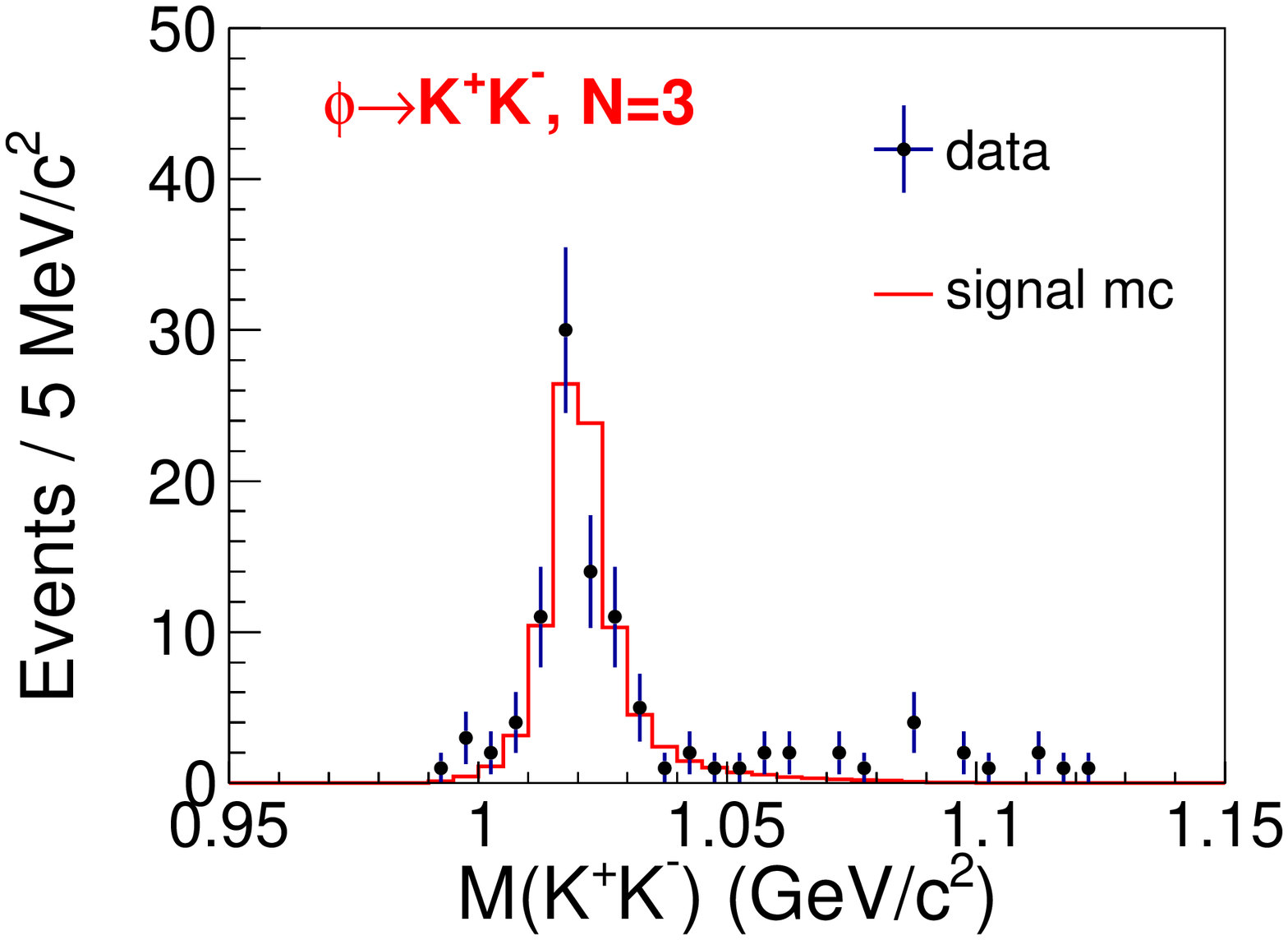}
	\vfill
	\includegraphics[width=0.48\linewidth]{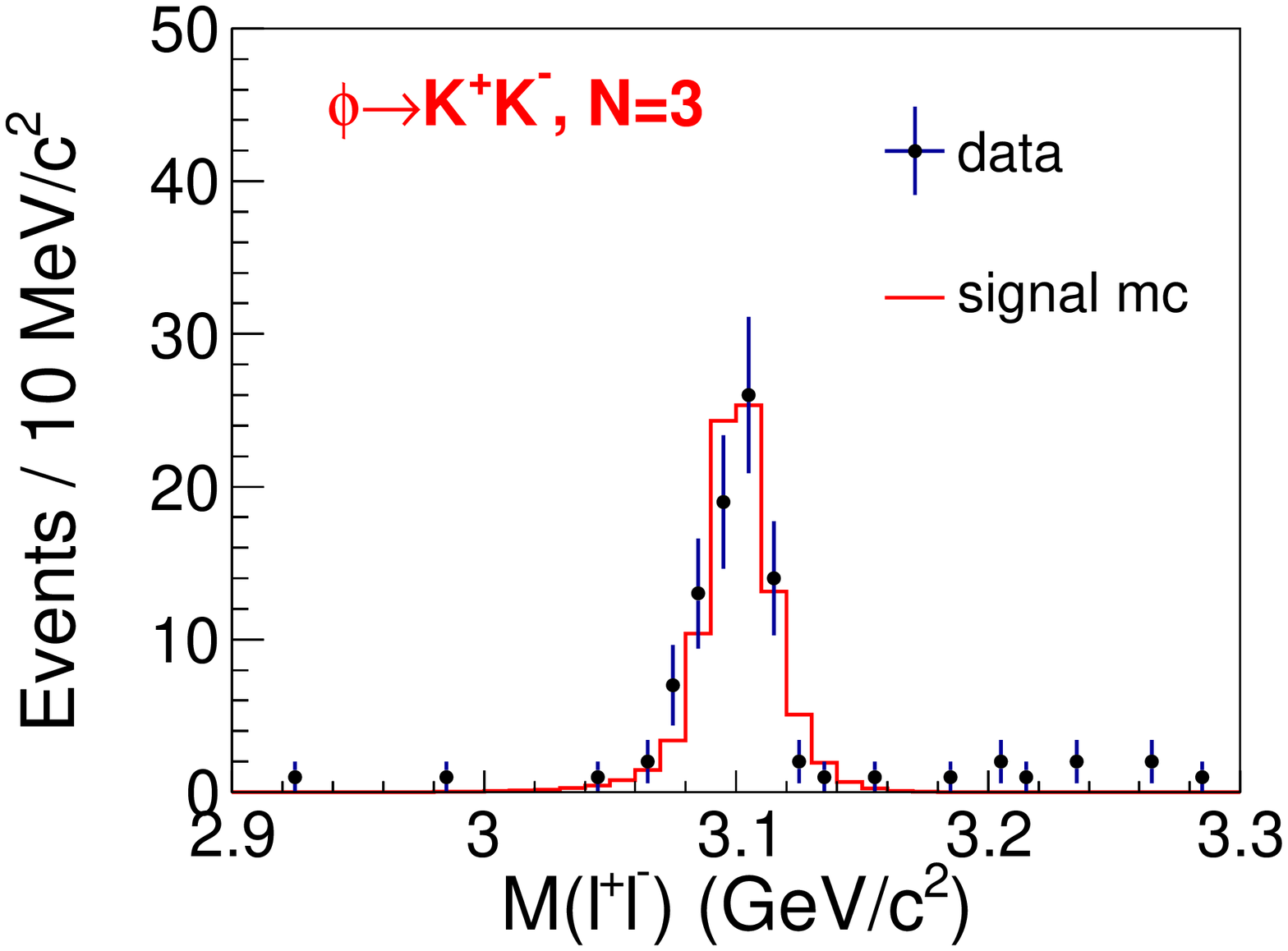}
	\includegraphics[width=0.48\linewidth]{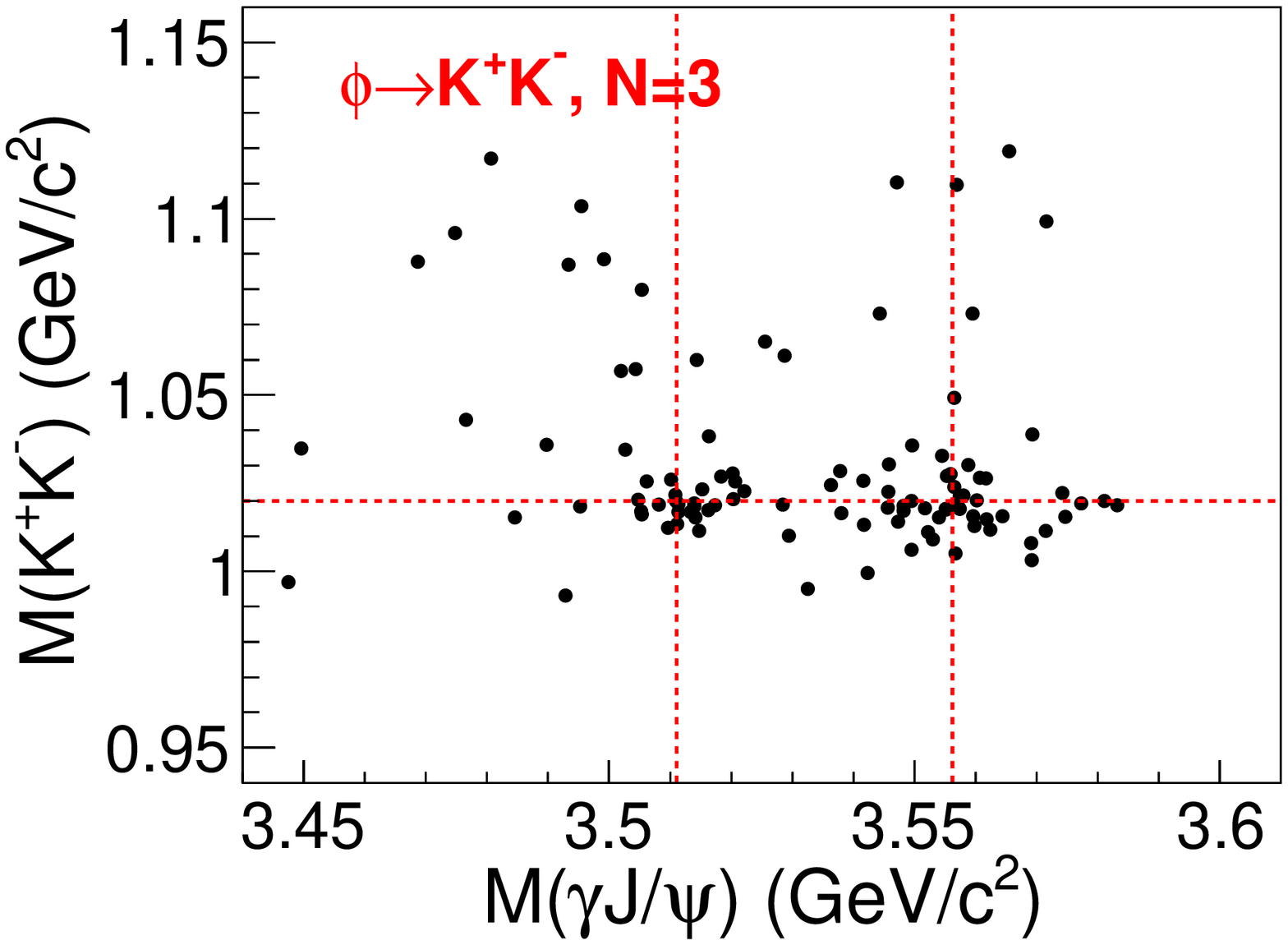}
	\caption{The 2-dimensional distribution of the $M(\kk)$ versus
          $M(\LL)$ (upper left), the invariant mass distributions of the $M(\kk)$ (upper right), $M(\LL)$ (bottom left) and the 2-dimensional
          distribution of the $M(\kk)$ versus $M(\gamma\jpsi)$ (bottom right) for the 3-track events in the $\phi\to\kk$ mode. Dots with and without error bars are the full data, the red histograms
          are the signal MC. In the upper left panel, the red solid box is the $\phi \jpsi$ signal region
          (S), the blue dashed, dotted and dash-dotted boxes indicate
          the $\phi$ non-$\jpsi$ (B1), $\jpsi$ non-$\phi$ (B2) and
          non-$\phi$ non-$\jpsi$ (B3) sideband regions,
          respectively. The vertical (horizontal) dashed lines in the bottom right panel 
          are central masses of $\chi_{c1}/\chi_{c2}$ ($\phi$). \label{fig:inv-kk3}}
\end{figure}

After imposing these selection criteria, there is a clear $\phi \jpsi$
event cluster in the 2-dimensional distribution of the $M(\kk)$ versus
$M(\LL)$ as shown in \cref{fig:inv-kk3}.  The $\phi$ and $\jpsi$
mass windows are defined as $0.995\gevcc<M(\kk)<1.050\gevcc$ (the mass
resolution is $10\mevcc$) and $3.045\gevcc<M(\LL)<3.155\gevcc$ (the
mass resolution is $17\mevcc$), respectively.  To estimate the
non-$\phi$ and non-$\jpsi$ backgrounds, the $\phi$ sideband region is
defined as $1.068\gevcc<M(\kk)<1.178\gevcc$, which is twice as wide as
the $\phi$ signal region, while $2.90\gevcc<M(\LL)<3.01\gevcc$ and
$3.19\gevcc<M(\LL)<3.30\gevcc$ are defined as the $\jpsi$ sideband
region, which is twice as wide as the $\jpsi$ signal region (see
\cref{fig:inv-kk3}).  \cref{fig:inv-kk3} also shows the
invariant mass distributions of the $M(\kk)$, $M(\LL)$ and the
2-dimensional distribution of the $M(\kk)$ versus $M(\gamma\jpsi)$,
where the events are clustered in the $\phi$ and $\chi_{c1,c2}$ mass
regions.

\subsubsection{\boldmath 4-track events with $\phi\to\kk$}

For a candidate event with $\kk\LL$ detected (referred to as 4-track
events), the photon candidate is always ignored and not required to be detected in order
to improve the efficiency.  At least four charged tracks are required,
two of which are assigned as the lepton pair and the remaining tracks
as kaons. Both kaons are required to be identified. A similar 1C
kinematic fit is performed by constraining the mass of the missing
particle to be a photon inferred from the four momentum conservation,
i.e. $M(\gamma_\mathrm{miss})=\sqrt{(P_{\EE}-P_{\kk\LL})^2}$. The
kinematic fit $\chi^2$ is required to be $\chi^2<35$. The same MUC
requirement as for 3-track events is applied to the muon candidates to
suppress pion background.

\begin{figure}[H]
	\centering
	\includegraphics[width=0.48\linewidth]{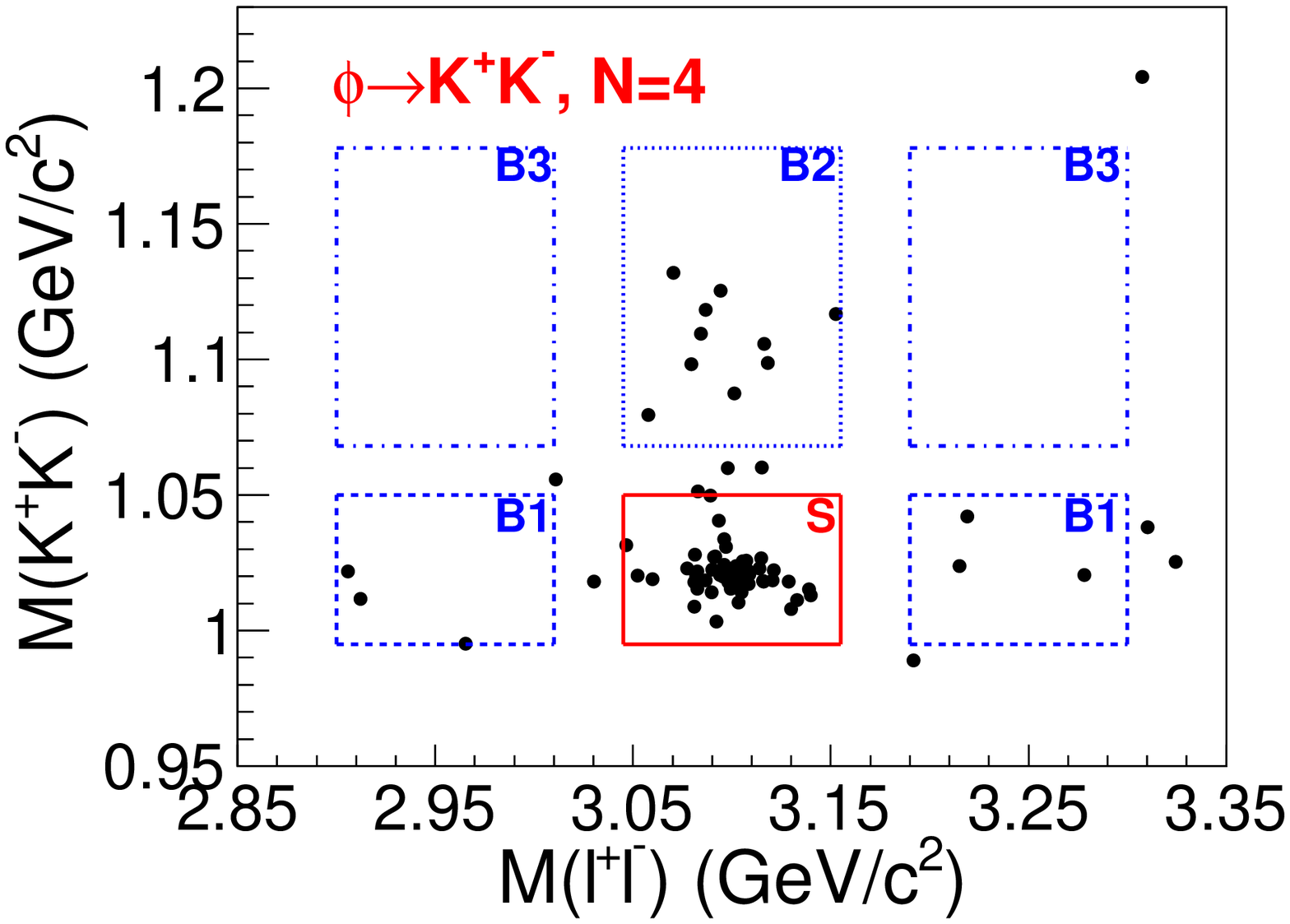}
	\includegraphics[width=0.48\linewidth]{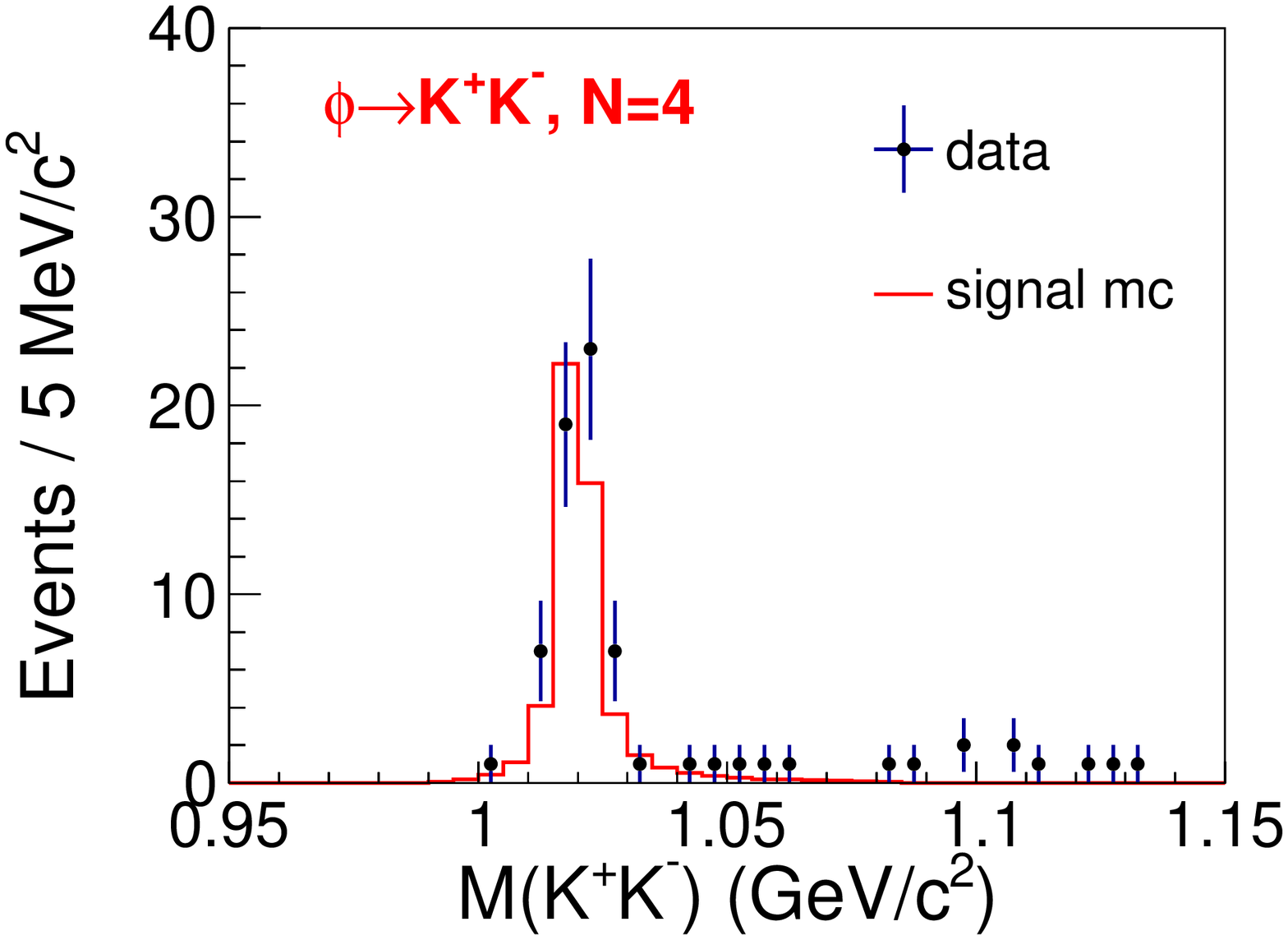}
	\vfill
	\includegraphics[width=0.48\linewidth]{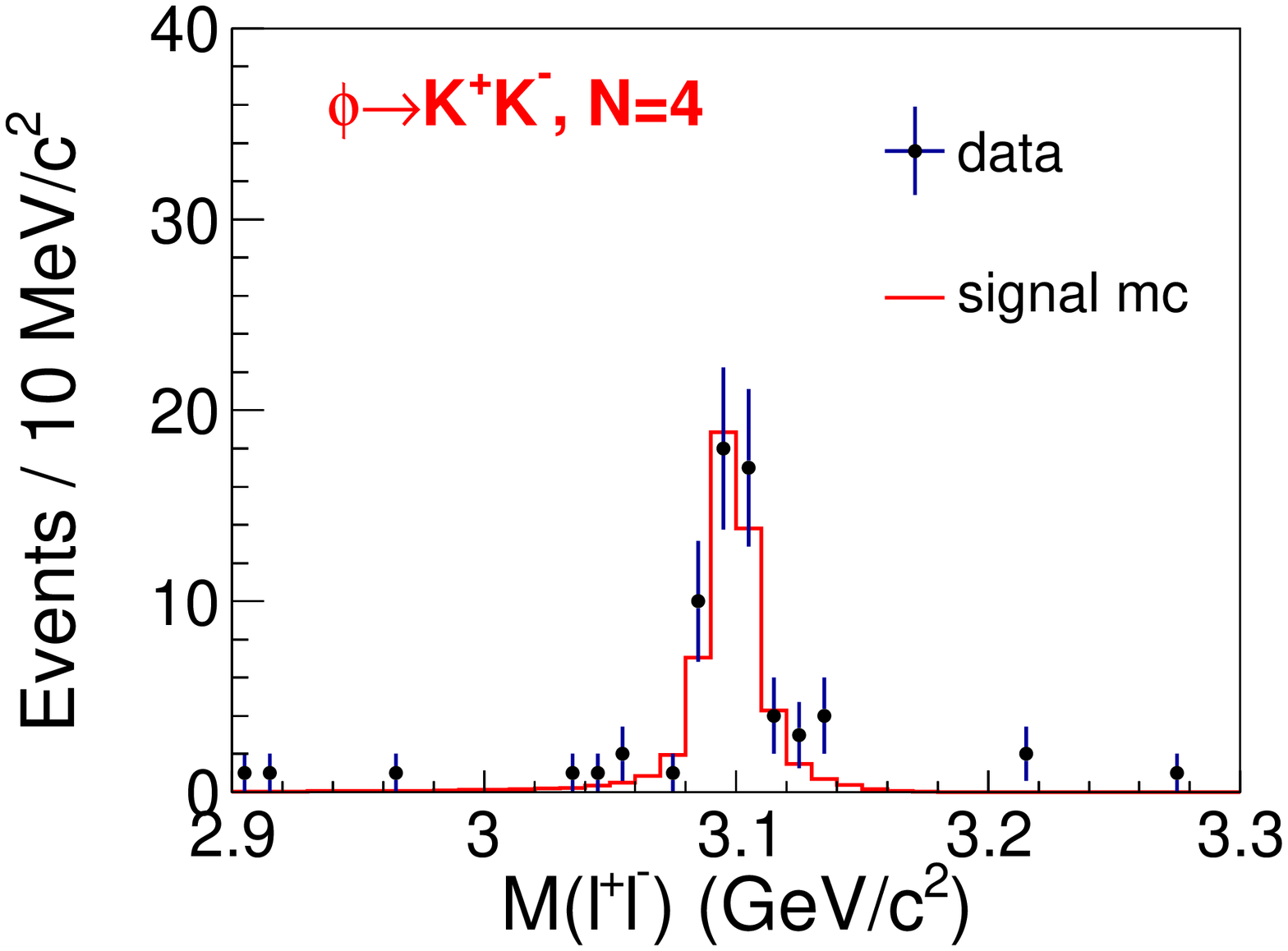}
	\includegraphics[width=0.48\linewidth]{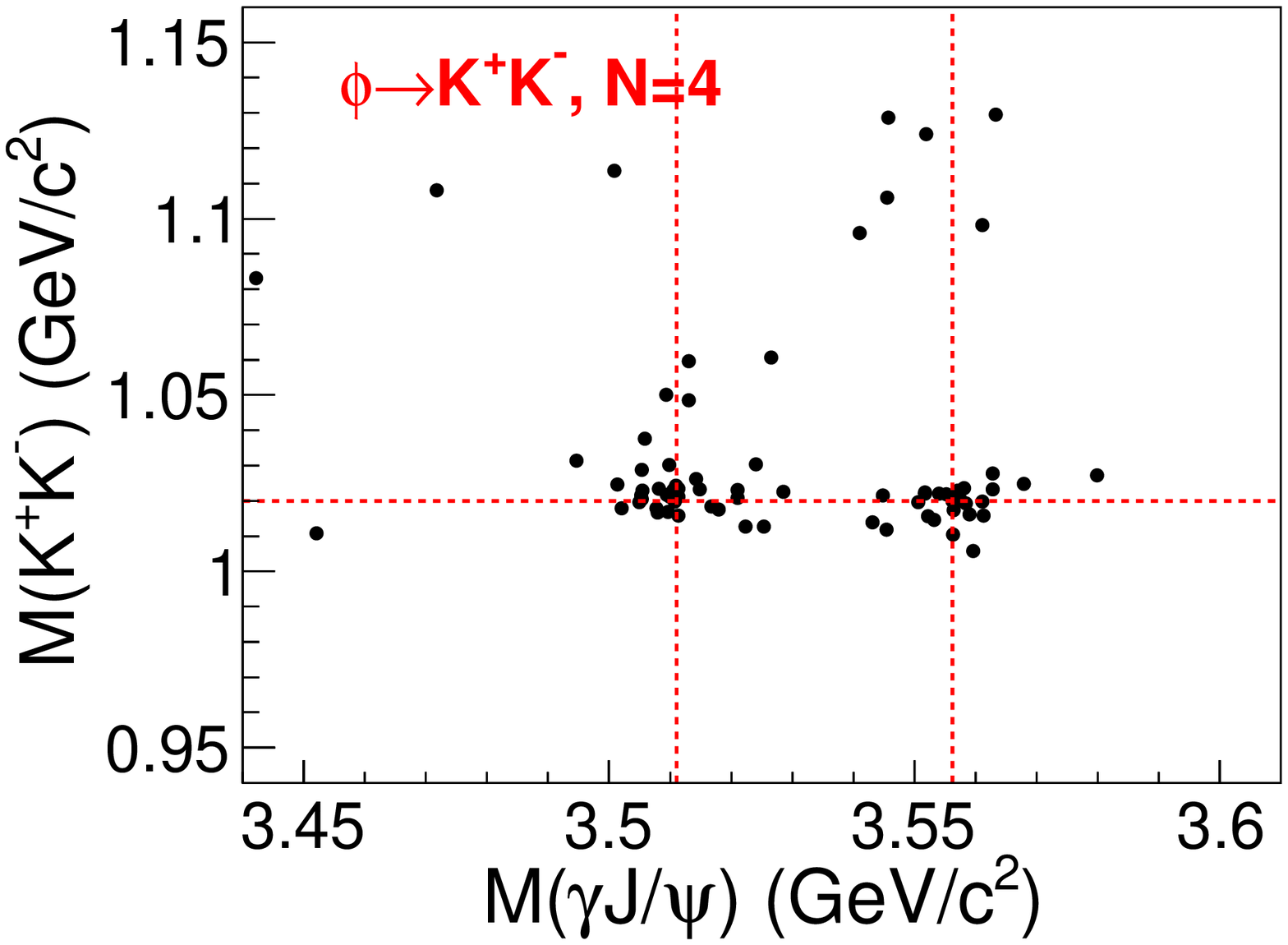}
	\caption{The 2-dimensional distribution of the $M(\kk)$ versus
		$M(\LL)$ (upper left), the invariant mass distributions of the $M(\kk)$ (upper right), $M(\LL)$ (bottom left) and the 2-dimensional
		distribution of the $M(\kk)$ versus $M(\gamma\jpsi)$ (bottom right) for the 4-track events in the $\phi\to\kk$ mode. Dots with and without error bars are the full data, the red histograms
		are the signal MC. In the upper left panel, the red solid box is the $\phi \jpsi$ signal region
		(S), the blue dashed, dotted and dash-dotted boxes indicate
		the $\phi$ non-$\jpsi$ (B1), $\jpsi$ non-$\phi$ (B2) and
		non-$\phi$ non-$\jpsi$ (B3) sideband regions,
		respectively. The vertical (horizontal) dashed lines in the bottom right panel 
		are central masses of $\chi_{c1}/\chi_{c2}$ ($\phi$). \label{fig:inv-kk4}}
\end{figure}

\cref{fig:inv-kk4} shows the 2-dimensional
distribution of the $M(\kk)$ versus $M(\LL)$, the invariant mass distributions of the $M(\kk)$, $M(\LL)$, and the 2-dimensional
distribution of the $M(\kk)$ versus $M(\gamma\jpsi)$ after the above selections. Clear $\phi$ and $\jpsi$ resonance peaks are shown in the
$M(\kk)$ and $M(\LL)$ distributions, and the events are clearly
clustered in the $\chico$ and $\chict$ mass regions in the
$M(\gamma\jpsi)$ distribution, where the same mass window
requirements defined in~\cref{subsec:3trk} have been applied.

\subsubsection{\boldmath Events with $\phi\to\ks\kl$}

The events from $\phi\to\kskl$ decay are reconstructed with
$\ks\to\pp$. The neutral $\kl$ candidate has a long lifetime 
and is not detected. We require at least
four charged tracks to be detected in each event, two of which are
assigned as the lepton pair and the remaining charged tracks are
pions. The $\ks$ candidates are reconstructed from two oppositely
charged pions satisfying $|V_{z}|<$ 20~cm.  There is no PID
requirement for the charged pions, and they are constrained to
originate from a common secondary decay vertex. The decay length of
the $K^0_S$ candidate is required to be greater than twice the vertex
resolution away from the IP to suppress the non-$\ks$ background. After
the vertex fit, we set a mass window of
$0.490\gevcc<M(\pp)<0.505\gevcc$ for the $\ks$ candidate (the mass
resolution is $5\mevcc$). At least one good photon candidate is also required in each event.

A 1C kinematic fit is applied to each event, with the mass of the
missing particle constrained to the $\kl$ nominal mass inferred from the four
momentum conservation,
i.e. $M({\kl}_\mathrm{miss})=\sqrt{(P_{\EE}-P_{\gamma\ks\LL})^2}$. The
kinematic fit $\chi^2$ is required to be $\chi^2<20$.

\begin{figure}[H]
	\centering
	\includegraphics[width=0.48\linewidth]{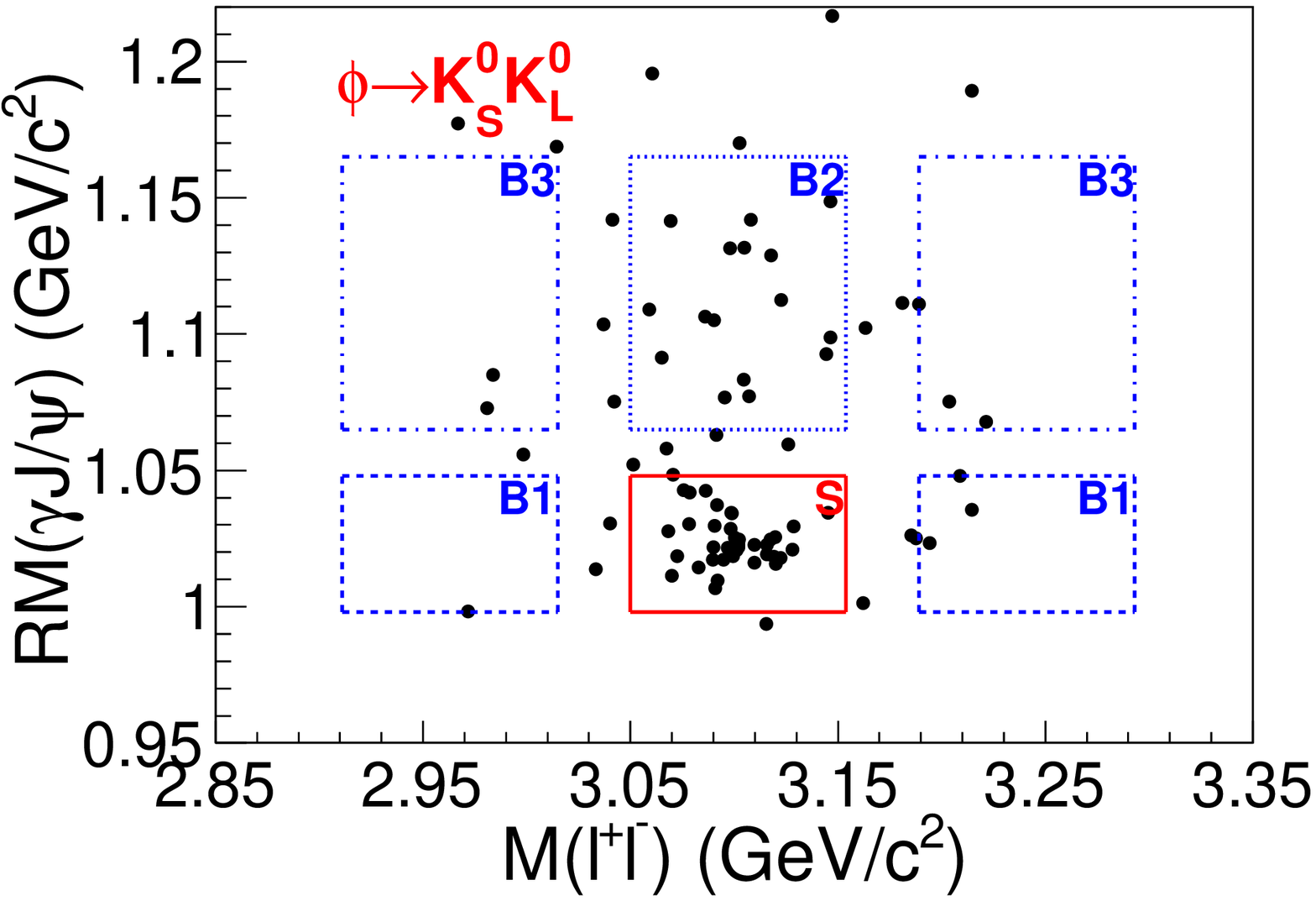}
	\includegraphics[width=0.48\linewidth]{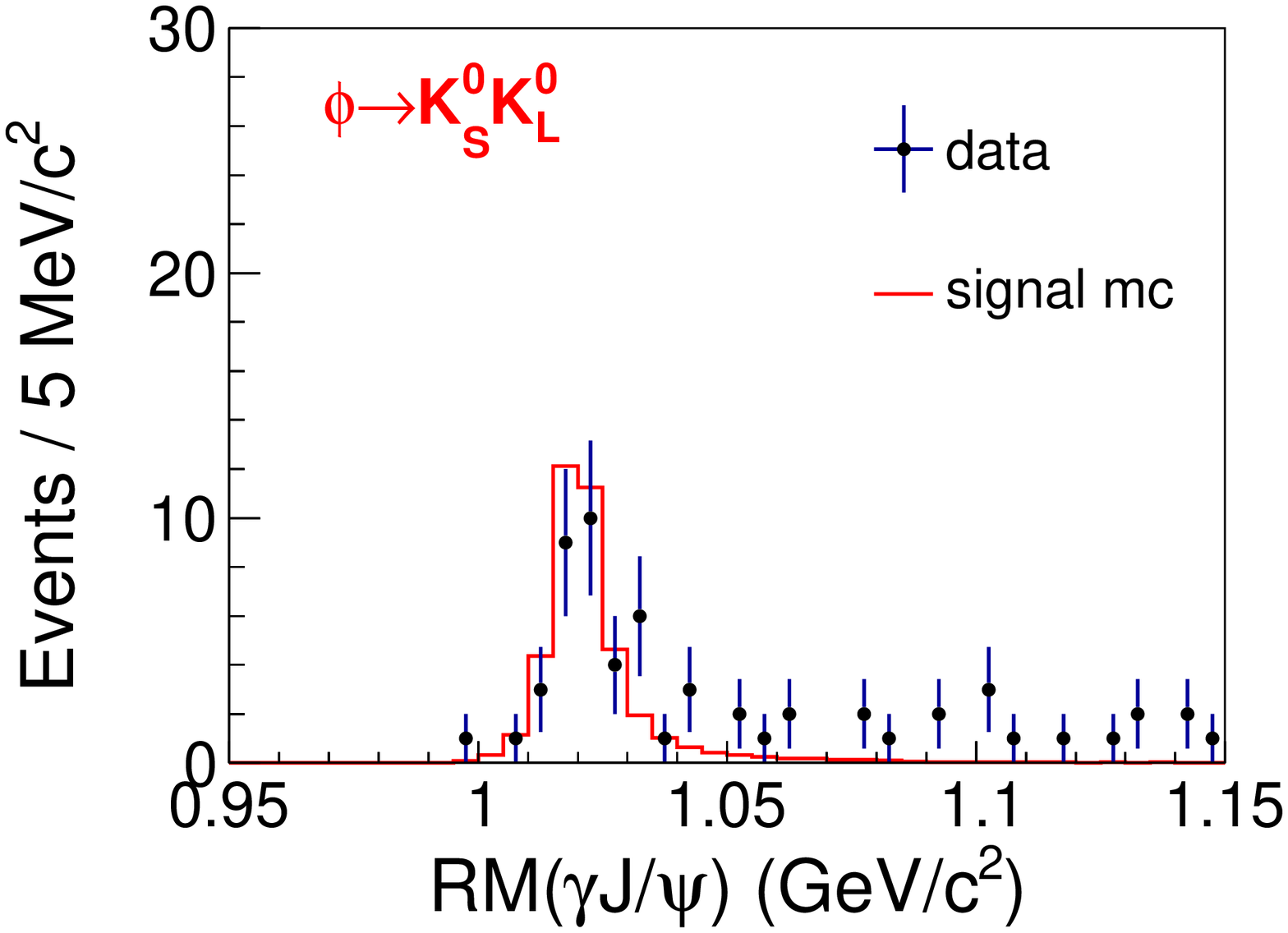}
	\vfill
	\includegraphics[width=0.48\linewidth]{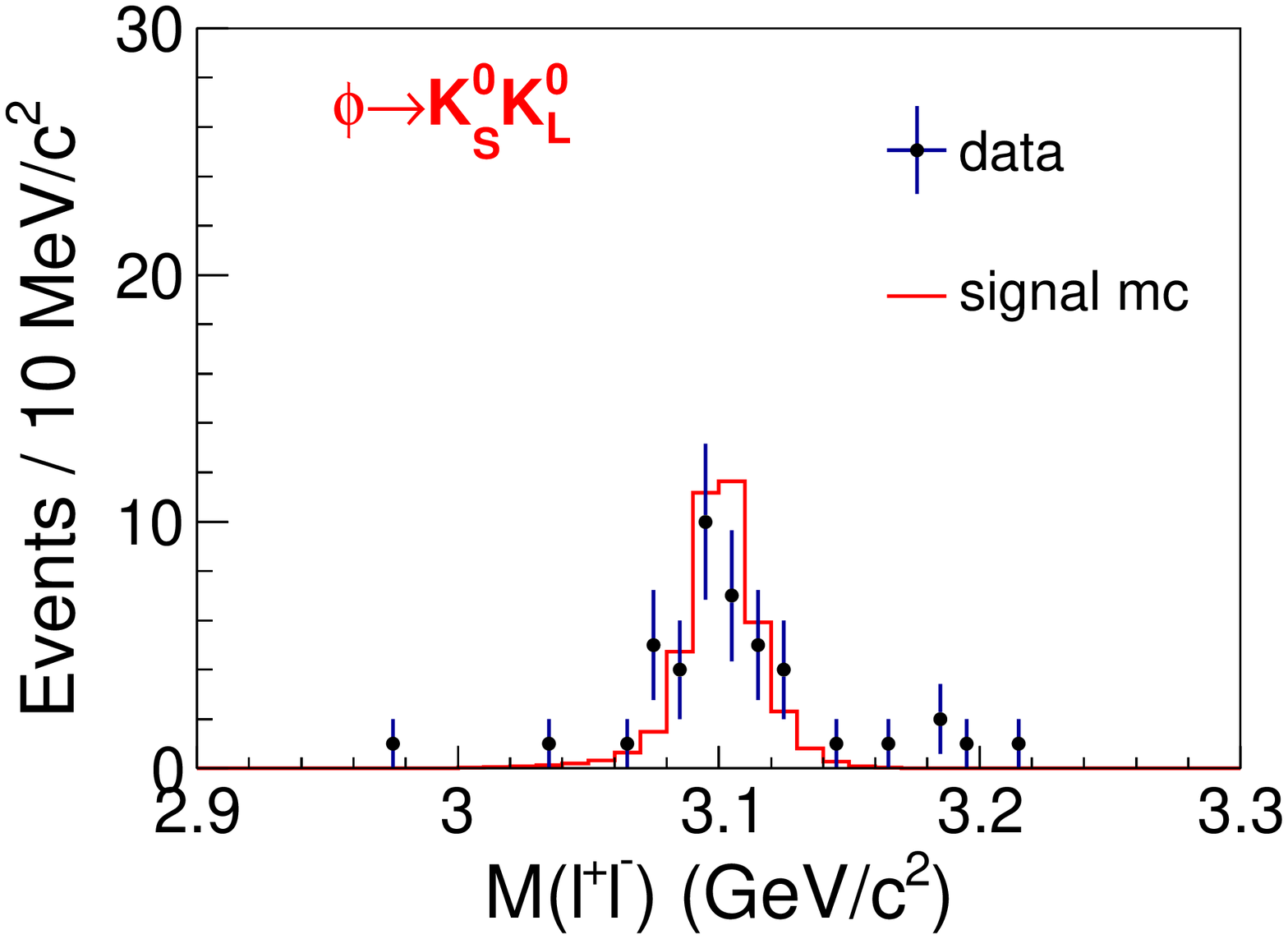}
	\includegraphics[width=0.48\linewidth]{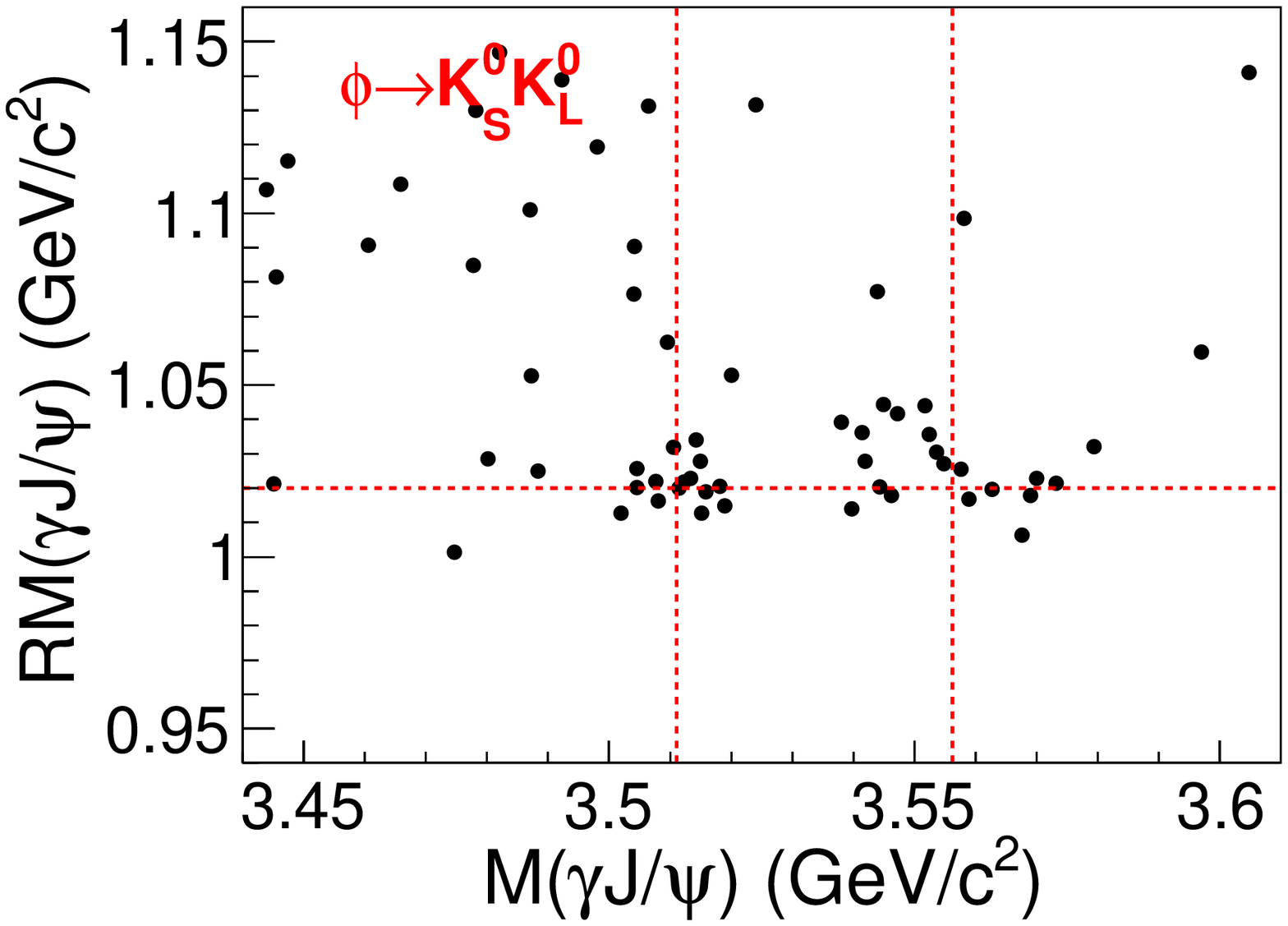}

	\caption{The 2-dimensional distribution of the $RM(\gamma\jpsi)$
          versus $M(\LL)$ (upper left), the invariant mass distributions of the $RM(\gamma\jpsi)$ (upper right), $M(\LL)$ (bottom left) and the 2-dimensional
          distribution of the $RM(\gamma\jpsi)$ versus $M(\gamma\jpsi)$ (bottom right) in the $\phi\to\kskl$ mode. Dots with and without error bars are the full data, the red histograms
          are the signal MC. In the upper left panel, the red solid box is the $\phi \jpsi$ signal region
          (S), the blue dashed, dotted and dash-dotted boxes indicate
          the $\phi$ non-$\jpsi$ (B1), $\jpsi$ non-$\phi$ (B2) and
          non-$\phi$ non-$\jpsi$ (B3) sideband regions,
          respectively. The vertical (horizontal) dashed lines in the bottom right panel 
          are central masses of $\chi_{c1}/\chi_{c2}$ ($\phi$). \label{fig:inv-kskl}}
\end{figure}

After applying these requirements, clear $\phi$ and $\jpsi$ resonance
peaks are observed in the $RM(\gamma\jpsi)$ and $M(\LL)$ mass
distributions as shown in \cref{fig:inv-kskl}, where
$RM(\gamma\jpsi)=\sqrt{(P_{\EE}-P_{\gamma\jpsi})^2}$ is the recoil
mass from the $\gamma\jpsi$ system.  We define the $\phi$ and $\jpsi$
mass windows as $0.998\gevcc<RM(\gamma\jpsi)<1.048\gevcc$ (the mass
resolution is $9\mevcc$) and $3.050\gevcc<M(\LL)<3.154\gevcc$ (the
mass resolution is $16\mevcc$), respectively, as shown in
\cref{fig:inv-kskl}.  The $\phi$ sideband is defined as
$1.065\gevcc<RM(\gamma\jpsi)<1.165\gevcc$, which is twice as wide as
the $\phi$ signal region, and the $\jpsi$ sidebands are defined as
$2.911\gevcc<M(\LL)<3.015\gevcc$ and $3.189\gevcc<M(\LL)<3.293\gevcc$,
which is twice as wide as the $\jpsi$ signal region.
\cref{fig:inv-kskl} also shows the 2-dimensional distribution
of the $RM(\gamma\jpsi)$ versus $M(\gamma\jpsi)$, where the events are clearly
clustered in the $\chico$ and $\chict$ mass regions in the
$M(\gamma\jpsi)$ distribution.

\subsection{Cross section measurement\label{xs}}

Based on the event selection, the $\chico$ and $\chict$ signals
are observed from both the $\phi\to\kk$ and $\kskl$ modes. To
determine the signal yields, an unbinned maximum likelihood fit is
performed to the $M(\gamma\jpsi)$ distribution in the $\phi\to\kk$ and
$\kskl$ modes simultaneously.  In the fit at each c.m.~energy, the
signal probability-density-function (PDF) is described by a MC-simulated shape
convolved with a Gaussian function, which models the resolution difference between data
and MC simulation. 
The MC-simulated shape is a weighted sum of the simulations at each c.m.~energy,
which has already taken into account the c.m.~energy and decay modes dependence for
the resolution.
The Gaussian parameters are determined from the fit to the full dataset
which has higher statistics. A linear function is used to describe the
background. The two modes share the same $\phi\chi_{c1,c2}$
production cross section at the same c.m.~energy. The selection
efficiencies and branching fractions of the $\phi\to\kk/\kskl$ modes at
each c.m.~energy are included in the fit. 

\begin{figure}[H]
	\centering
	\includegraphics[width=\linewidth]{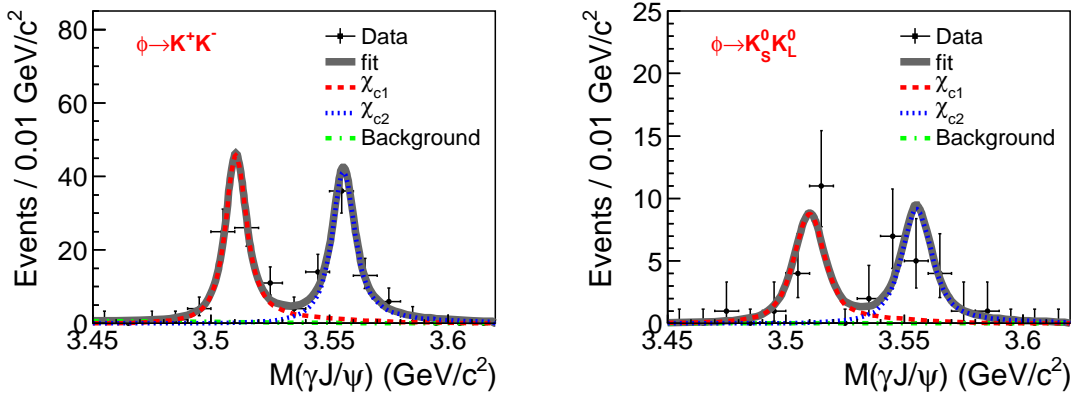}
	\caption{Sum of the simultaneous fits to $M(\gamma\jpsi)$
          distribution for the full data sets. Dots with
          error bars are the data, the solid curves are the fit results, the
          red dashed, blue dotted, and green dash-dotted lines are the
          $\chico$, $\chict$ and the background shape,
          respectively.\label{fig:sim-fit}}
\end{figure}

\cref{fig:sim-fit} shows the fit result for the full dataset from
$\sqrt{s}=4.600\gev$ to $4.951\gev$, and the corresponding plots
at each individual c.m.~energy are shown in~\cref{fig:fit-result} of Appendix~\ref{app-a}.
The statistical significance is
estimated by comparing the fit likelihoods with and without the
$\chi_{c1,c2}$ signal. In addition to the nominal fit, the fits by
changing the background shape and the fit range have also been
performed. In all the cases, the significance of the $\chi_{c1,c2}$ is
found to be greater than $10\sigma$, by comparing the difference of
log-likelihoods $\Delta(-2\ln\mathcal{L})=137(131)$ for the
$\chico(\chict)$ and taking into account the change of the number of
degrees of freedom ($\Delta \rm{d.o.f.}=5$).

The Born cross section of $\EE\to\phi\chi_{c1,c2}$ at c.m.~energy
$\sqrt{s}$ is calculated with

\begin{equation}
	\sigma^\mathrm{B}(\sqrt{s})  = \frac{N^\mathrm{fit}}{\mathcal{L}_\mathrm{int}(1+\delta)\frac{1}{|1-\Pi|^2}\mathcal{B}},
\end{equation}
where $N^\mathrm{fit}$ is the number of fitted events for the
$\phi\chi_{c1,c2}$, which is equal to the number of the $\phi\chi_{c1,c2}$
events in data divided by the efficiency and branching fraction of $\phi$, $\mathcal{L}_\mathrm{int}$ is the integrated luminosity,
$(1+\delta)$ is the ISR correction factor obtained from {\sc kkmc},
$\frac{1}{|1-\Pi|^2}$ is the vacuum polarization factor~\cite{vacuum},
and $\mathcal{B}$ is the product of the branching fraction for
$\chi_{c1,c2}\to\gamma\jpsi$ and $\jpsi\to\LL$. The Born cross
sections of $\EE\to\phi\chico$ and $\EE\to\phi\chict$ at each
c.m.~energy are listed in \cref{tab:xsec-c1,tab:xsec-c2}, respectively. 
In case the signal significance is less than $3\sigma$,
an upper limit of the Born cross section ($\sigma^\mathrm{U.L.}$) at the 90\% confidence level (C.L.) is also reported.
The upper limits of $\phi\chico$ and $\phi\chict$ yields are
estimated via a Bayesian approach~\cite{Zyla:2020zbs}. A likelihood scan $L(n)$ is performed with various assumptions for the number of 
signal events ($n$) in the fit. The systematic uncertainty is also considered by smearing the likelihood distribution with a Gaussian function with
width equal to the systematic uncertainty. The upper limit of
$N^\mathrm{U.L.}$ at the 90\% C.L.
corresponds to $\int_{0}^{N^\mathrm{U.L.}} L(x)
dx/\int_{0}^{\infty} L(x) dx=0.9$.
The detection efficiencies of $\phi\chi_{c1,c2}$ events depend on the $\EE$ c.m.~energy.
With the increasing c.m.~energy, charged kaons have higher momentum and are thus much more efficient
to be detected, while for the $\ks\kl$ channel, due to more ISR events the reconstruction efficiency
whereas decreases (the $\pp$ from $\ks$ decay already have sufficient momentum to be detected and
are not sensitive to c.m.~energy).

\begin{table}[H]
	\renewcommand\arraystretch{1.5}
	\tabcolsep=0.5pt
	\centering

		\begin{tabular}{cccccccc}
			\hline
			\hline

			$\sqrt{s}~(\mathrm{GeV})$  & $\mathcal{L}_\mathrm{int}~(\mathrm{pb}^{-1})$ & $\epsilon_{\kk}^3$ & $\epsilon_{\kk}^4$ & $\epsilon_{\ks\kl}$ & $\frac{1+\delta}{|1-\Pi|^2}$ & $N^\mathrm{fit}$ & $\sigma^\mathrm{B}~(\mathrm{pb})$ \\
			
			\hline
			
			4.600  & 586.9 & 0.261 & 0.092 & 0.229 & 0.88 & $56.0^{+18.2}_{-15.1}$ &  $2.63^{+0.86}_{-0.71}\pm0.20~(5.8\sigma)$  \\
			
			4.612  & 103.8 & 0.257 & 0.101 & 0.223 & 0.90 & $13.3^{+9.4}_{-6.3}~(<29.8)$ &  $3.45^{+2.43}_{-1.64}\pm0.25~(<7.7)$  \\
			
			4.628  & 521.5 & 0.247 & 0.120 & 0.224 & 0.92 & $54.7^{+17.3}_{-14.3}$ &  $2.77^{+0.88}_{-0.73}\pm0.20~(3.3\sigma)$  \\
			
			4.641  & 552.4 & 0.245 & 0.133 & 0.222 & 0.94 & $60.4^{+18.6}_{-15.4}$ &  $2.83^{+0.87}_{-0.72}\pm0.20~(5.3\sigma)$  \\
			
			4.661  & 529.6 & 0.233 & 0.156 & 0.220 & 0.97 & $21.5^{+11.3}_{-8.4}$ &  $1.02^{+0.53}_{-0.39}\pm0.07~(3.6\sigma)$  \\
			
			4.682  & 1669.3 & 0.219 & 0.176 & 0.218 & 0.99 & $79.4^{+21.5}_{-18.5}$ &  $1.17^{+0.32}_{-0.27}\pm0.08~(5.6\sigma)$  \\
			
			4.699  & 536.5 & 0.208 & 0.188 & 0.213 & 1.02 & $34.7^{+14.3}_{-11.3}$ &  $1.54^{+0.63}_{-0.50}\pm0.11~(4.4\sigma)$  \\

			4.740  & 164.3 & 0.188 & 0.215 & 0.210 & 1.07 & $20.2^{+10.4}_{-9.8}~(<37.5)$ &  $2.80^{+1.44}_{-1.35}\pm0.19~(<5.2)$  \\
			
			4.750  & 367.2 & 0.181 & 0.214 & 0.208 & 1.09 & $22.2^{+12.2}_{-9.3}~(<42.0)$ &  $1.35^{+0.74}_{-0.57}\pm0.10~(<2.5)$  \\

			4.781  & 512.8 & 0.163 & 0.222 & 0.201 & 1.13 & $0.0^{+1.3}_{-0.0}~(<13.5)$&  $0.0^{+0.23}_{-0.0}\pm0.02~(<0.6)$  \\
			
			4.843  & 527.3 & 0.142 & 0.228 & 0.188 & 1.24 & $4.5^{+6.0}_{-3.1}~(<17.2)$ &  $0.17^{+0.22}_{-0.12}\pm0.01~(<0.6)$  \\

			4.918  & 208.1 & 0.115 & 0.214 & 0.167 & 1.41 & $15.3^{+10.7}_{-7.3}~(<34.3)$ &  $1.27^{+0.89}_{-0.61}\pm0.09~(<2.8)$  \\
			
			4.951  & 160.4 & 0.106 & 0.208 & 0.155 & 1.50 & $5.3^{+7.1}_{-3.7}~(<20.4)$ &  $0.53^{+0.72}_{-0.37}\pm0.04~(<2.1)$  \\
			
			\hline
			\hline
						
		\end{tabular}
	\caption{The Born cross section $\sigma^\mathrm{B}$ for
          $\EE\to\phi\chico$ at each c.m.~energy ($\sqrt{s}$).
          The numbers in the brackets are the signal significances or
          upper limits $\sigma^\mathrm{U.L.}$ at the 90\% C.L. in case
          the signal significance is less than $3\sigma$.
          The table also includes integrated luminosity
          $\mathcal{L}_\mathrm{int}$, detection efficiency
          $\epsilon_{\kk}^3$, $\epsilon_{\kk}^4$ and
          $\epsilon_{\ks\kl}$ for the 3-track events of the $\phi\to\kk$
          mode, 4-track events of the $\phi\to\kk$ mode and the events of
          the $\phi\to\ks\kl$ mode, respectively, the product of radiative
          correction factor and vacuum polarization factor
          $\frac{1+\delta}{|1-\Pi|^2}$ and the number of fitted
          events $N^\mathrm{fit}$ (also the corresponding upper limit 
          $N^\mathrm{U.L.}$ at the 90\% C.L. in case the signal significance is less than $3\sigma$). 
          The first uncertainty is statistical and the
          second is systematic. \label{tab:xsec-c1}}
\end{table}


\begin{table}[H]
	\renewcommand\arraystretch{1.5}
	\tabcolsep=0.5pt
	\centering

			\begin{tabular}{cccccccc}
				\hline
				\hline

$\sqrt{s}~(\mathrm{GeV})$ & $\mathcal{L}_\mathrm{int}(\mathrm{pb}^{-1})$ & $\epsilon_{\kk}^3$ & $\epsilon_{\kk}^4$ & $\epsilon_{\ks\kl}$ & $\frac{1+\delta}{|1-\Pi|^2}$ & $N^\mathrm{fit}$ & $\sigma^\mathrm{B}~(\mathrm{pb})$ \\
				
				\hline
4.600  & 586.9 & 0.253 & 0.031 & 0.226 & 0.73 & $26.7^{+14.6}_{-11.0}$ &  $2.73^{+1.49}_{-1.13}\pm0.27~(3.6\sigma)$  \\
				
4.612  & 103.8 & 0.257 & 0.047 & 0.215 & 0.75 & $9.8^{+8.9}_{-5.6}~(<26.6)$ &  $5.50^{+5.02}_{-3.14}\pm0.61~(<15.0)$  \\
				
4.628  & 521.5 & 0.261 & 0.070 & 0.222 & 0.76 & $15.1^{+11.0}_{-7.8}~(<34.0)$ &  $1.67^{+1.22}_{-0.86}\pm0.17~(<3.8)$  \\
				
4.641  & 552.4 & 0.263 & 0.086 & 0.225 & 0.77 & $24.4^{+13.9}_{-10.9}$ &  $2.52^{+1.44}_{-1.12}\pm0.27~(3.6\sigma)$  \\
				
4.661  & 529.6 & 0.259 & 0.112 & 0.230 & 0.80 & $45.5^{+15.6}_{-12.7}$ &  $4.71^{+1.61}_{-1.32}\pm0.42~(6.4\sigma)$  \\
				
4.682  & 1669.3 & 0.255 & 0.137 & 0.234 & 0.84 & $136.3^{+26.9}_{-24.2}$ &  $4.26^{+0.84}_{-0.76}\pm0.42~(9.5\sigma)$  \\
				
4.699  & 536.5 & 0.245 & 0.152 & 0.232 & 0.88 & $81.9^{+20.0}_{-17.3}$ &  $7.61^{+1.86}_{-1.61}\pm1.02~(8.2\sigma)$  \\

4.740  & 164.3 & 0.219 & 0.181 & 0.226 & 1.01 &  $0.0^{+1.3}_{-0.0}~(<9.9)$ &  $0.0^{+1.36}_{-0.0}\pm0.26~(<2.6)$  \\
				
4.750  & 367.2 & 0.208 & 0.184 & 0.221 & 1.04 & $6.5^{+8.9}_{-5.3}~(<23.5)$ &  $0.75^{+1.02}_{-0.61}\pm0.13~(<2.7)$  \\
				
4.781  & 512.8 & 0.179 & 0.194 & 0.209 & 1.12 & $17.2^{+10.1}_{-7.2}~(<34.5)$ &  $1.31^{+0.77}_{-0.55}\pm0.13~(<2.6)$  \\
				
4.843  & 527.3 & 0.145 & 0.196 & 0.180 & 1.28 &  $0.0^{+1.3}_{-0.0}~(<11.2)$ &  $0.0^{+0.40}_{-0.0}\pm0.03~(<0.7)$  \\
				
4.918  & 208.1 & 0.113 & 0.189 & 0.160 & 1.44 & $5.0^{+7.6}_{-3.9}~(<21.1)$ &  $0.73^{+1.11}_{-0.57}\pm0.06~(<3.1)$  \\
				
4.951  & 160.4 & 0.107 & 0.183 & 0.151 & 1.51 &  $0.0^{+1.3}_{-0.0}~(<13.0)$ &  $0.0^{+1.31}_{-0.0}\pm0.11~(<2.4)$  \\

				\hline
				\hline

			\end{tabular}
\caption{The Born cross section $\sigma^\mathrm{B}$ for $\EE\to\phi\chict$ at each
  c.m.~energy ($\sqrt{s}$). The numbers in the brackets are the signal significances or
          upper limits $\sigma^\mathrm{U.L.}$ at the 90\% C.L. in case
          the signal significance is less than $3\sigma$.
  The table also includes integrated
  luminosity $\mathcal{L}_\mathrm{int}$, detection efficiency
  $\epsilon_{\kk}^3$, $\epsilon_{\kk}^4$ and $\epsilon_{\ks\kl}$ for the
  3-track events of the $\phi\to\kk$ mode, 4-track events of the
  $\phi\to\kk$ mode and the events of the $\phi\to\ks\kl$ mode,
  respectively, the product of radiative correction factor and vacuum
  polarization factor $\frac{1+\delta}{|1-\Pi|^2}$ and the number
  of fitted events $N^\mathrm{fit}$ (also the corresponding upper limit $N^\mathrm{U.L.}$ at the 90\% C.L.
  in case the signal significance is less than $3\sigma$). 
  The first uncertainty is statistical and the second
  is systematic. \label{tab:xsec-c2}}
\end{table}

To investigate the $\sqrt{s}$-dependent cross section line shape of
$\EE\to\phi\chi_{c1,c2}$, a maximum likelihood fit is performed to the
dressed cross section
($\sigma^\mathrm{B}(\sqrt{s})\frac{1}{|1-\Pi|^2}$).  Due to the small
numbers of events at each single c.m.~energy, the likelihood function
is constructed as

\begin{equation}
	\mathcal{L}=\prod_i P(N^\mathrm{obs}_i;N^\mathrm{exp}_i+N^\mathrm{bkg}_i)
\end{equation}
where $P$ represents a Poisson distribution, $N^\mathrm{obs}_i$,
$N^\mathrm{exp}_i$ and $N^\mathrm{bkg}_i$ are the number of observed
events, the number of expected $\chi_{c1,c2}$ signal events and the
background events in the $\chi_{c1,c2}$ signal region for the $i$-th
dataset, respectively. Here in the fit, only statistical uncertainties are considered.

\begin{figure}[H]
	
	\includegraphics[width=0.49\linewidth]{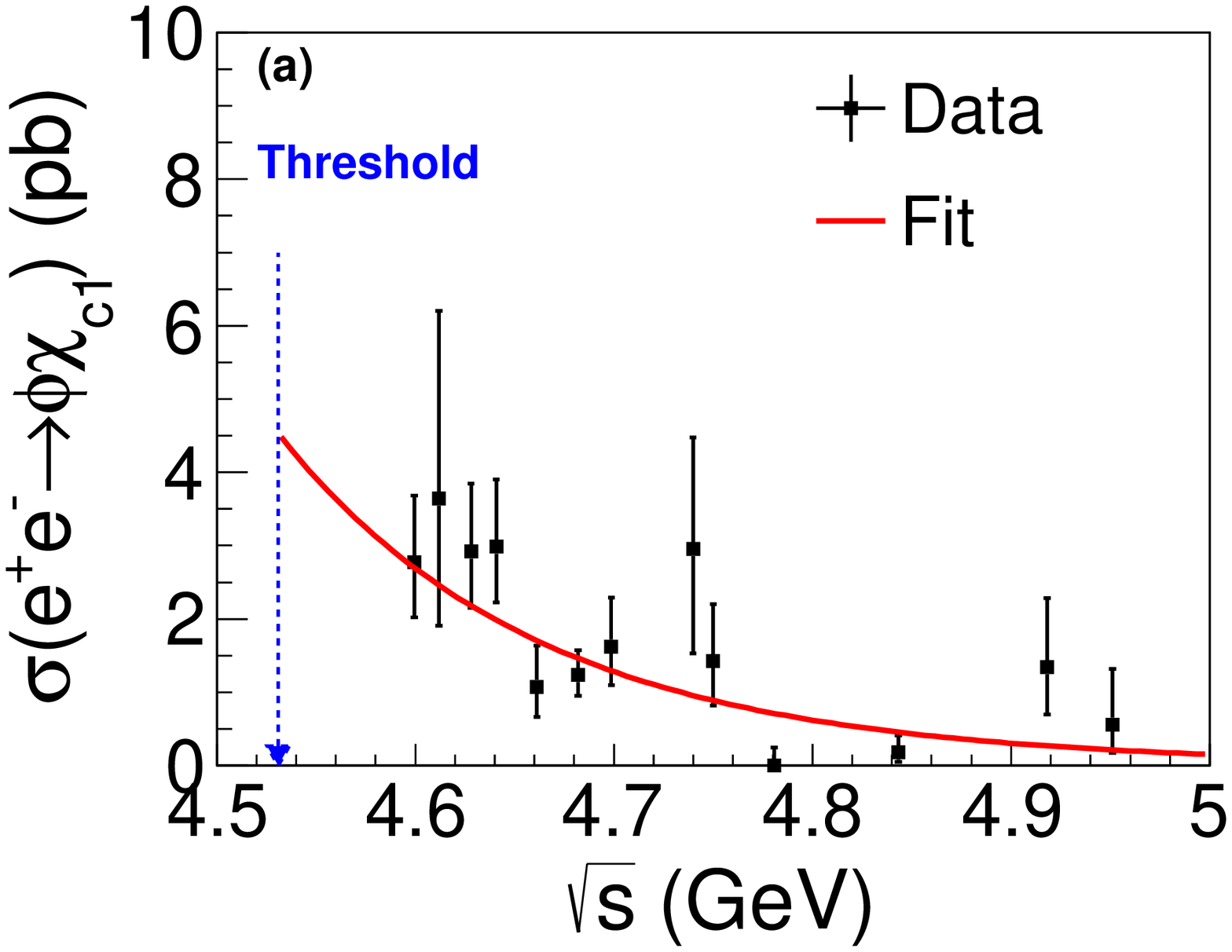}
	\includegraphics[width=0.49\linewidth]{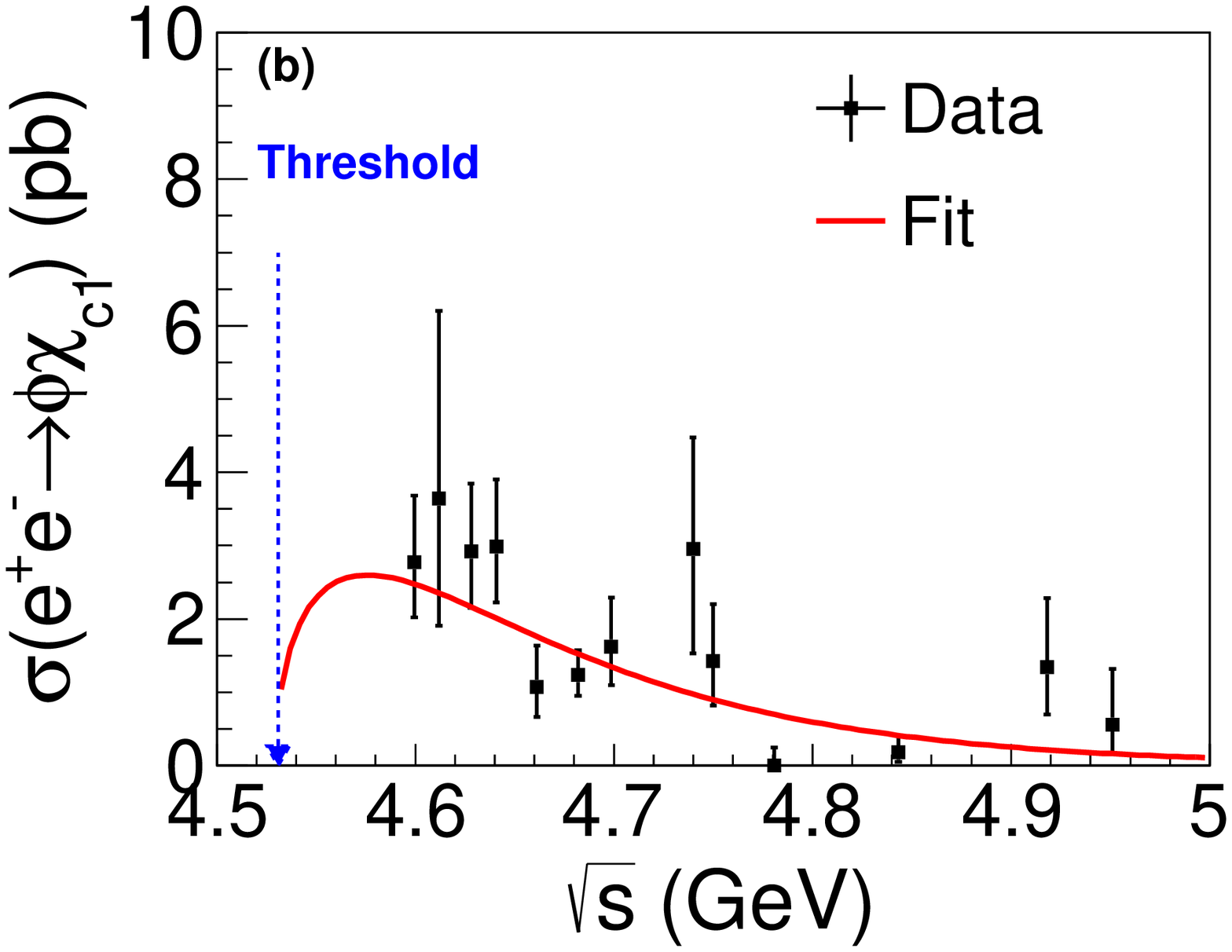}
	\caption{Fit to the cross section of $\EE\to\phi\chico$ with
          (a) the continuum amplitude  and (b) the PHSP corrected continuum
          amplitude.\label{fig:xs-c1}}
\end{figure}

For the $\EE\to\phi\chico$ process, a continuum amplitude is used to
fit the cross section,
\begin{equation}\label{eq:pure-cont}
	A_\mathrm{cont}(\sqrt{s}) = \sqrt{\frac{f_\mathrm{cont}}{(\sqrt{s}/4.682)^n}},
\end{equation}
where $f_\mathrm{cont}$ and $n$ are free parameters in the fit. We
also use a phase space (PHSP) shape corrected continuum amplitude
$A_\mathrm{cont}(\sqrt{s})\sqrt{\Phi(\sqrt{s})}$ to fit the cross
section, where $\Phi(\sqrt{s})$ is the two-body PHSP
factor. \cref{fig:xs-c1} shows the fit results with both models, and
the numerical results are listed in Table~\ref{tab:fitxs-c1}. We also
fit the cross section data with a Breit-Wigner (BW) function and the
coherent sum of a BW and a continuum amplitude, and no significant
resonance structures are found.

\begin{table}[H]
	\centering
	
	\setlength{\tabcolsep}{5mm}{
		\begin{tabular}{ccc}
			\hline\hline
			Parameter & $|A_\mathrm{cont}|^2$ & $|A_\mathrm{cont}\sqrt{\Phi}|^2$  \\
			\hline
			$f_\text{cont}$ 	& $1.47\pm0.16$	&	$14.26\pm1.59$	\\
			$n$	& $34.52\pm8.34$	&	$48.94\pm8.74$	\\
			$\chi^2/\text{d.o.f.}$ & $21.6/11$ & $21.9/11$ \\
			\hline\hline
		\end{tabular}
	}
\caption{The numerical results for the fit to the cross section of
  $\EE\to\phi\chico$ with the pure continuum amplitude (2nd column) and
  PHSP corrected continuum amplitude (3rd column). The errors are
  statistical.\label{tab:fitxs-c1}}
\end{table}

\begin{figure}[H]
	\centering
	\includegraphics[width=0.49\linewidth]{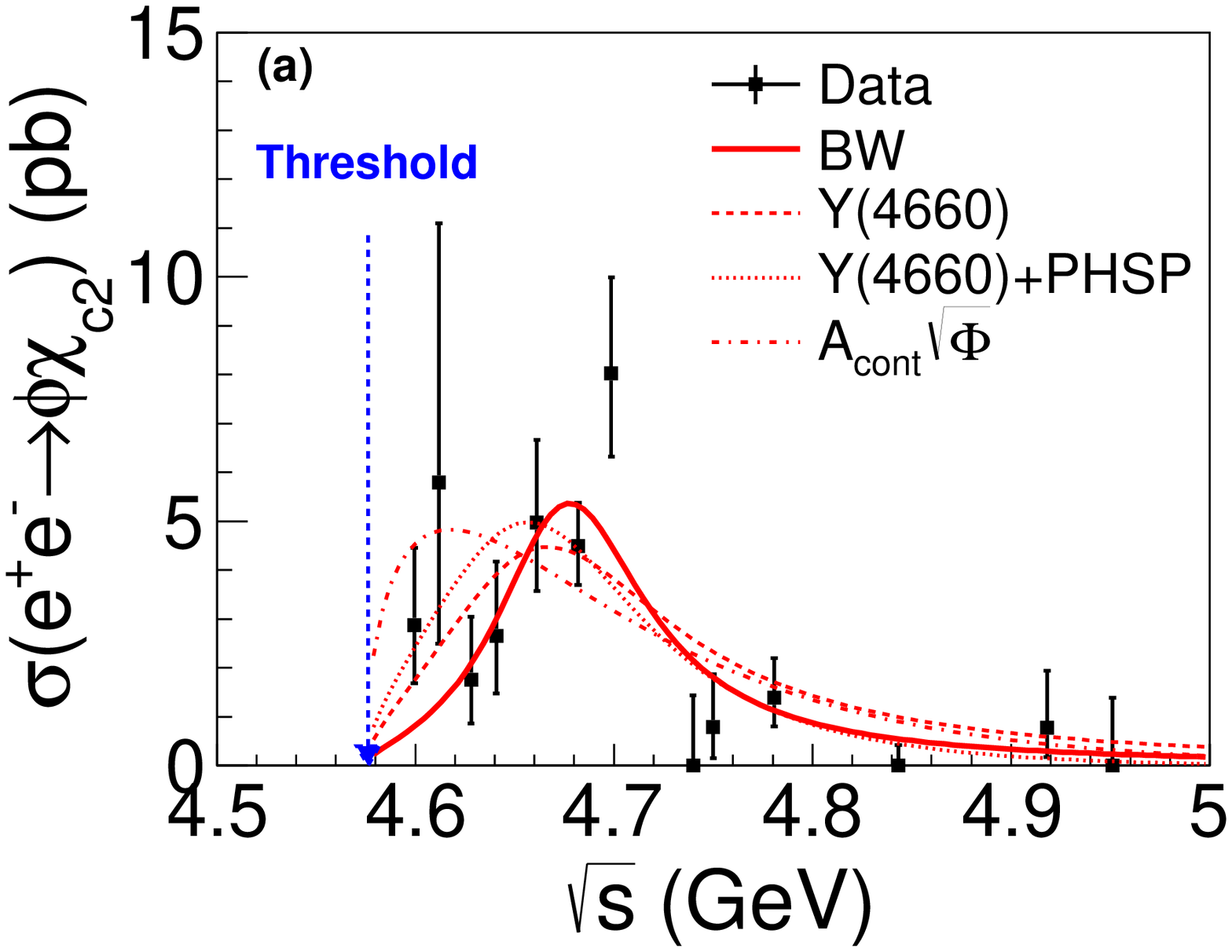}
	\includegraphics[width=0.49\linewidth]{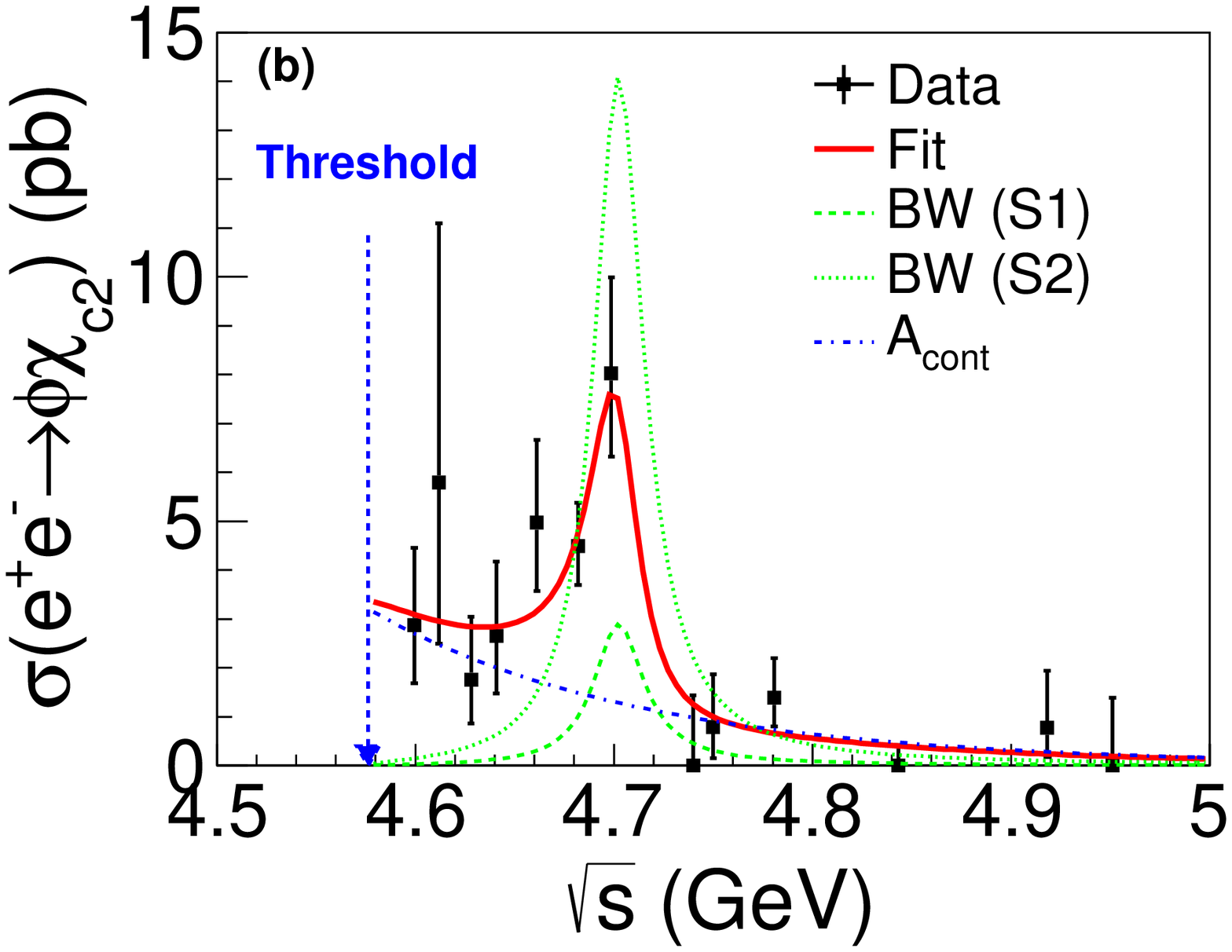}

	\caption{(a) Fit to the cross section of $\EE\to\phi\chict$
          with a single BW (solid line), the $Y(4660)$ resonance
          hypothesis (dashed line), the coherent sum of $Y(4660)$ and
          PHSP (dotted line), and the PHSP corrected continuum
          amplitude as the non-resonance hypothesis (dash-dotted
          line). (b) Fit to the cross section of $\EE\to\phi\chict$
          with the coherent sum of a BW and continuum amplitude. The
          solid line is the fit result, the dashed and dotted lines
          correspond to the BW with constructive (S1) and destructive (S2)
          solutions of interference, and the dash-dotted line is the
          continuum term.  \label{fig:xs-c2}}
\end{figure}

For the $\EE\to\phi\chict$ process, there is a possible resonance
structure around $4.7\gev$ in the cross section line shape as shown in Fig.~\ref{fig:xs-c2}, which is
fitted with a BW function:
\begin{equation}
	{\rm BW}(\sqrt{s})=\frac{M}{\sqrt{s}}\cdot\frac{\sqrt{12\pi\Gamma_\mathrm{tot}\Gamma_{\EE}\mathcal{B}(Y\to\phi\chict)}}{s-M^2+iM\Gamma_\mathrm{tot}}\cdot\sqrt{\frac{\Phi(\sqrt{s})}{\Phi(M)}}
\end{equation}
where $M$, $\Gamma_\mathrm{tot}$ and $\Gamma_{\EE}$ are the mass, full
width, and electric width of the potential resonance $Y$,
respectively, and $\mathcal{B}(Y\to\phi\chict)$ is the branching
fraction of $Y\to\phi\chict$. Figure~\ref{fig:xs-c2} (a) shows the fit
results, which yields
\begin{equation}\label{eq:bw1-c2}
	M=(4672.7\pm10.8)\mevcc,~\Gamma_{\rm tot}=(93.2\pm19.8)\mev,
\end{equation}
for the resonance. A $\chi^2$ test method is used to estimate the fit
quality, which gives $\chi^2/\text{d.o.f.}=15.9/10$. The significance
for the resonance hypothesis over the continuum hypothesis is
estimated to be $3.1\sigma$, by comparing the difference of
log-likelihoods $\Delta(-2\ln\mathcal{L})=10.0$ and taking into
account the change of number of degree of freedom ($\Delta
\rm{d.o.f.}=1$). Here the continuum hypothesis follows
$A_\mathrm{cont}(\sqrt{s})\sqrt{\Phi(\sqrt{s})}$. The fit result for the continuum hypothesis is
shown in \cref{fig:xs-c2} (a) (dash-dotted line) and listed in~\cref{tab:fitxs-c2} (last column).

The potential resonance (solid line in \cref{fig:xs-c2} (a)) is found
to be consistent with the $Y(4660)$ reported in
$\EE\to\pp\psip$~\cite{belley4360,BaBar:2012hpr,Belle:2014wyt,BESIII:2021njb}.
Next, we fit the $\EE\to\phi\chict$ cross section with the fixed mass
and width of the $Y(4660)$ \cite{BESIII:2021njb}. Two fit models are
considered: one is the single BW model, which gives
$\Gamma_{\EE}\mathcal{B}[Y(4660)\to\phi\chict]=1.0\pm0.1$~eV with a fit quality
$\chi^2/\text{d.o.f.}=21.5/12$ (the dashed line in \cref{fig:xs-c2}
(a)), and the other is the coherent sum of a BW and PHSP model (${\rm
  BW}+f\sqrt{\Phi}e^{i\phi}$), which gives
$\Gamma_{\EE}\mathcal{B}[Y(4660)\to\phi\chict]=1.2\pm0.4$~eV with a fit quality
$\chi^2/\text{d.o.f.}=17.9/10$ (the dotted line in \cref{fig:xs-c2}
(a)).  Since the fit quality with the fixed $Y(4660)$ is close to the
one with a single free BW model ($\chi^2/\text{d.o.f.}=15.9/10$), we
cannot distinguish between these two models.

To improve the fit quality, the fit model is parameterized as the
coherent sum of a BW resonance and a possible continuum term (${\rm
  BW}+A_{\rm cont}e^{i\phi}$).  The fit result is shown in
\cref{fig:xs-c2} (b), which gives
\begin{equation}\label{eq:bw2-c2}
	M=(4701.8\pm10.9)\mevcc,~\Gamma_{\rm tot}=(30.5\pm22.3)\mev 
\end{equation}
for the resonance.  The fit quality is $\chi^2/\text{d.o.f.}=7.3/7$,
and the significance for the resonance hypothesis is estimated using
the same method, which gives $3.6\sigma$
($\Delta(-2\ln\mathcal{L})=20.7$, $\Delta \rm{d.o.f.}=4$).  All the
numerical results of the fits are summarized in \cref{tab:fitxs-c2}.

The significance for the coherent sum of a BW and continuum model
(${\rm BW}+A_{\rm cont}e^{i\phi}$) over the single BW model is
estimated to be $2.3\sigma$. Thus, we are not able to distinguish
these two models based on the current data.

\begin{table}[H]
	\centering

		\scalebox{0.9}{
		\begin{tabular}{ccccc}
			\hline\hline
			Parameter & $|{\rm BW}|^2$ &  $|{\rm BW}+A_{\rm cont}e^{i\phi}|^2$ (S1) & $|{\rm BW}+A_{\rm cont}e^{i\phi}|^2$ (S2) & $|A_{\rm cont}\sqrt{\Phi}|^2$\\
			\hline
			$M(\mevcc)$	& $4672.75\pm10.80$	&	\multicolumn{2}{c}{$4701.77\pm10.89$}  &	-	\\
			$\Gamma_\mathrm{tot}(\mev)$	&	$93.15\pm19.78$	&	\multicolumn{2}{c}{$30.50\pm22.33$}	&	-	\\
			$\mathcal{B}\Gamma_{\EE}(\ev)$	&	$0.74\pm0.13$	&	$0.13\pm0.13$ & $0.66\pm0.41$	&	-	\\
			$f_\text{cont}$ 	& -	&	\multicolumn{2}{c}{$1.48\pm0.72$}	&	$40.61\pm4.57$	\\
			$n$		& -	&	\multicolumn{2}{c}{$33.95\pm22.24$}	&	$54.28\pm8.87$ \\
			$\phi(^\circ)$	& -	&	$240.20\pm40.53$& $109.77\pm13.57$	&	-	\\	
			$\chi^2/\text{d.o.f}$ & $15.9/10$ & \multicolumn{2}{c}{$7.3/7$}	&	$26.9/11$ \\
			Significance	&	3.1$\sigma$	& \multicolumn{2}{c}{3.6$\sigma$}	&	- \\
			\hline\hline
		\end{tabular}
	}
\caption{The numerical results for the fits to the cross section of
  $\EE\to\phi\chict$ with the single BW model (2nd column), the
  coherent sum of a BW and continuum model (3rd and 4th columns
  correspond to the constructive (S1) and destructive (S2) solutions
  of the interference), and PHSP corrected continuum model (5th
  column). The errors are statistical.\label{tab:fitxs-c2}}
\end{table}

Since no obvious structures are observed in the $\phi\chico$ mode, the upper limit of $\Gamma_{\EE}\mathcal{B}(Y\to\phi\chico)$ 
is also determined for the possible structures observed in the $\phi\chict$ mode. A similar method by scanning the
$\Gamma_{\EE}\mathcal{B}(Y\to\phi\chico)$ dependent likelihood distribution is used, and 
the results at 90\% C.L. are listed in~\cref{tab:ul-gee}.

\begin{table}[H]
	\centering
	\begin{tabular}{cc}
		\hline\hline
		Resonance & $\Gamma_{\EE}\mathcal{B}(Y\to\phi\chico)$ (eV) \\
		\hline
		$\mathrm{BW}_1$ & $<0.07$ at 90\% C.L.\\
		$\mathrm{BW}_2$ & $<0.04$ at 90\% C.L.\\
		$Y(4660)$~\cite{BESIII:2021njb} & $<0.36$ at 90\% C.L.\\
		\hline\hline
		
	\end{tabular}
	\caption{The upper limit of $\Gamma_{\EE}\mathcal{B}(Y\to\phi\chico)$ at 90\% C.L. for the possible structures in $\phi\chict$, where $\mathrm{BW}_1$ and $\mathrm{BW}_2$ correspond to \cref{eq:bw1-c2,eq:bw2-c2}, respectively. \label{tab:ul-gee}}
\end{table}

\subsection{Systematic uncertainty\label{subsec:sys-cj}}
\subsubsection{Systematic uncertainty for cross section measurement}

The sources of systematic uncertainties in the cross section
measurement of $\EE\to\phi\chi_{c1,c2}$ include the luminosity
measurement, tracking efficiency, PID efficiency, $\ks$
reconstruction, photon reconstruction, kinematic fit, radiative
correction, MC model, MUC response, branching ratios, and the fit.

The uncertainty of the integrated luminosity measurement is 0.6\% by
analyzing the large angle Bhabha events at
BESIII~\cite{BESIII:2022ulv}.  The uncertainty of the tracking
efficiency for high momentum leptons is 1\% per track, and thus 2\% by
adding both leptons linearly~\cite{BESIII:2016bnd} since we require both
leptons detected.  For the $\phi\to\kk$ mode, both one kaon events and
two kaon events are reconstructed. Assuming $p~(q)$ is the corresponding
tracking efficiency for a single kaon from data (MC), the efficiency to
reconstruct both one and two kaon candidates is $2p(1-p)+p^2=1-(1-p)^2$ [$1-(1-q)^2$]
for data (MC). Considering $p\approx 85\%$ and the tracking efficiency
uncertainty $p/q-1=1\%$ at BESIII, the uncertainty due to the detection of
both one and two kaon candidates for the
tracking efficiency can be calculated as $\left|1-\frac{1-(1-p)^2}{1-(1-q)^2}\right|$, 
which is negligible. The same calculation can be applied to the kaon PID
uncertainty, which is also negligible.
For the $\phi\to\kskl$ mode, the uncertainty of tracking
efficiency is 1\% per pion.  The uncertainty of $\ks$ reconstruction
is estimated to be 1.2\% by studying the $\jpsi\to\kstarpm
K^\mp\to\ks\pi^\pm K^\mp$ and $\jpsi\to\phi\ks K^\mp\pi^\pm$ control
samples~\cite{ks-sys-err}.  The uncertainty from photon reconstruction
is estimated to be 1\% per photon by studying the $\jpsi\to\rho^0\piz$
events~\cite{c02pi0-bes3}.

The systematic uncertainty associated with kinematic fitting is
estimated by comparing the efficiency difference with or
without the helix parameters correction in MC simulations~\cite{KF}.  
The radiative correction factor and efficiency depend on
the input cross section line shape in {\sc kkmc}. Using
different cross section line-shapes as studied in \cref{xs}, the
difference in $(1+\delta)\epsilon$ between different models is taken
as the systematic uncertainty.  In the signal MC simulation, a
phase space model is used. To estimate the uncertainty due to the MC
model, the angular distribution of $\EE\to\phi\chi_{c1,c2}$ is modelled
by a $1\pm\cos^2\theta$ distribution, and the efficiency difference is
taken as the systematic uncertainty.

The uncertainty from the MUC response is studied with a control sample
of $\EE\to\MM$ events. The difference in
efficiency between the data and MC simulation due to the requirement
of $\mu$ hit depth in the MUC is taken as the systematic uncertainty. The
uncertainties of branching fractions of the intermediate states are
taken from the PDG~\cite{Zyla:2020zbs}. The uncertainties related to
the fit are investigated by changing the fit range and changing
the background shape from a free 1st-order polynomial to a fixed flat
shape with the number of events estimated from $\phi$ and $\jpsi$
sidebands. The largest difference in signal yields is taken as the
systematic uncertainty.

In \cref{sec:event-selection}, three data samples, which are the 3-track
events, 4-track events with $\phi\to\kk$ and the events with
$\phi\to\kskl$, are reconstructed. A source of systematic uncertainty
can contribute differently to the three data samples. To propagate the
systematic uncertainty to the cross section, we take the weighted
average of the systematic uncertainties in the three data samples,
which follows

\begin{equation}
	\sigma^2_\mathrm{tot}=\sum_{i=1}^{3}\omega_i^2\sigma_i^2+2\sum_{i\neq j}^{3}cov(i,j),
\end{equation}
\begin{equation}
	\omega_i = \frac{\epsilon_i\mathcal{B}_i}{\sum_{i=1}^3\epsilon_i\mathcal{B}_i},~cov(i,j)=\rho_{ij}\omega_i\omega_j\sigma_i\sigma_j,
\end{equation}
where $\sigma_\mathrm{tot}$ is the average systematic uncertainty to
the cross section as listed in \cref{tab:sys-pcj}, $\omega_i$ and
$\sigma_i$ are the weight and systematic uncertainty for $i{\rm th}$
data sample, $\epsilon_i$ and $\mathcal{B}_i$ are the efficiency and
branching ratio of $\phi$ for the $i{\rm th}$ data sample, $\rho_{ij}$
is the correlation parameter between the $i{\rm th}$ and $j{\rm th}$
data samples, and $\rho_{ij}=1$ if the systematic uncertainty is
correlated between the $i{\rm th}$ and $j{\rm th}$ data samples,
otherwise $\rho_{ij}=0$.

Assuming all these sources are independent, the total systematic
uncertainty in the cross section measurement is obtained by adding
them in quadrature. \cref{tab:sys-pcj} summarizes all the systematic
sources and their contributions at 4.68~\gev, and the systematic
uncertainties at other energy points are listed in
\cref{tab:sys-pc1,tab:sys-pc2} of Appendix~\ref{app:sys-pcj}.

\begin{table}[H]
	\setlength\tabcolsep{10pt}
	\centering
	
	\begin{tabular}{lcc}
		\hline\hline
		Source & $\phi\chico$ & $\phi\chict$ \\
		\hline
		Luminosity			 & 0.60				& 0.60	\\
		Tracking             & 2.42             & 2.44	\\
		Photon               & 0.65             & 0.73	\\
		$\ks$ reconstrcution & 0.25             & 0.27	\\
		Kinematic fit        & 0.49             & 0.52	\\
		$\mathcal{B}(\phi)$  & 0.83             & 0.82	\\
		$\mathcal{B}(\chicJ)$& 2.90             & 2.60	\\
		$\mathcal{B}(\jpsi)$ & 0.60             & 0.60	\\
		Radiative correction & 0.40				& 5.31	\\
		MC model			 & 0.18				& 0.16	\\
		Muon hit depth       & 0.86             & 0.85	\\
		Fit related			 & 5.54				& 7.14	\\
		\hline
		Total                & 6.93             & 9.74	\\  
		\hline\hline  
	\end{tabular}
\caption{The systematic uncertainty sources and their contributions (in \%) for the cross section of $\EE\to\phi\chi_{c1,c2}$ at 4.68~\gev.\label{tab:sys-pcj}}
\end{table}

\subsubsection{Systematic uncertainties for the resonance parameters}

The systematic uncertainties for the resonance parameters mainly come
from the absolute c.m.~energy calibration, the parameterization of the BW
function, and the cross section measurement.

The c.m.~energies of the data sets used in this work are measured with
$\Lambda_c$ events, with an uncertainty of
$\pm0.6\mev$~\cite{BESIII:2022ulv,BESIII:2015zbz}. This common
uncertainty for all the data samples could shift the cross section
line-shape globally, and is thus the systematic uncertainty to the
mass of the resonance.

In the fit to the cross section of $\EE\to\phi\chict$
(\cref{fig:xs-c2}), a constant full width BW function is employed. We
also use an alternative BW function, where the constant width is replaced
by an energy dependent width $\Gamma(\sqrt{s}) =
\Gamma_0\cdot\frac{\sqrt{s}}{M}$. Here $\Gamma_0$ is the full width at
$\sqrt{s}=M$. The difference in the resonance parameters between the
two BWs is taken as the systematic uncertainty.

In the fit to the cross section of $\EE\to\phi\chict$
(\cref{fig:xs-c2} (b)), a continuum amplitude (\cref{eq:pure-cont}) is
used to describe the non-resonance contribution. We also use a PHSP
corrected continuum amplitude ($A_\mathrm{cont}\sqrt{\Phi}$) in the
fit. The difference in the resonance parameters is taken as the systematic
uncertainty.

The uncertainty from the cross section measurement can be divided into
two parts, one is the correlated systematic uncertainty for all the
energy points, including tracking, photon reconstruction, $\ks$
reconstruction, luminosity, branching fraction, muon hit depth,
background shape, and fit range. They are propagated to
$\Gamma_{\EE}\mathcal{B}(Y\to\phi\chict)$ directly.  The other is the
uncorrelated systematic uncertainty, which is dominated by the radiation
correction according to the previous section. This uncertainty can be
considered in the fit to the cross section.  The two types of
uncertainties are added in quadrature assuming they are independent.

\cref{tab:sys-c2-res-bw,tab:sys-c2-res-bwcon} summarize the sources of
systematic uncertainty for the resonance parameters and their
contributions, and the total systematic uncertainty is obtained by
adding them in quadrature.

\begin{table}[H]
	\centering

		\begin{tabular}{cccc}
			\hline\hline
			Source & Mass (${\rm MeV}/c^2$) & Width (MeV) & $\mathcal{B}\Gamma_{\EE}$ (eV) \\
			\hline
			c.m.~energy 		& 0.6	&	-	&	-	\\
			Parameterization of BW	& 0.04	&	0.70&	0.01	\\
			Cross section	&	3.81	&	9.39	&	0.07	\\
			\hline
			Total	&	3.86	&	9.42	&	0.07	\\
			\hline\hline
		\end{tabular}
\caption{The systematic uncertainties for the resonance parameters
  with the single BW model.\label{tab:sys-c2-res-bw}}
	
\end{table}

\begin{table}[H]
	\centering
	\tabcolsep=1pt

		\begin{tabular}{ccccc}
			\hline\hline
			Source & Mass (${\rm MeV}/c^2$) & Width (MeV) & $\mathcal{B}\Gamma_{\EE}$ [S1] (eV) & $\mathcal{B}\Gamma_{\EE}$ [S2] (eV)\\
			\hline
			c.m.~energy 		& 0.6	&	-	&	-	&	-	\\

			Parameterization of BW	& 0.05	&	0.06&	0.0	&	0.01	\\
			
			Parameterization of $A_{\rm cont}$	& 2.12	&	13.51&	0.05	&	0.27	\\
		
			Cross section	&	1.63	&	5.52	&	0.01	&	0.09 	\\
			\hline
			Total	&	2.74	&	14.59	&	0.05	&	0.29	\\
			\hline\hline
		\end{tabular}
\caption{The systematic uncertainties for resonance parameters with
  the coherent sum of a BW and continuum. \label{tab:sys-c2-res-bwcon}}
\end{table}

\section{\texorpdfstring{Study of \bm{$\EE\to\gamma X$} with \bm{$X\to\phi\jpsi$}}{Study of ee->gamma X with X->phi J/psi}}
	
\begin{figure}[H]
	\centering
	\includegraphics[width=0.49\columnwidth]{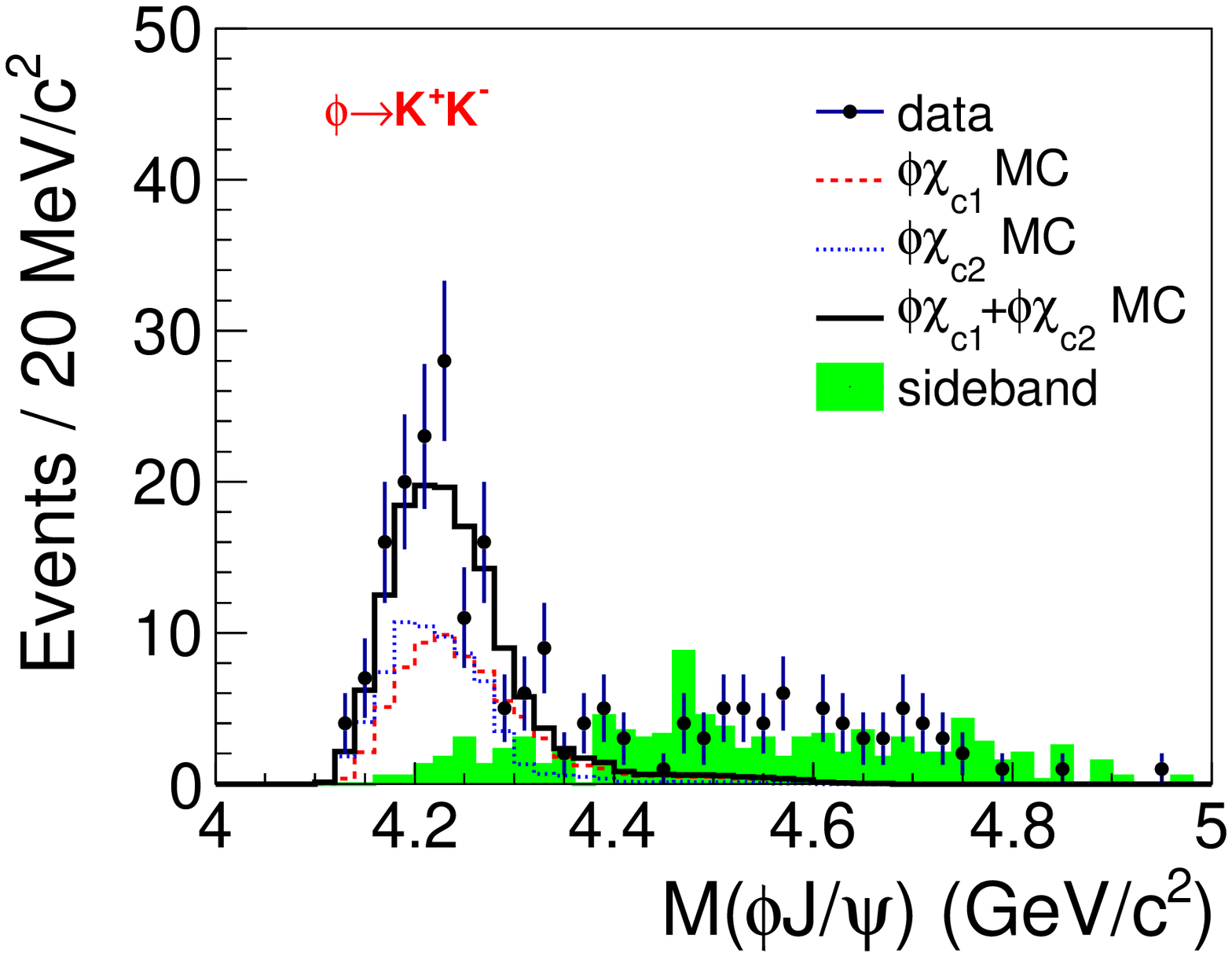}
	\includegraphics[width=0.49\columnwidth]{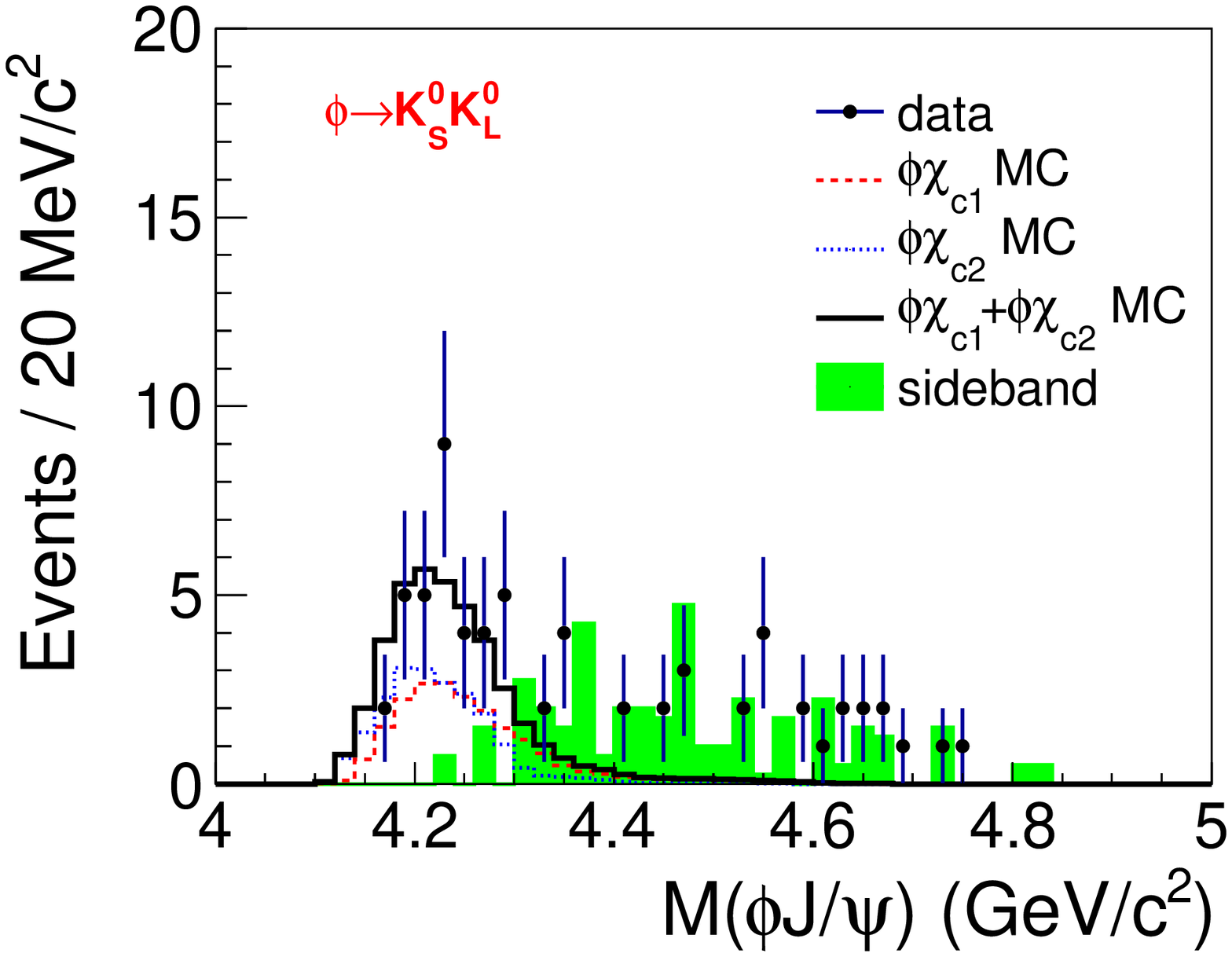}
	\caption{The invariant mass distribution of $M(\phi\jpsi)$ in the
          $\phi\to\kk$ and $\phi\to\kskl$ modes. Dots with error bars
          are the full data, the red dashed and blue dotted
          histograms are from $\phi\chico$ and $\phi\chict$ MC, which
          have been normalized to the data, the black solid histograms are
          the sum of $\phi\chico$ and $\phi\chict$, and the green filled
          histograms are the $\phi-\jpsi$ 2-dimensional
          sideband.\label{fig:mphijpsi}}
\end{figure}

The process of $\EE\to\gamma X\to\gamma\phi\jpsi$ shares the same
final states as that of $\EE\to\phi\chi_{c1,c2}$, thus the same event
selection criteria are applied to the $\EE\to\gamma X$ process.  The
$M(\phi\jpsi)$ invariant mass distribution, shown in
\cref{fig:mphijpsi}, is well described by the $\phi\chi_{c1,c2}$
events, together with the non-$\gamma\phi\jpsi$ background events
estimated from the $\phi$-$\jpsi$ 2-dimensional sidebands (${\rm
  B1/2+B2/2-B3/4}$ as exhibited in
\cref{fig:inv-kk3,fig:inv-kk4,fig:inv-kskl}). No other structure is
observed in the $M(\phi\jpsi)$ mass distribution.

\subsection{Upper limit of $\EE\to\gamma X$ cross section}
The product of Born cross section of $\EE\to\gamma X$ and the branching fraction of $X\to\phi\jpsi$ is calculated by

\begin{equation}\label{eq:xs-gx}
	\sigma^\mathrm{B}_{\gamma X}\mathcal{B}(X\to\phi\jpsi) = \frac{N^\textrm{fit}_{\gamma X}}{\mathcal{L}_\mathrm{int}(1+\delta)\frac{1}{|1-\Pi|^2}\mathcal{B}},
\end{equation}
where $N^\textrm{fit}_{\gamma X}$ is the number of fitted events for
$\gamma X$, which is equal to the number of $\gamma X$ events in data
divided by the efficiency and branching fraction of $\phi$,
$\mathcal{L}_\mathrm{int}$ is the integrated luminosity,
$1+\delta$ is the ISR correction factor, $\frac{1}{|1-\Pi|^2}$ is the
vacuum polarization factor, and $\mathcal{B}$ is the branching
fraction of $\jpsi\to\LL$.

\begin{figure}[H]
	\centering
	\includegraphics[width=0.5\linewidth]{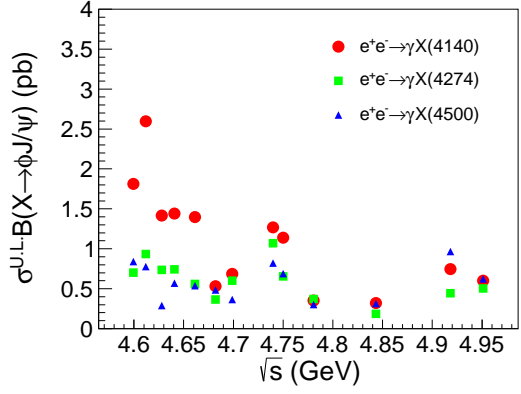}
	\caption{The upper limit of Born cross section product
          branching fraction at the 90\% C.L. versus c.m.~energy for
          $\EE\to\gamma X(4140)/\gamma X(4274)/\gamma
          X(4500)$. \label{fig:ul-gx}}
\end{figure}

Since no significant structures are observed, we determine the upper
limit of the production cross section for $\EE\to\gamma
X\to\gamma\phi\jpsi$ using the same method as described in \cref{xs}.  An unbinned maximum likelihood fit is performed
to the $M(\phi\jpsi)$ distribution simultaneously for the $\phi\to\kk$ and
$\kskl$ modes.  In the fit, the signal PDF is described by MC-simulated shapes, where the mass and width of $X$ are fixed to
LHCb's measurements~\cite{LHCb:2021uow}. The background is composed of
$\phi\chi_{c1,c2}$ and a smooth polynomial shape
(including both the non-$\gamma\phi\jpsi$ and the continuum $\gamma\phi\jpsi$ contribution).
The $\phi\chi_{c1,c2}$ background shapes are from the MC simulation,
and their yields are normalized to the cross section measurement
described in \cref{xs}. The contribution for the sum of non-$\gamma\phi\jpsi$ 
and continuum $\gamma\phi\jpsi$ backgrounds is free.
The selection efficiencies and
branching fractions of $\phi\to\kk/\kskl$ modes are also included in
the fit procedure.
Figure~\ref{fig:ul-gx} shows the
upper limit of the Born cross section at the 90\% C.L. for
$\EE\to\gamma X\to\gamma\phi\jpsi$ at each c.m.~energy, and the
numerical results are listed in
\cref{tab:xsec-gx4140,tab:xsec-gx4274,tab:xsec-gx4500}.

\begin{table}[H]
	\centering

		\begin{tabular}{cccccccc}
			\hline
			\hline

			$\sqrt{s}~(\mathrm{GeV})$  & $\mathcal{L}_\mathrm{int}~(\mathrm{pb}^{-1})$ & $\epsilon_{\kk}^3$ & $\epsilon_{\kk}^4$ & $\epsilon_{\ks\kl}$ & $\frac{1+\delta}{|1-\Pi|^2}$ & $N_{\gamma X(4140)}^\mathrm{U.L.}$ & $\sigma^\mathrm{U.L.}\mathcal{B}~(\mathrm{pb})$ \\
			
			\hline
			
			4.600  & 586.9 & 0.214 & 0.101 & 0.221 & 0.91 &  $116.2$ &  $1.81$  \\
			
			4.612  & 103.8 & 0.215 &  0.097 & 0.212 & 0.92 &  $29.8$ &  $2.60$  \\
			
			4.628  & 521.5 & 0.213 &  0.098 & 0.212 & 0.92 &  $81.5$ &  $1.42$  \\
			
			4.641  & 552.4 & 0.213 &  0.099 & 0.216 & 0.92 &  $88.0$  &  $1.44$  \\
			
			4.661  & 529.6 & 0.216 & 0.101 & 0.216 & 0.92 &  $81.7$ &  $1.40$  \\
			
			4.682  & 1669.3 & 0.219 & 0.101 & 0.213 & 0.92 & $98.0$ &  $0.53$  \\
			
			4.699  & 536.5 & 0.218 & 0.102 & 0.213 & 0.93 &  $41.1$ &  $0.69$  \\
			
			4.740  & 164.3 & 0.213 & 0.109 & 0.221 & 0.93 &  $23.2$ &  $1.27$  \\
			
			4.750  & 367.2 & 0.210 & 0.107 & 0.220 & 0.93 &  $46.6$ &  $1.14$  \\
			
			4.781  & 512.8 & 0.213 & 0.108 & 0.219 & 0.93 &  $20.1$ &  $0.35$  \\
			
			4.843  & 527.3 & 0.213 & 0.120 & 0.224 & 0.94 &  $19.2$ &  $0.32$  \\
			
			4.918  & 208.1 & 0.213 & 0.122 & 0.223 & 0.95 &  $17.7$ &  $0.75$  \\
			
			4.951  & 160.4 & 0.214 & 0.122 & 0.218 & 0.95 &  $11.0$ &  $0.60$  \\
			
			\hline
			\hline

		\end{tabular}
\caption{The upper limit of Born cross section at 90\% C.L.~$\sigma^\mathrm{U.L.}\mathcal{B}(X\to\phi\jpsi)$ for $\EE\to\gamma
  X(4140)$ at each c.m.~energy $\sqrt{s}$. The table also includes
  integrated luminosity $\mathcal{L}_\mathrm{int}$, detection
  efficiency $\epsilon_{\kk}^3$, $\epsilon_{\kk}^4$ and
  $\epsilon_{\ks\kl}$ for the 3-track events in the $\phi\to \kk$ mode,
  4-track events in the $\phi\to \kk$ mode and the events in the
  $\phi\to\ks\kl$ mode, respectively, the product of radiative
  correction factor and vacuum polarization factor
  $\frac{1+\delta}{|1-\Pi|^2}$, and the 90\% C.L.~upper limit of the
  number of fitted events for $\gamma X(4140)$ $N_{\gamma
    X(4140)}^\mathrm{U.L.}$.  \label{tab:xsec-gx4140}}
\end{table}
\begin{table}[H]
	\centering

		\begin{tabular}{cccccccc}
			\hline
			\hline

			$\sqrt{s}~(\mathrm{GeV})$  & $\mathcal{L}_\mathrm{int}~(\mathrm{pb}^{-1})$ & $\epsilon_{\kk}^3$ & $\epsilon_{\kk}^4$ & $\epsilon_{\ks\kl}$  & $\frac{1+\delta}{|1-\Pi|^2}$ & $N_{\gamma X(4274)}^\mathrm{U.L.}$ & $\sigma^\mathrm{U.L.}\mathcal{B}~(\mathrm{pb})$ \\
			
			\hline
			
			4.600  & 586.9 & 0.217 & 0.216 & 0.242 & 0.88 & $43.5$ &  $0.70$  \\
			
			4.612  & 103.8 & 0.219 & 0.211 & 0.236 & 0.88 & $10.3$ &  $0.93$  \\
			
			4.628  & 521.5 & 0.220 & 0.208 & 0.230 & 0.89 & $41.0$ &  $0.74$  \\
			
			4.641  & 552.4 & 0.221 & 0.209 & 0.238 & 0.89 & $43.9$ &  $0.74$  \\
			
			4.661  & 529.6 & 0.218 & 0.207 & 0.231 & 0.89 & $31.6$ &  $0.56$  \\
			
			4.682  &1669.3 & 0.218 & 0.209 & 0.232 & 0.90 & $66.0$ &  $0.37$  \\
			
			4.699  & 536.5 & 0.217 & 0.209 & 0.232 & 0.90 & $34.8$ &  $0.60$  \\
			
			4.740  & 164.3 & 0.221 & 0.202 & 0.237 & 0.91 &  $19.2$ &  $1.07$  \\
			
			4.750  & 367.2 & 0.218 & 0.208 & 0.239 & 0.91 &  $26.3$ &  $0.65$  \\
			
			4.781  & 512.8 & 0.217 & 0.201 & 0.239 & 0.91 &  $20.8$ &  $0.37$  \\
			
			4.843  & 527.3 & 0.220 & 0.205 & 0.239 & 0.92 &  $10.9$ &  $0.19$  \\
			
			4.918  & 208.1 & 0.214 & 0.205 & 0.239 & 0.93 &  $10.3$ &  $0.45$  \\
			
			4.951  & 160.4 & 0.215 & 0.202 & 0.232 & 0.93 &  $9.0$ &  $0.50$  \\

			\hline
			\hline

		\end{tabular}
\caption{The upper limit of Born cross section at 90\% C.L.~$\sigma^\mathrm{U.L.}\mathcal{B}(X\to\phi\jpsi)$ for $\EE\to\gamma
  X(4274)$ at each c.m.~energy $\sqrt{s}$. The table also includes
  integrated luminosity $\mathcal{L}_\mathrm{int}$, detection
  efficiency $\epsilon_{\kk}^3$, $\epsilon_{\kk}^4$ and
  $\epsilon_{\ks\kl}$ for the 3-track events in the $\phi\to \kk$ mode,
  4-track events in the $\phi\to \kk$ mode and the events in the
  $\phi\to\ks\kl$ mode, respectively, the product of radiative correction factor and
  vacuum polarization factor $\frac{1+\delta}{|1-\Pi|^2}$ and the
  90\% C.L.~upper limit of the number of fitted events for $\gamma
  X(4274)$ $N_{\gamma
    X(4274)}^\mathrm{U.L.}$.  \label{tab:xsec-gx4274}}
\end{table}
\begin{table}[H]
	\centering

		\begin{tabular}{cccccccc}
			\hline
			\hline

			$\sqrt{s}~(\mathrm{GeV})$  & $\mathcal{L}_\mathrm{int}~(\mathrm{pb}^{-1})$ & $\epsilon_{\kk}^3$ & $\epsilon_{\kk}^4$ & $\epsilon_{\ks\kl}$  & $\frac{1+\delta}{|1-\Pi|^2}$ & $N_{\gamma X(4500)}^\mathrm{U.L.}$ & $\sigma^\mathrm{U.L.}\mathcal{B}~(\mathrm{pb})$ \\
			
			\hline
			
			4.600  & 586.9 & 0.181 & 0.311 & 0.258 & 0.82 & $48.5$ &  $0.84$  \\
			
			4.612  & 103.8 & 0.181 & 0.303 & 0.246 & 0.83 & $8.0$ &  $0.78$  \\
			
			4.628  & 521.5 & 0.182 & 0.302 & 0.244 & 0.83 & $15.1$ &  $0.29$  \\
			
			4.641  & 552.4 & 0.180  & 0.304 & 0.241 & 0.84 & $31.7$ &  $0.57$  \\
			
			4.661  & 529.6 & 0.178 & 0.303 & 0.241 & 0.85 & $29.1$ &  $0.54$  \\
			
			4.682  &1669.3 & 0.178 & 0.296 & 0.233 & 0.86 & $83.7$ &  $0.49$  \\
			
			4.699  & 536.5 & 0.174 & 0.293 & 0.236 & 0.86 & $20.1$ &  $0.36$  \\
			
			4.740  & 164.3 & 0.169 & 0.306 & 0.231 & 0.87 &  $14.1$ &  $0.82$  \\
			
			4.750  & 367.2 & 0.166 & 0.305 & 0.232 & 0.87 &  $26.4$ &  $0.69$  \\
			
			4.781  & 512.8 & 0.164 & 0.298 & 0.231 & 0.88 &  $16.3$ &  $0.30$  \\
			
			4.843  & 527.3 & 0.164 & 0.301 & 0.227 & 0.89 &  $17.7$ &  $0.31$  \\
			
			4.918  & 208.1 & 0.162 & 0.299 & 0.228 & 0.90 &  $21.7$ &  $0.96$  \\
			
			4.951  & 160.4 & 0.161 & 0.293 & 0.223 & 0.91 &  $10.9$ &  $0.62$  \\
			
			\hline
			\hline

		\end{tabular}
\caption{The upper limit of Born cross section at 90\% C.L.~$\sigma^\mathrm{U.L.}\mathcal{B}(X\to\phi\jpsi)$ for $\EE\to\gamma
  X(4500)$ at each c.m.~energy $\sqrt{s}$. The table also includes
  integrated luminosity $\mathcal{L}_\mathrm{int}$, detection
  efficiency $\epsilon_{\kk}^3$, $\epsilon_{\kk}^3$ and
  $\epsilon_{\ks\kl}$ for the 3-track events in the $\phi\to \kk$ mode,
  4-track events in the $\phi\to \kk$ mode and the events in the
  $\phi\to\ks\kl$ mode, respectively, the product of radiative correction factor and
  vacuum polarization factor $\frac{1+\delta}{|1-\Pi|^2}$ and the
  90\% C.L.~upper limit of the number of fitted events for $\gamma
  X(4500)$ $N_{\gamma
    X(4500)}^\mathrm{U.L.}$.  \label{tab:xsec-gx4500}}
\end{table}

\subsection{Systematic uncertainty}

Since the same selection criteria have been applied to the
$\EE\to\phi\chi_{c1,c2}$ and $\EE\to\gamma X$ processes, they share
most of the systematic uncertainties, such as the tracking efficiency, PID
efficiency etc. (cf.~\cref{subsec:sys-cj}), and their contributions are listed in~\cref{tab:sys-err-gx}. The
systematic uncertainties specifically for the $\EE\to\gamma X$ process are
described below.

The uncertainty due to the signal shape is considered by varying the mass
and width of $X$ states within $\pm1\sigma$, and changing the signal shape
to a MC shape convolved with a $2\mev$ Gaussian resolution
function~\cite{BESIII:2013fnz}. For the uncertainty due to background,
the number of $\phi\chi_{c1,c2}$ background events is varied within
$\pm1\sigma$, 
and the smooth polynomial background is studied
by varying the order of the polynomial or replacing it with a shape estimated from the sideband data in the fit.
The uncertainty associated with the fit range is determined by
varying the fit range within $\pm10\mev$. By taking these sources into
consideration in the fit, the most conservative upper limit
for $\EE\to\gamma X$ is reported.

\begin{table}[H]
	\setlength\tabcolsep{3pt}
	\centering
	
	\begin{tabular}{cccc}
		\hline\hline
		Source               & $\gamma X(4140)$ & $\gamma X(4274)$ & $\gamma X(4500)$ \\
		\hline
		Luminosity			 & 0.6				& 0.6				& 0.6 \\
		Tracking             & 2.5              & 2.5              & 2.4              \\
		Photon               & 0.8              & 0.6              & 0.5              \\
		$\ks$ reconstrcution & 0.3              & 0.3              & 0.3              \\
		Kinematic fit        & 0.6              & 0.5              & 0.5              \\
		$\mathcal{B}(\phi)$  & 1.1              & 1.1              & 1.1              \\
		$\mathcal{B}(\jpsi)$ & 0.6              & 0.6              & 0.6              \\
		MUC                  & 1.1              & 1.1              & 1.2              \\
		\hline
		Total                & 3.2              & 3.1              & 3.1              \\    
		\hline\hline       
	\end{tabular}
\caption{Systematic uncertainty sources and their
	contributions (in \%) for the cross section of $\EE\to\gamma
	X(4140)/\gamma X(4274)/\gamma
	X(4500)$.  \label{tab:sys-err-gx}}
\end{table}

\section{Summary}
In summary, with $6.4~\invfb$ of data taken from $\sqrt{s}=4.600$ to
4.951$\gev$, the process of $\EE\to\gamma\phi\jpsi$ is studied at
BESIII.  The $\EE\to\phi\chi_{c1,c2}$ processes with
$\chi_{c1,c2}\to\gamma\jpsi$ are observed with significances over
10$\sigma$. The $\sqrt{s}$-dependent Born cross sections of $\EE \to
\phi\chi_{c1,c2}$ are also measured from 4.600 to 4.951$\gev$.

We search for potential vector $Y$-states in the cross section line
shape of $\EE\to\phi\chi_{c1,c2}$, which might contain
$c\bar{c}s\bar{s}$ components in their internal structure.  For the
$\EE\to\phi\chico$ process, we find no obvious structure in the cross
section line shape, and a continuum amplitude can well describe it. For
the $\EE\to\phi\chict$ process, there is an enhancement in the cross
section line shape. A fit to the cross section with a single BW
resonance gives $M=(4672.8\pm10.8\pm3.9)\mevcc$ and
$\Gamma=(93.2\pm19.8\pm9.4)\mev$ for the mass and width of the
structure. The significance of the resonance hypothesis over
non-resonance hypothesis is estimated to be 3.1$\sigma$. The mass and
width of the resonance are consistent with the $Y(4660)$ reported in
$\EE\to\pp\psip$~\cite{belley4360,BaBar:2012hpr,Belle:2014wyt,BESIII:2021njb}.
An alternative fit to the cross section with the coherent sum of a BW
and a continuum amplitude gives $M=(4701.8\pm10.9\pm2.7)\mevcc$ and
$\Gamma=(30.5\pm22.3\pm14.6)\mev$ for the mass and width of the
structure, which has a higher mass and narrower width. The
significance for the resonance hypothesis in this model is estimated
to be 3.6$\sigma$. However, within the current uncertainties, we are
not able to distinguish whether it is the same structure as the
$Y(4660)$, and the significance for the second fit over the first one is
only $2.3\sigma$. This is the first evident structure observed
in the $\phi\chict$ system.

We also search for a possible $X$-state in the $\phi\jpsi$ system
through the radiative process $\EE\to\gamma X\to\gamma\phi\jpsi$. The
$\phi\jpsi$ spectrum can be well described by the $\phi\chi_{c1,c2}$ and
background events, and no other structure is evident in the
$M(\phi\jpsi)$ mass distribution. The $X(4140)$, $X(4274)$ and
$X(4500)$ resonances reported by the LHCb
Collaboration~\cite{LHCb:2021uow} are not observed, and the upper
limits on the Born cross sections for $\EE\to\gamma X(4140)$, $\gamma
X(4274)$, $\gamma X(4500)\to\gamma\phi\jpsi$ at the 90\% C.L. are
determined.

\acknowledgments
The BESIII Collaboration thanks the staff of BEPCII and the IHEP computing center for their strong support. This work is supported in part by National Key R\&D Program of China under Contracts Nos. 2020YFA0406300, 2020YFA0406400; National Natural Science Foundation of China (NSFC) under Contracts Nos. 11975141, 11635010, 11735014, 11835012, 11935015, 11935016, 11935018, 11961141012, 12022510, 12025502, 12035009, 12035013, 12192260, 12192261, 12192262, 12192263, 12192264, 12192265; the Chinese Academy of Sciences (CAS) Large-Scale Scientific Facility Program; Joint Large-Scale Scientific Facility Funds of the NSFC and CAS under Contract No. U1832207; the CAS Center for Excellence in Particle Physics 
(CCEPP); 100 Talents Program of CAS; Project ZR2022JQ02 supported by Shandong Provincial Natural Science Foundation; The Institute of Nuclear and Particle Physics (INPAC) and Shanghai Key Laboratory for Particle Physics and Cosmology; ERC under Contract No. 758462; European Union's Horizon 2020 research and innovation programme under Marie Sklodowska-Curie grant agreement under Contract No. 894790; German Research Foundation DFG under Contracts Nos. 443159800, 455635585, Collaborative Research Center CRC 1044, FOR5327, GRK 2149; Istituto Nazionale di Fisica Nucleare, Italy; Ministry of Development of Turkey under Contract No. DPT2006K-120470; National Science and Technology fund; National Science Research and Innovation Fund (NSRF) via the Program Management Unit for Human Resources \& Institutional Development, Research and Innovation under Contract No. B16F640076; Olle Engkvist Foundation under Contract No. 200-0605; STFC (United Kingdom); Suranaree University of Technology (SUT), Thailand Science Research and Innovation (TSRI), and National Science Research and Innovation Fund (NSRF) under Contract No. 160355; The Royal Society, UK under Contracts Nos. DH140054, DH160214; The Swedish Research Council; U. S. Department of Energy under Contract No. DE-FG02-05ER41374.

\appendix
\section{\texorpdfstring{Fit result for \bm{$M(\gamma\jpsi)$}}{Fit result for M(gamma J/psi)}}\label{app-a}
\begin{figure}[H]
	
	\begin{flushleft}
		
		\includegraphics[width=0.48\linewidth]{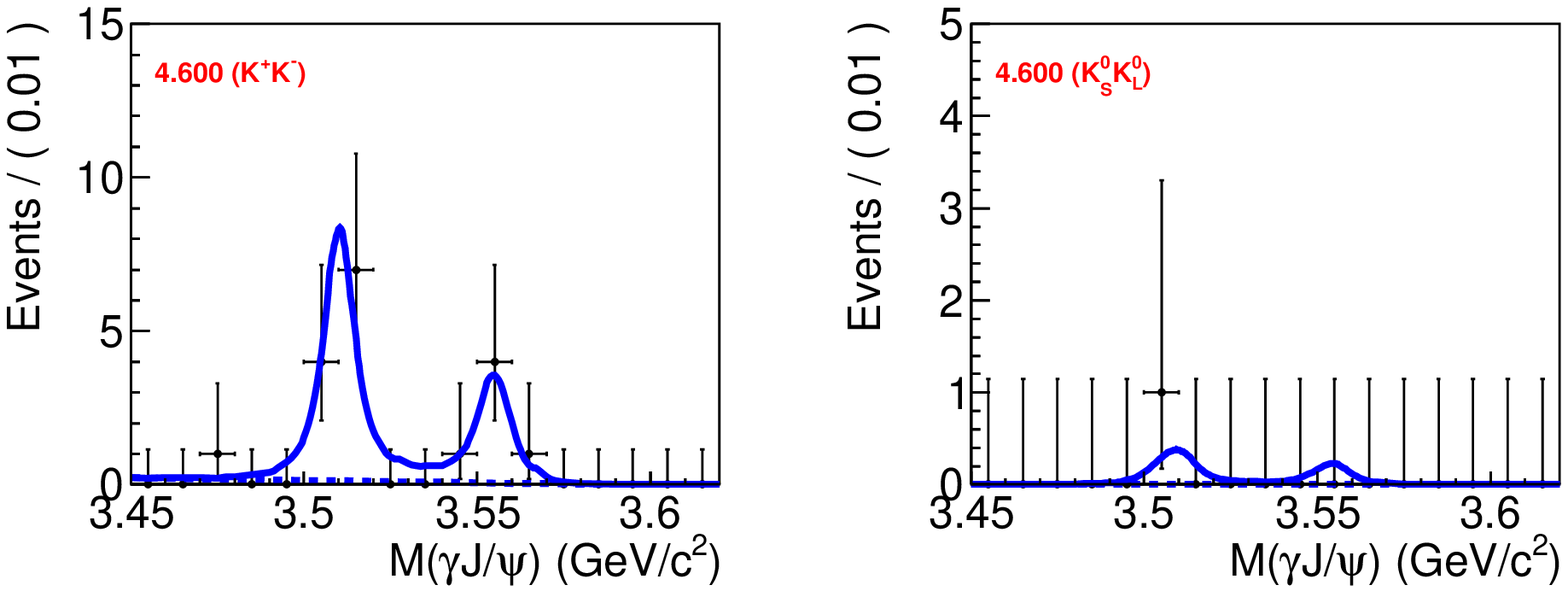}
		\includegraphics[width=0.48\linewidth]{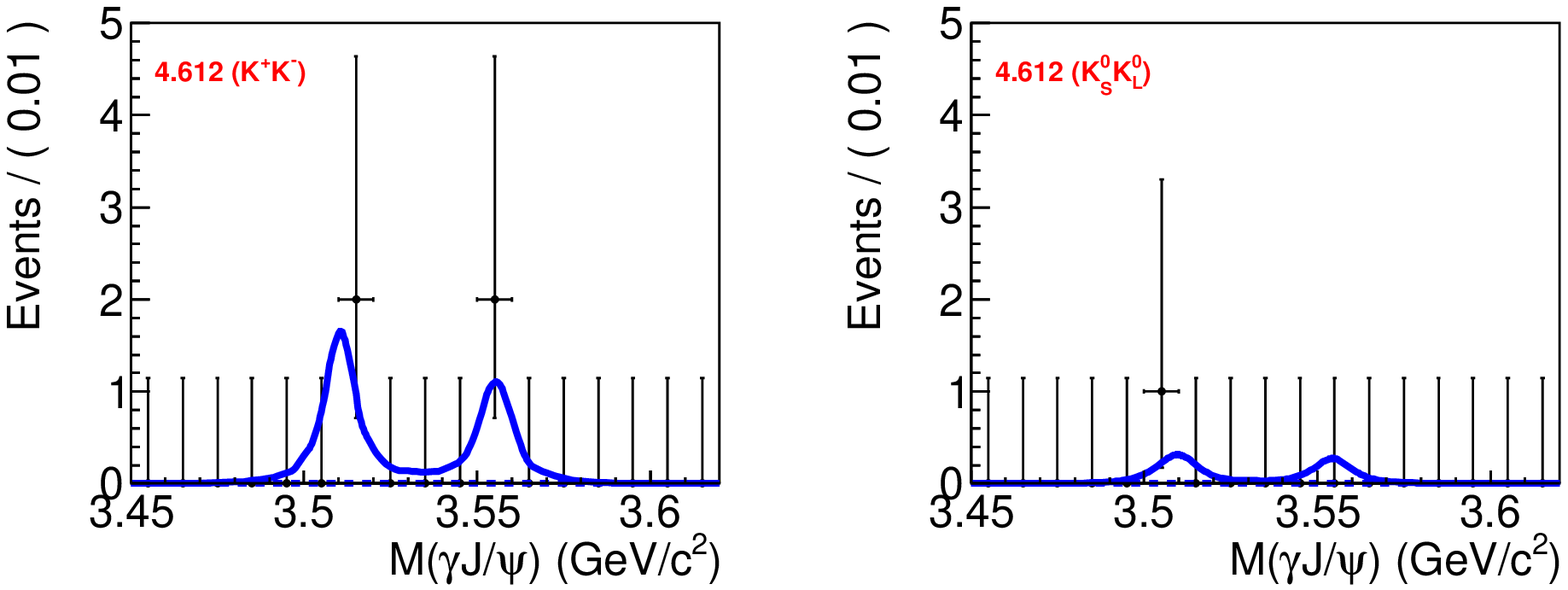}
		\vfill
		\includegraphics[width=0.48\linewidth]{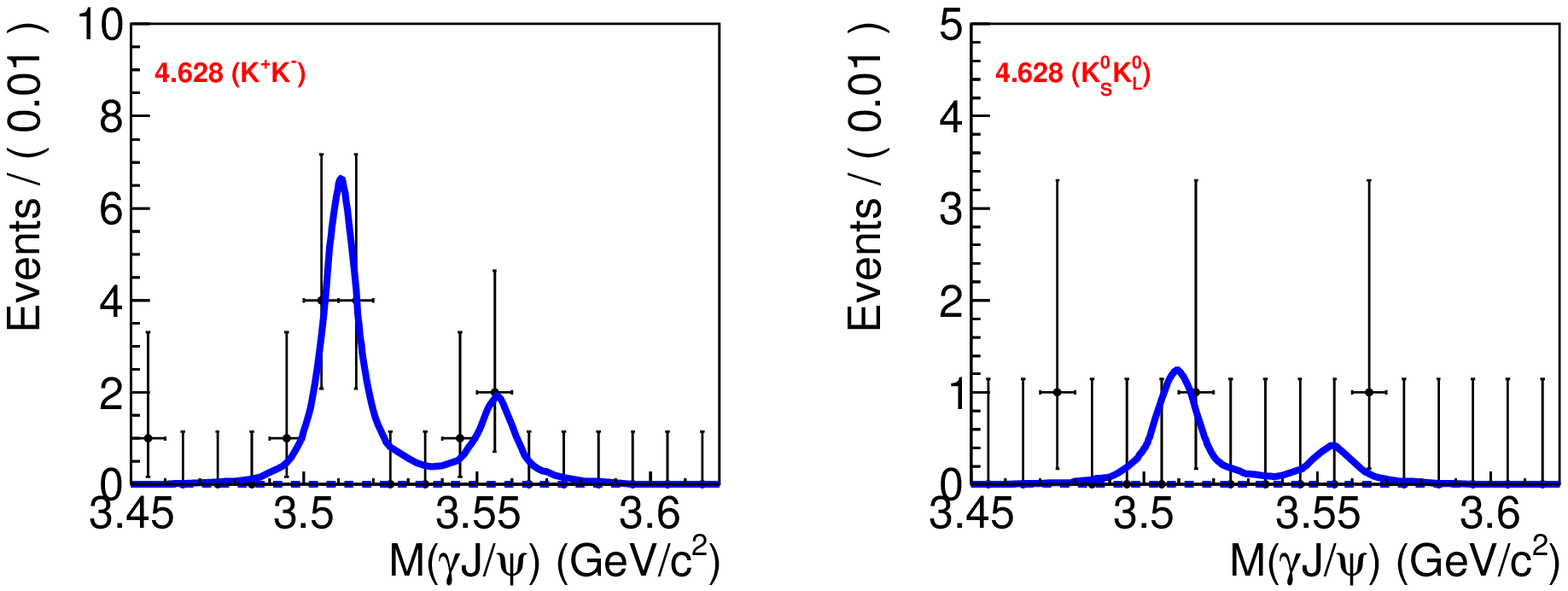}
		\includegraphics[width=0.48\linewidth]{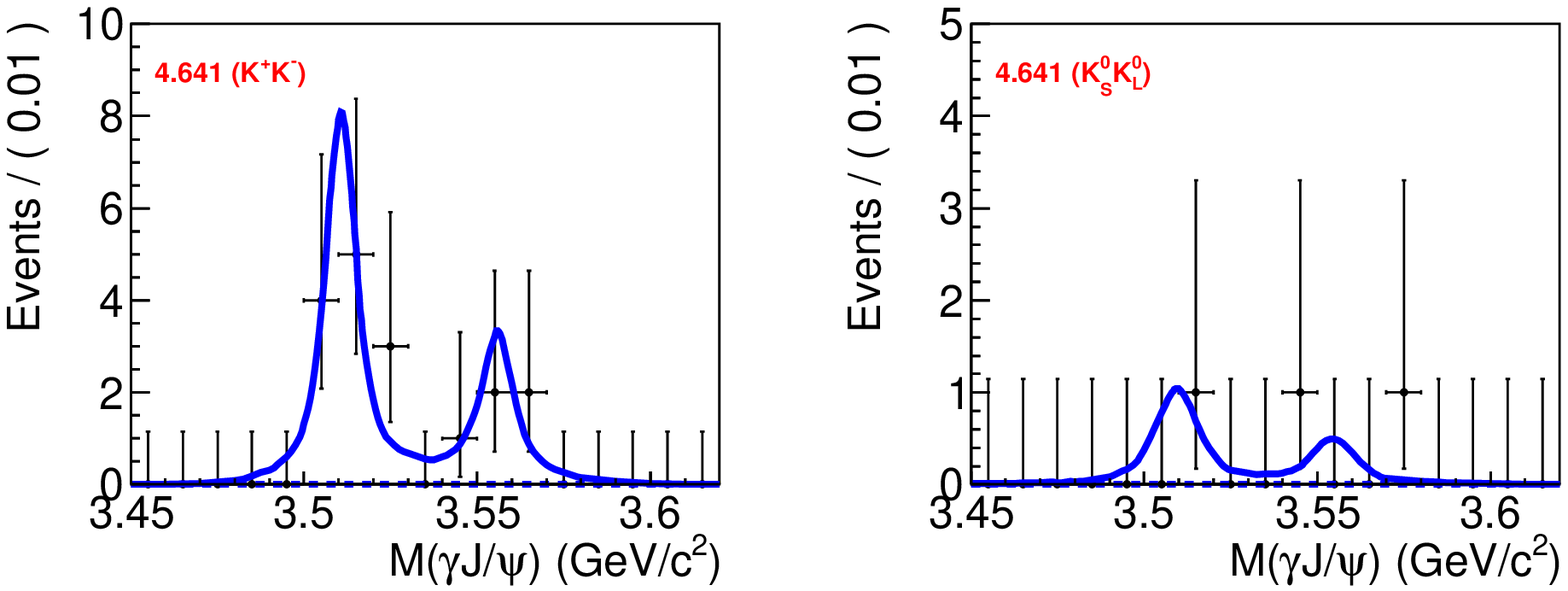}
		\vfill
		\includegraphics[width=0.48\linewidth]{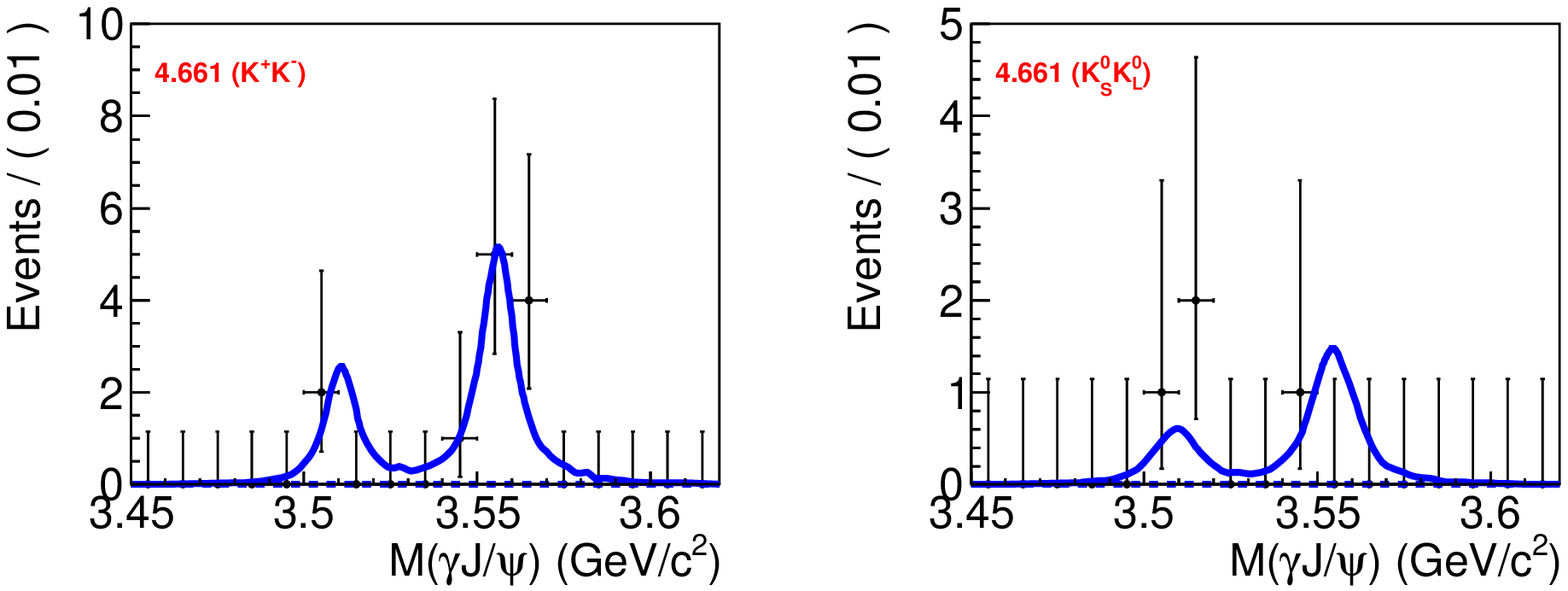}
		\includegraphics[width=0.48\linewidth]{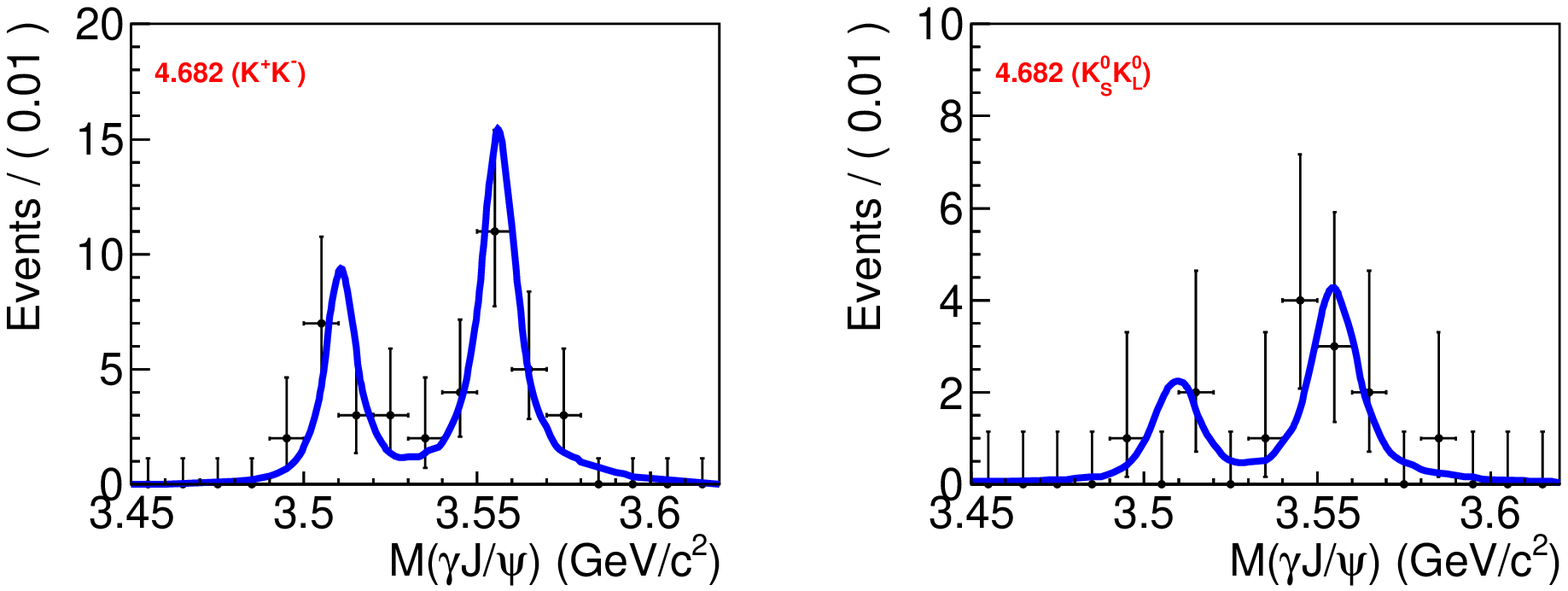}
		\vfill
		\includegraphics[width=0.48\linewidth]{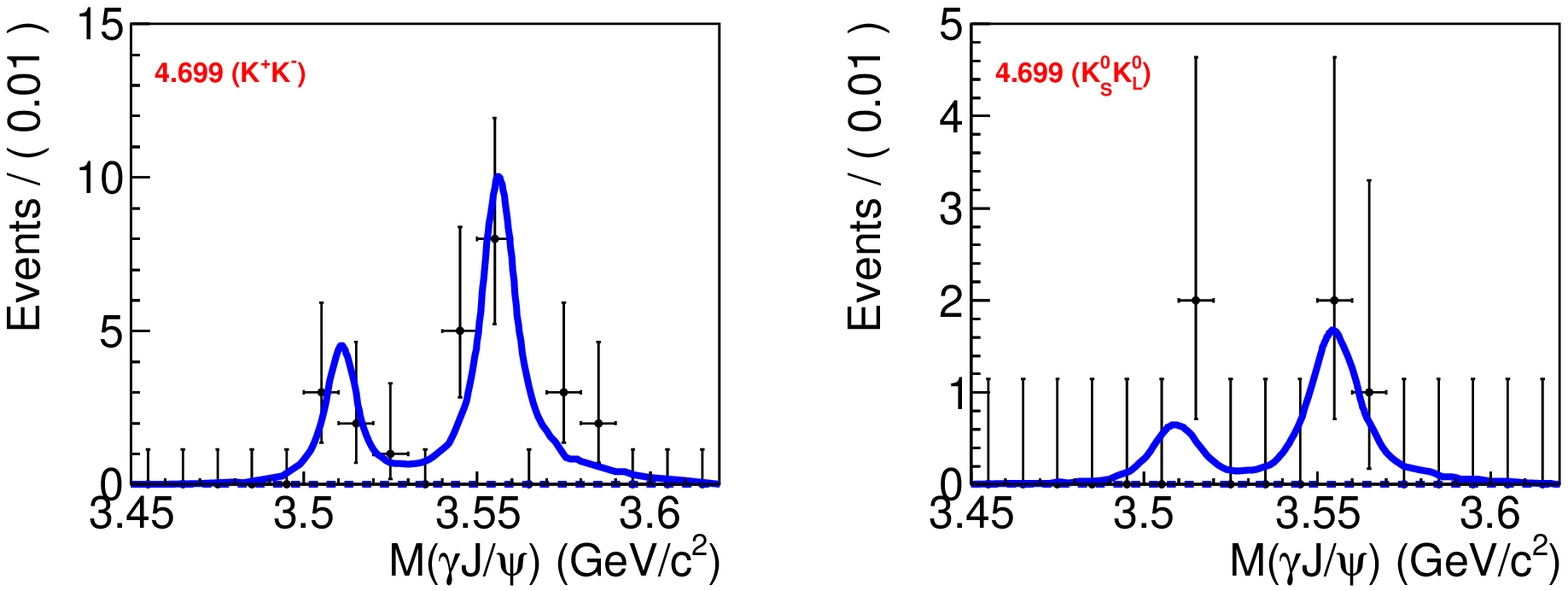}
		\includegraphics[width=0.48\linewidth]{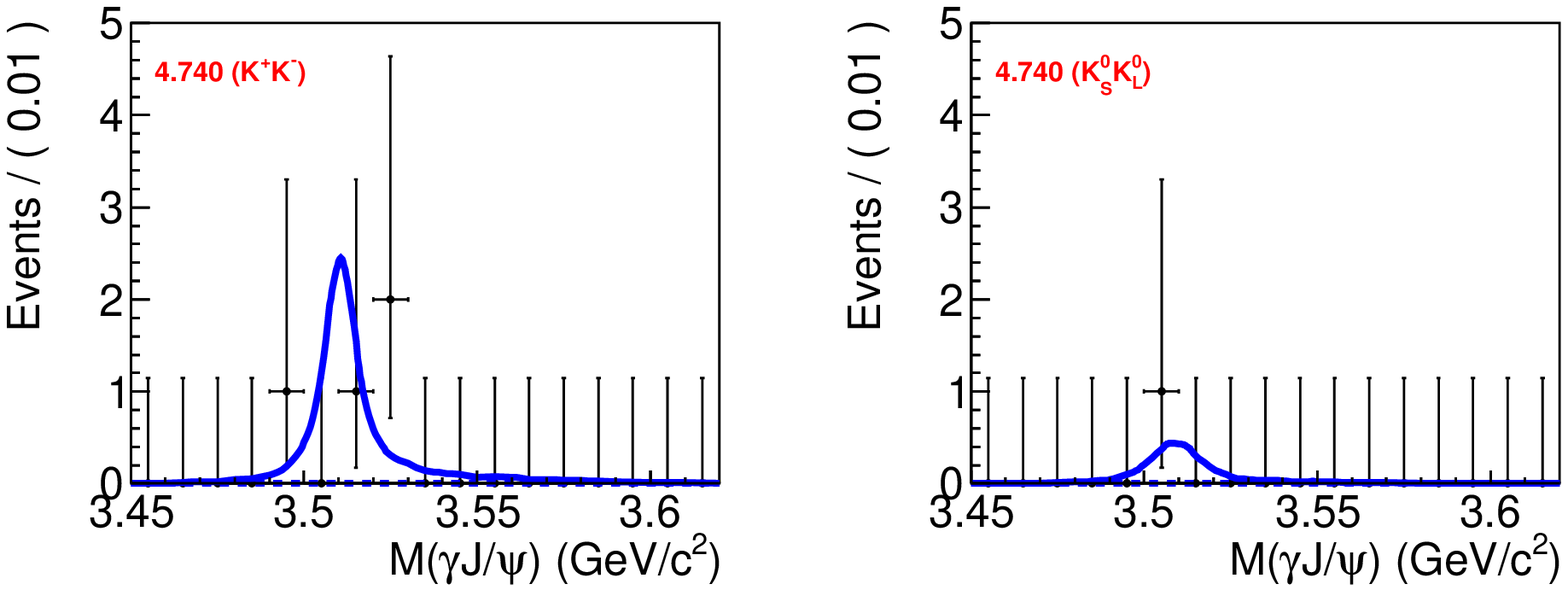}
		\includegraphics[width=0.48\linewidth]{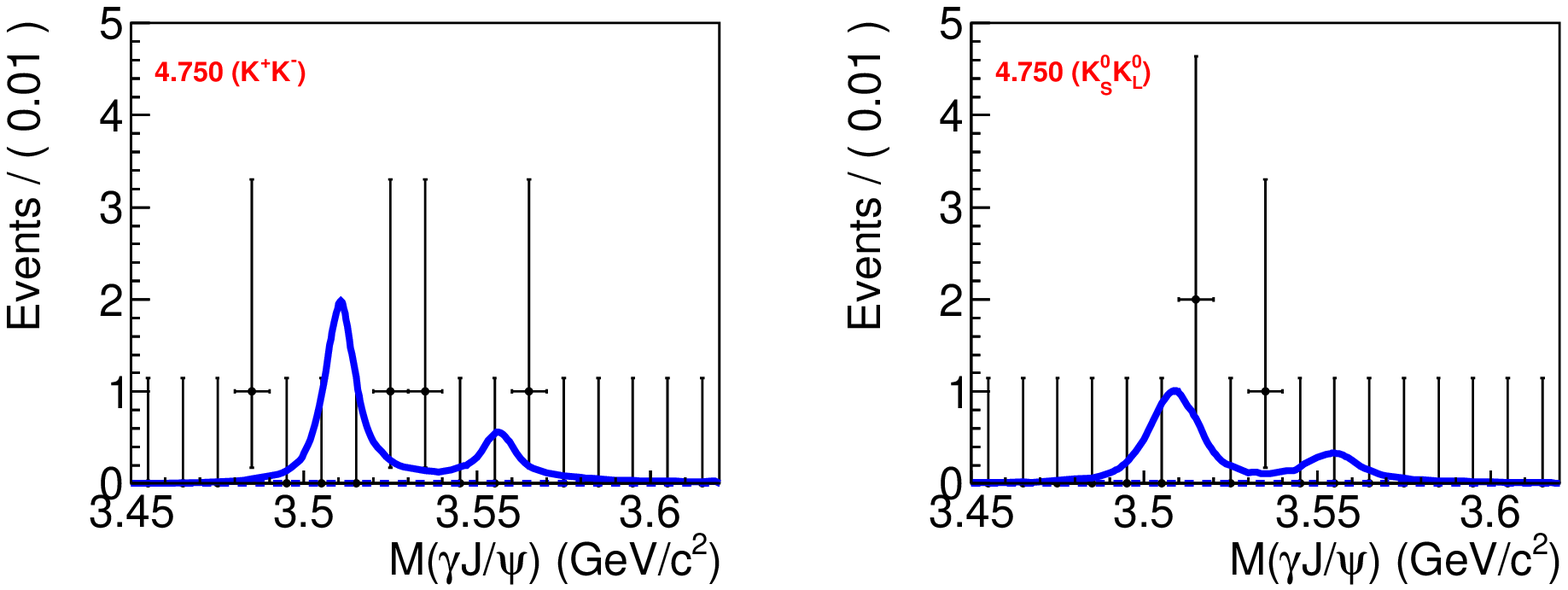}
		\includegraphics[width=0.48\linewidth]{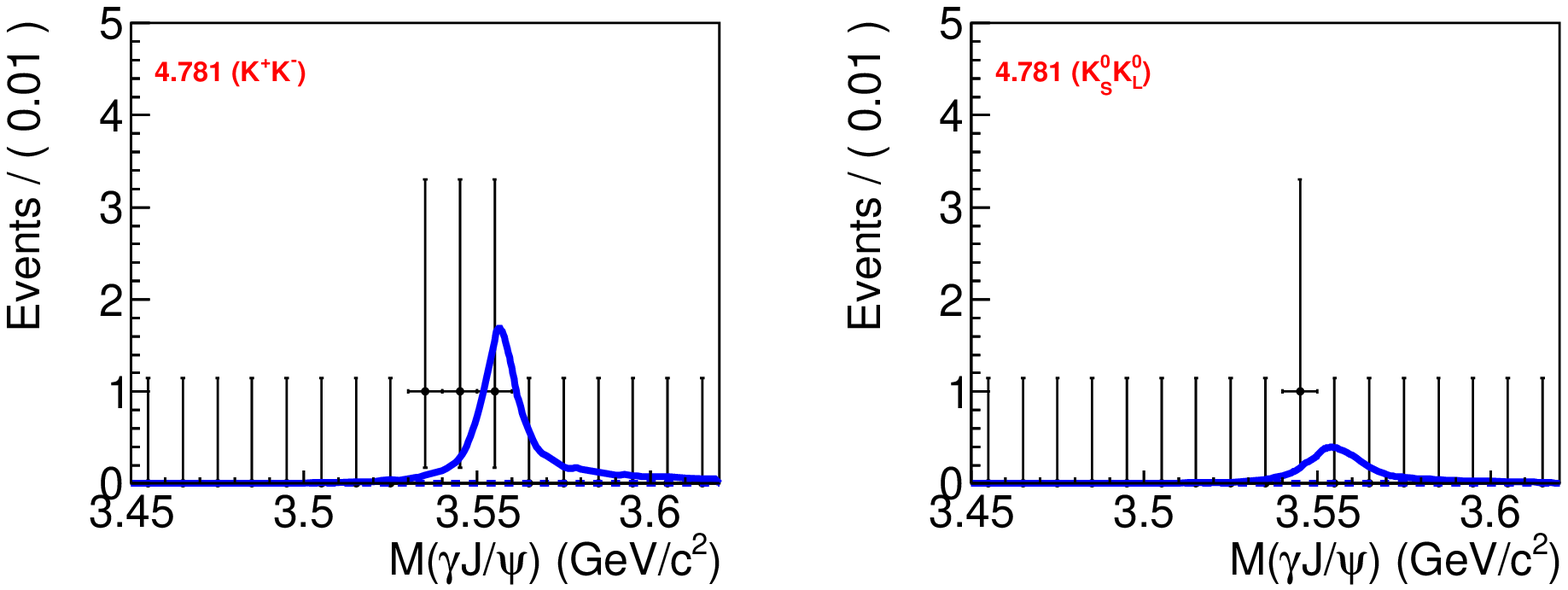}
		\includegraphics[width=0.48\linewidth]{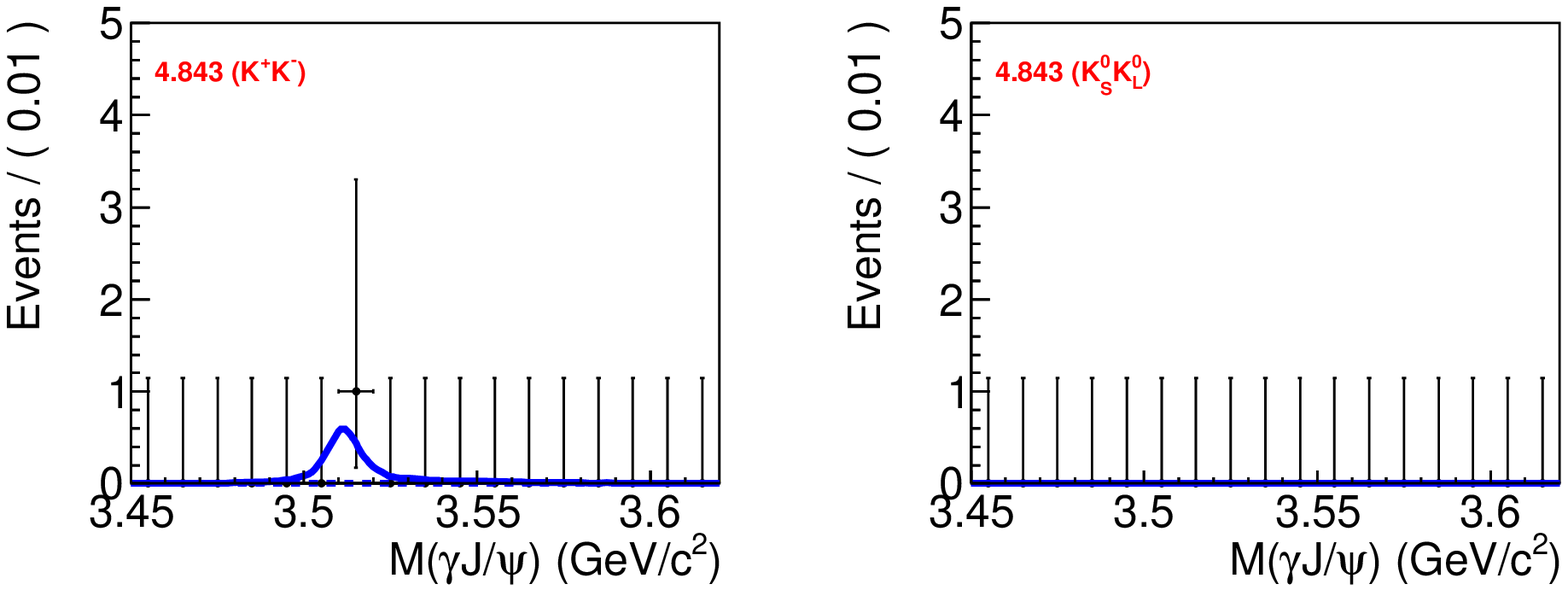}
		\includegraphics[width=0.48\linewidth]{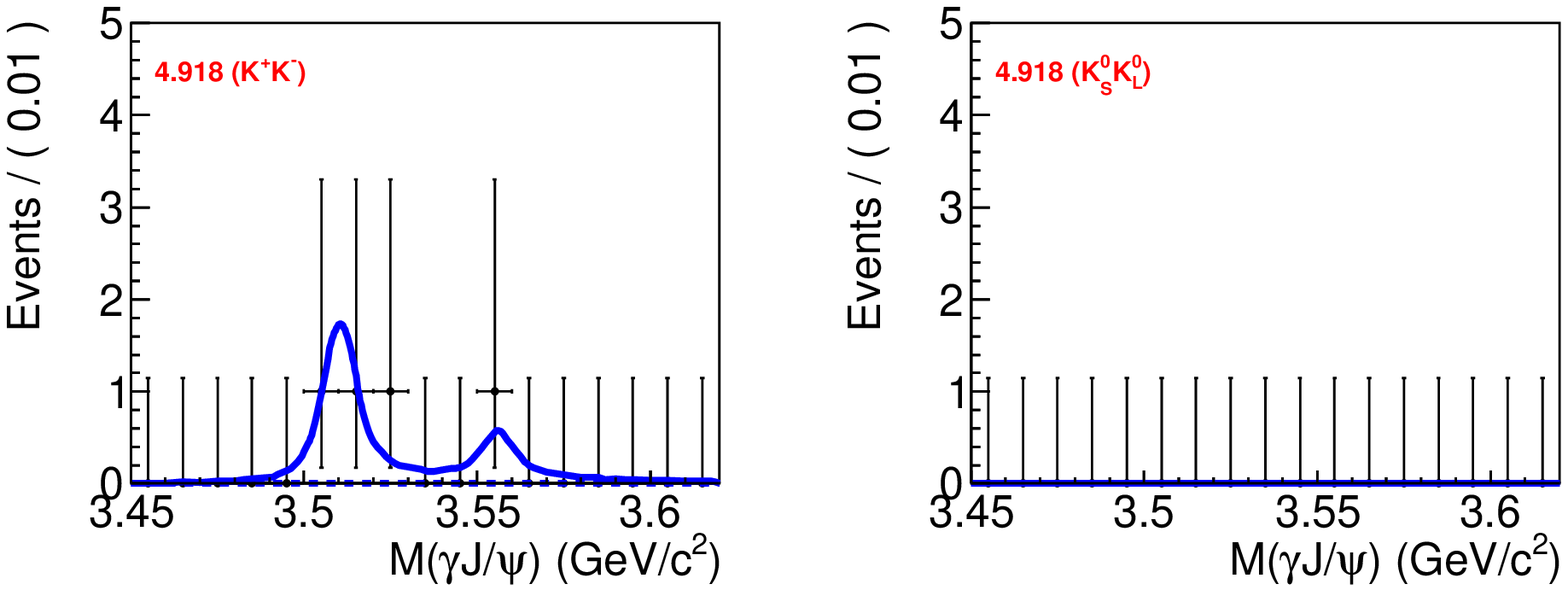}
		\includegraphics[width=0.48\linewidth]{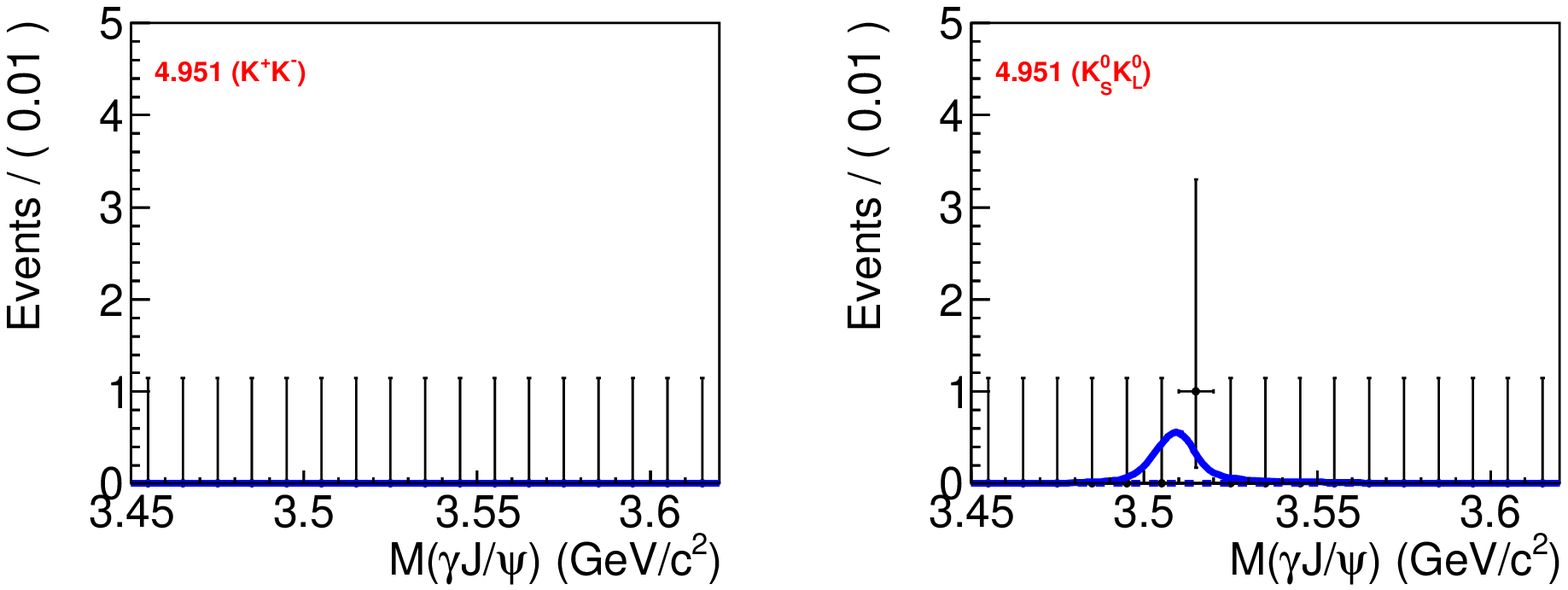}
		
		\caption{The simultaneous fit to $M(\gamma\jpsi)$ for $\phi\to\kk$ and $\phi\to\ks\kl$ modes from 4.600 to 4.951$\gev$. Dots with error bars are data, blue lines are the fit results.\label{fig:fit-result}}
	\end{flushleft}
\end{figure}

\section{Systematic uncertainty in cross section measurement \label{app:sys-pcj}}

\begin{table}[H]
	\centering

	\scalebox{0.75}{
		\begin{tabular}{cccccccccccccc}
			\hline\hline
			Source & 4.600 & 4.612 & 4.628 & 4.641 & 4.661 & 4.682 & 4.699 & 4.740 & 4.750 & 4.781 & 4.843 & 4.918 & 4.951 \\
			\hline
			
			Luminosity & 0.60 & 0.60 & 0.60 & 0.60 & 0.60 & 0.60 & 0.60 & 0.60 & 0.60 & 0.60 & 0.60 & 0.60 & 0.60 \\
			
			Tracking & 2.47 & 2.46 & 2.45 & 2.44 & 2.42 & 2.42 & 2.41 & 2.40 & 2.40 & 2.40 & 2.39 & 2.39 & 2.38 \\
						
			Photon & 0.80 & 0.78 & 0.75 & 0.73 & 0.68 & 0.65 & 0.62 & 0.57 & 0.57 & 0.54 & 0.50 & 0.48 & 0.46 \\

			$\ks$ & 0.28 & 0.28 & 0.27 & 0.26 & 0.25 & 0.25 & 0.25 & 0.24 & 0.24 & 0.24 & 0.23 & 0.23 & 0.23 \\
			
			Kinematic fit & 0.54 & 0.53 & 0.52 & 0.52 & 0.50 & 0.49 & 0.48 & 0.46 & 0.46 & 0.45 & 0.44 & 0.43 & 0.43 \\
						
			$\mathcal{B}(\phi)$ & 0.81 & 0.82 & 0.82 & 0.82 & 0.83 & 0.83 & 0.83 & 0.84 & 0.83 & 0.84 & 0.84 & 0.84 & 0.84 \\
			
			$\mathcal{B}(\chico)$ & 2.90 & 2.90 & 2.90 & 2.90 & 2.90 & 2.90 & 2.90 & 2.90 & 2.90 & 2.90 & 2.90 & 2.90 & 2.90 \\
			
			$\mathcal{B}(\jpsi)$ & 0.60 & 0.60 & 0.60 & 0.60 & 0.60 & 0.60 & 0.60 & 0.60 & 0.60 & 0.60 & 0.60 & 0.60 & 0.60 \\
			
			Radiative correction & 3.16 & 2.48 & 1.62 & 1.90 & 1.42 & 0.40 & 0.49 & 0.57 & 0.70 & 1.60 & 1.18 & 2.46 & 1.37 \\
			
			MC model & 0.30 & 0.46 & 0.21 & 0.23 & 0.10 & 0.18 & 0.28 & 0.48 & 0.43 & 0.57 & 0.49 & 0.48 & 0.51 \\
			
			Muon hit depth & 1.51 & 0.87 & 1.15 & 1.09 & 1.06 & 0.86 & 0.92 & 0.97 & 1.44 & 1.28 & 1.39 & 0.95 & 1.34 \\
			
			Fit related & 5.54 & 5.54 & 5.54 & 5.54 & 5.54 & 5.54 & 5.54 & 5.54 & 5.54 & 5.54 & 5.54 & 5.54 & 5.54 \\
			
			\hline

			Total & 7.75 & 7.39 & 7.17 & 7.22 & 7.09 & 6.93 & 6.94 & 6.96 & 7.04 & 7.16 & 7.09 & 7.34 & 7.11 \\
			\hline
			\hline
			
		\end{tabular}
	}
\caption{The systematic uncertainties (in \%) for $\EE\to \phi\chico$ cross sections at each energy point.\label{tab:sys-pc1}}
\end{table}

\begin{table}[H]
	\centering

		\scalebox{0.75}{
		\begin{tabular}{cccccccccccccc}
			\hline\hline
			Source & 4.600 & 4.612 & 4.628 & 4.641 & 4.661 & 4.682 & 4.699 & 4.740 & 4.750 & 4.781 & 4.843 & 4.918 & 4.951 \\
			\hline
			
			Luminosity & 0.60 & 0.60 & 0.60 & 0.60 & 0.60 & 0.60 & 0.60 & 0.60 & 0.60 & 0.60 & 0.60 & 0.60 & 0.60 \\
			
			Tracking & 2.55 & 2.50 & 2.49 & 2.47 & 2.46 & 2.44 & 2.44 & 2.42 & 2.42 & 2.42 & 2.40 & 2.40 & 2.40 \\

			Photon & 0.92 & 0.88 & 0.84 & 0.81 & 0.77 & 0.73 & 0.70 & 0.64 & 0.63 & 0.59 & 0.54 & 0.50 & 0.49 \\

			$\ks$ & 0.33 & 0.30 & 0.29 & 0.28 & 0.27 & 0.27 & 0.26 & 0.25 & 0.25 & 0.25 & 0.24 & 0.24 & 0.24 \\
			
			Kinematic fit & 0.58 & 0.57 & 0.55 & 0.54 & 0.53 & 0.52 & 0.51 & 0.49 & 0.48 & 0.47 & 0.45 & 0.44 & 0.44 \\

			$\mathcal{B}(\phi)$ & 0.80 & 0.81 & 0.81 & 0.81 & 0.82 & 0.82 & 0.82 & 0.83 & 0.83 & 0.83 & 0.83 & 0.83 & 0.84 \\
			
			$\mathcal{B}(\chict)$ & 2.60 & 2.60 & 2.60 & 2.60 & 2.60 & 2.60 & 2.60 & 2.60 & 2.60 & 2.60 & 2.60 & 2.60 & 2.60 \\
			
			$\mathcal{B}(\jpsi)$ & 0.60 & 0.60 & 0.60 & 0.60 & 0.60 & 0.60 & 0.60 & 0.60 & 0.60 & 0.60 & 0.60 & 0.60 & 0.60 \\

			Radiative correction & 5.17 & 7.42 & 6.23 & 6.57 & 3.27 & 5.31 & 10.57 & 17.33 & 15.53 & 5.61 & 1.28 & 0.69 & 2.93 \\

			MC model & 0.38 & 0.43 & 0.37 & 0.38 & 0.27 & 0.16 & 0.11 & 0.24 & 0.34 & 0.39 & 0.47 & 0.51 & 0.44 \\
			
			Muon hit depth & 1.45 & 0.85 & 1.12 & 1.08 & 1.05 & 0.85 & 0.91 & 0.96 & 1.43 & 1.26 & 1.37 & 0.95 & 1.33 \\
			
			Fit related & 7.14 & 7.14 & 7.14 & 7.14 & 7.14 & 7.14 & 7.14 & 7.14 & 7.14 & 7.14 & 7.14 & 7.14 & 7.14\\
			
			\hline

			Total & 9.79 & 11.07 & 10.33 & 10.53 & 8.83 & 9.74 & 13.36 & 19.16 & 17.58 & 9.94 & 8.32 & 8.19 & 8.72\\
			\hline
			\hline
			
		\end{tabular}
	}
\caption{The systematic uncertainties (in \%) for $\EE\to \phi\chict$ cross sections at each energy point.\label{tab:sys-pc2}}
\end{table}


\clearpage


\newpage
\begin{thebibliography}{99}
	
\bibitem{babary4260}
B.~Aubert et al. [BaBar Collaboration],
\emph{Observation of a broad structure in the $\pi^+ \pi^- J/\psi$ mass spectrum around 4.26 GeV/c$^2$},
\emph{Phys. Rev. Lett.} \textbf{95} (2005), 142001.

\bibitem{cleoy4260}
T.~E.~Coan et al. [CLEO Collaboration],
\emph{Charmonium decays of Y(4260), psi(4160) and psi(4040)},
\emph{Phys. Rev. Lett.} \textbf{96} (2006), 162003.

\bibitem{belley4260}
C.~Z.~Yuan et al. [Belle Collaboration],
\emph{Measurement of $\EE\to\pp\jpsi$ cross section via initial state radiation at Belle},
\emph{Phys. Rev. Lett.} \textbf{99} (2007), 182004.

\bibitem{babary4360}
B.~Aubert et al. [BaBar Collaboration],
\emph{Evidence of a broad structure at an invariant mass of 4.32  $GeV/c^{2}$ in the reaction $e^{+} e^{-} \to \pi^{+} \pi^{-} \psi(2S)$ measured at BaBar},
\emph{Phys. Rev. Lett.} \textbf{98} (2007), 212001.

\bibitem{belley4360}
X.~L.~Wang et al. [Belle Collaboration],
\emph{Observation of Two Resonant Structures in $\EE\to\pp\psi(2S)$ via Initial State Radiation at Belle},
\emph{Phys. Rev. Lett.} \textbf{99} (2007), 142002.


\bibitem{BaBar:2012hpr}
J.~P.~Lees et al. [BaBar Collaboration],
\emph{Study of the reaction $e^{+}e^{-}\to \psi(2S)\pi^{+}\pi^{-}$ via initial-state radiation at BaBar},
\emph{Phys. Rev. D} \textbf{89} (2014) no.11, 111103. 

\bibitem{Belle:2014wyt}
X.~L.~Wang et al. [Belle Collaboration],
\emph{Measurement of $e^+e^- \to \pi^+\pi^-\psi(2S)$ via Initial State Radiation at Belle},
\emph{Phys. Rev. D} \textbf{91} (2015), 112007. 



\bibitem{BESIII:2021njb}
M.~Ablikim et al. [BESIII Collaboration],
\emph{Cross section measurement of $e^+e^-\rightarrow\pi^+\pi^-(3686)$ from $\sqrt{s}=$4.0076 to 4.6984~GeV},
\emph{Phys. Rev. D} \textbf{104} (2021), no.5, 052012. 

\bibitem{potential-mode}
W.~Kwong, J.~L.~Rosner and C.~Quigg,
\emph{Heavy Quark Systems},
\emph{Ann. Rev. Nucl. Part. Sci.} \textbf{37} (1987), 325-382.


\bibitem{Zyla:2020zbs}
P.A.~Zyla et al. [Particle Data Group],
\emph{Review of Particle Physics},
PTEP \textbf{2020}, no.8, 083C01 (2020) and 2021 update


\bibitem{Brambilla:2019esw}
N.~Brambilla, S.~Eidelman, C.~Hanhart, A.~Nefediev, C.~P.~Shen, C.~E.~Thomas, A.~Vairo and C.~Z.~Yuan,
\emph{The $XYZ$ states: experimental and theoretical status and perspectives},
\emph{Phys. Rept.} \textbf{873} (2020), 1-154.

\bibitem{Chen:2016qju}
H.~X.~Chen, W.~Chen, X.~Liu and S.~L.~Zhu,
\emph{The hidden-charm pentaquark and tetraquark states},
\emph{Phys. Rept.} \textbf{639} (2016), 1-121.


\bibitem{Jia:2019gfe}
S.~Jia et al. [Belle Collaboration],
\emph{Observation of a vector charmoniumlike state in $e^+e^- \to D^+_sD_{s1}(2536)^-+c.c.$},
\emph{Phys. Rev. D} \textbf{100} (2019) no.11, 111103. 

\bibitem{Jia:2020epr}
S.~Jia et al. [Belle Collaboration],
\emph{Evidence for a vector charmoniumlike state in $e^+e^- \to D^+_sD^*_{s2}(2573)^-+c.c.$},
\emph{Phys. Rev. D} \textbf{101} (2020) no.9, 091101. 

\bibitem{Karliner:2016ith}
M.~Karliner and J.~L.~Rosner,
\emph{Exotic resonances due to $\eta$ exchange},
\emph{Nucl. Phys. A} \textbf{954} (2016), 365-370. 

\bibitem{He:2019csk}
J.~He, Y.~Liu, J.~T.~Zhu and D.~Y.~Chen,
\emph{Y(4626) as a molecular state from interaction ${D}^*_s{\bar{D}}_{s1}(2536)-{D}_s{\bar{D}}_{s1}(2536)$},
\emph{Eur. Phys. J. C} \textbf{80} (2020) no.3, 246.

\bibitem{Deng:2019dbg}
C.~Deng, H.~Chen and J.~Ping,
\emph{Can the state $Y(4626)$ be a $P$-wave tetraquark state $[cs][\bar{c}\bar{s}]$?},
\emph{Phys. Rev. D} \textbf{101} (2020) no.5, 054039. 

\bibitem{BESIII:2017qtm}
M.~Ablikim et al. [BESIII Collaboration],
\emph{Observation of $e^{+}e^{-} \to \phi\chi_{c1}$ and $\phi\chi_{c2}$ at $\sqrt{s}$=4.600 GeV},
\emph{Phys. Rev. D} \textbf{97} (2018) no.3, 032008. 

\bibitem{cdfy4140}
T.~Aaltonen et al. [CDF Collaboration],
\emph{Evidence for a Narrow Near-Threshold Structure in the $J/\psi\phi$ Mass Spectrum in $B^+\to J/\psi\phi K^+$ Decays},
\emph{Phys. Rev. Lett.} \textbf{102} (2009), 242002. 




\bibitem{cdf-X2}
T.~Aaltonen et al. [CDF Collaboration],
\emph{Observation of the $Y(4140)$ Structure in the $J/\psi\phi$ Mass Spectrum in $B^\pm\to J/\psi\phi K^\pm$ Decays},
\emph{Mod. Phys. Lett. A} \textbf{32} (2017) no.26, 1750139. 

\bibitem{ChengPing:2009vu}
C. P. Shen [Belle Collaboration],
\emph{XYZ particles at Belle},
\emph{Chin. Phys. C} \textbf{34} (2010), 615-620. 



\bibitem{lhcb-X1}
R.~Aaij et al. [LHCb Collaboration],
\emph{Search for the $X(4140)$ state in $B^+ \to J/\psi \phi K^+$ decays},
\emph{Phys. Rev. D} \textbf{85} (2012), 091103.

\bibitem{lhcb-X2}
R.~Aaij et al. [LHCb Collaboration],
\emph{Observation of $J/\psi\phi$ structures consistent with exotic states from amplitude analysis of $B^+\to J/\psi \phi K^+$ decays},
\emph{Phys. Rev. Lett.} \textbf{118} (2017) no.2, 022003.

\bibitem{lhcb-X3}
R.~Aaij et al. [LHCb Collaboration],
\emph{Amplitude analysis of $B^+\to J/\psi \phi K^+$ decays},
\emph{Phys. Rev. D} \textbf{95} (2017) no.1, 012002.


\bibitem{LHCb:2021uow}
R.~Aaij et al. [LHCb Collaboration],
\emph{Observation of New Resonances Decaying to $J/\psi K^+$ and $J/\psi \phi$},
\emph{Phys. Rev. Lett.} \textbf{127} (2021) no.8, 082001.

\bibitem{cms-X}
S.~Chatrchyan et al. [CMS Collaboration],
\emph{Observation of a Peaking Structure in the $J/\psi \phi$ Mass Spectrum from $B^{\pm} \to J/\psi \phi K^{\pm}$ Decays},
\emph{Phys. Lett. B} \textbf{734} (2014), 261-281.

\bibitem{d0-X1}
V.~M.~Abazov et al. [D0 Collaboration],
\emph{A Quasi model independent search for new physics at large transverse momentum},
\emph{Phys. Rev. D} \textbf{64} (2001), 012004.

\bibitem{d0-X2}
V.~M.~Abazov et al. [D0 Collaboration],
\emph{Inclusive Production of the X(4140) State in $p \overline p$ Collisions at D0},
\emph{Phys. Rev. Lett.} \textbf{115} (2015) no.23, 232001.

\bibitem{babar-X}
J.~P.~Lees et al. [BaBar Collaboration],
\emph{Study of $B^{\pm,0} \to J/\psi K^+ K^- K^{\pm,0}$ and search for $B^0 \to J/\psi\phi$ at BABAR},
\emph{Phys. Rev. D} \textbf{91} (2015) no.1, 012003.





\bibitem{BESIII:2014fob}
M.~Ablikim et al. [BESIII Collaboration],
\emph{Search for the Y(4140) via $\EE\to\gamma\phi\jpsi$ at $\sqrt{s}$=4.23 , 4.26 and 4.36 GeV},
\emph{Phys. Rev. D} \textbf{91} (2015) no.3, 032002. 




\bibitem{Ebert:2008kb}
D.~Ebert, R.~N.~Faustov and V.~O.~Galkin,
\emph{Excited heavy tetraquarks with hidden charm},
\emph{Eur. Phys. J. C} \textbf{58} (2008), 399-405.

\bibitem{Chen:2010ze}
W.~Chen and S.~L.~Zhu,
\emph{The Vector and Axial-Vector Charmonium-like States},
\emph{Phys. Rev. D} \textbf{83} (2011), 034010. 

\bibitem{Lu:2016cwr}
Q.~F.~L\"u and Y.~B.~Dong,
\emph{X(4140), X(4274), X(4500), and X(4700) in the relativized quark model},
\emph{Phys. Rev. D} \textbf{94} (2016) no.7, 074007.

\bibitem{Wu:2016gas}
J.~Wu, Y.~R.~Liu, K.~Chen, X.~Liu and S.~L.~Zhu,
\emph{X(4140), X(4270), X(4500) and X(4700) and their $cs\bar{c}\bar{s}$ tetraquark partners},
\emph{Phys. Rev. D} \textbf{94} (2016) no.9, 094031. 

\bibitem{Wang:2016gxp}
Z.~G.~Wang,
\emph{Scalar tetraquark state candidates: $X(3915)$, $X(4500)$ and $X(4700)$},
\emph{Eur. Phys. J. C} \textbf{77} (2017) no.2, 78.

\bibitem{Chen:2016oma}
H.~X.~Chen, E.~L.~Cui, W.~Chen, X.~Liu and S.~L.~Zhu,
\emph{Understanding the internal structures of the $X(4140)$, $X(4274)$, $X(4500)$ and $X(4700)$},
\emph{Eur. Phys. J. C} \textbf{77} (2017) no.3, 160.

\bibitem{Deng:2017xlb}
C.~Deng, J.~Ping, H.~Huang and F.~Wang,
\emph{Hidden charmed states and multibody color flux-tube dynamics},
\emph{Phys. Rev. D} \textbf{98} (2018) no.1, 014026. 

\bibitem{Wang:2018djr}
E.~Wang, J.~J.~Xie, L.~S.~Geng and E.~Oset,
\emph{The $X(4140)$ and $X(4160)$ resonances in the $e^+e^-\to \gamma J/\psi \phi $ reaction},
\emph{Chin. Phys. C} \textbf{43} (2019) no.11, 113101.

\bibitem{Stancu:2009ka}
F.~Stancu,
\emph{Can Y(4140) be a c anti-c s anti-s tetraquark?},
\emph{J. Phys. G} \textbf{37} (2010), 075017. 
[erratum: \emph{J. Phys. G} \textbf{46} (2019) no.1, 019501.]

\bibitem{Liu:2021xje}
X.~Liu, H.~Huang, J.~Ping, D.~Chen and X.~Zhu,
\emph{The explanation of some exotic states in the $cs{\bar{c}}{\bar{s}}$ tetraquark system},
\emph{Eur. Phys. J. C} \textbf{81} (2021) no.10, 950.




\bibitem{lum-4600}
M.~Ablikim et al. [BESIII Collaboration],
\emph{Measurement of the integrated luminosities at BESIII for data samples at collision energies around 4 GeV},
arXiv:2203.03133 [hep-ex].


\bibitem{BESIII:2022ulv}
M.~Ablikim et al. [BESIII Collaboration],
\emph{Luminosities and energies of $e^+e^-$ collision data taken between $\sqrt{s}$=4.612 GeV and 4.946 GeV at BESIII},
arXiv:2205.04809 [hep-ex].

\bibitem{BESIII:2015zbz}
M.~Ablikim et al. [BESIII Collaboration],
\emph{Measurement of the center-of-mass energies at BESIII via the di-muon process},
\emph{Chin. Phys. C} \textbf{40} (2016) no.6, 063001.


\bibitem{Ablikim:2009aa}
M.~Ablikim et al. [BESIII Collaboration],
\emph{Design and Construction of the BESIII Detector},
\emph{Nucl. Instrum. Meth. A} {\bf 614} (2010), 345.

\bibitem{Yu:IPAC2016-TUYA01}
C.~H.~Yu et al.,
\emph{BEPCII Performance and Beam Dynamics Studies on Luminosity},
Proceedings of IPAC2016, Busan, Korea, 2016,
doi:10.18429/JACoW-IPAC2016-TUYA01.


\bibitem{Ablikim:2019hff}
M.~Ablikim et al. [BESIII Collaboration],
\emph{White Paper on the Future Physics Programme of BESIII},
\emph{Chin. Phys. C} {\bf 44} (2020), 040001.

\bibitem{etof}
X.~Li et al., \emph{Study of MRPC technology for BESIII endcap-TOF upgrade}, \emph{Radiat. Detect. Technol. Methods} {\bf 1} (2017), 13;
Y.~X.~Guo et al., \emph{The study of time calibration for upgraded end cap TOF of BESIII}, \emph{Radiat. Detect. Technol. Methods} {\bf 1} (2017), 15;
P.~Cao et al., \emph{Design and construction of the new BESIII endcap Time-of-Flight system with MRPC Technology}, \emph{Nucl.\ Instrum.\ Meth.\ A} {\bf 953} (2020), 163053.

\bibitem{geant4}
S.~Agostinelli et al. [GEANT4 Collaboration],
\emph{GEANT4: A Simulation toolkit},
\emph{Nucl.\ Instrum.\ Meth.\ A} {\bf 506} (2003), 250.

\bibitem{ref:kkmc}
S.~Jadach, B.~F.~L.~Ward and Z.~Was,
\emph{Coherent exclusive exponentiation for precision Monte Carlo calculations},
\emph{Phys. Rev. D} {\bf 63} (2001), 113009;
\emph{Comput. Phys. Commun.}  {\bf 130} (2000), 260.

\bibitem{ref:evtgen}
D.~J.~Lange, \emph{The EvtGen particle decay simulation package},
\emph{Nucl. Instrum. Meth. A} {\bf 462} (2001), 152;
R.~G.~Ping, \emph{Event generators at BESIII},
\emph{Chin. Phys. C} {\bf 32} (2008), 599.



\bibitem{ref:lundcharm}
J.~C.~Chen, G.~S.~Huang, X.~R.~Qi, D.~H.~Zhang and Y.~S.~Zhu,
\emph{Event generator for $\jpsi$ and $\psi(2S)$ decay},
\emph{Phys. Rev. D} {\bf 62} (2000), 034003;
R.~L.~Yang, R.~G.~Ping and H.~Chen,
\emph{Tuning and Validation of the Lundcharm Model with $J/\psi$ Decays},
\emph{Chin. Phys. Lett.}  {\bf 31} (2014), 061301.

\bibitem{photos}
E.~Richter-Was,
\emph{QED bremsstrahlung in semileptonic B and leptonic tau decays},
\emph{Phys. Lett. B} {\bf 303} (1993), 163.

\bibitem{vacuum}
S.~Actis et al. [Working Group on Radiative Corrections and Monte Carlo Generators for Low Energies],
\emph{Quest for precision in hadronic cross sections at low energy: Monte Carlo tools vs. experimental data},
\emph{Eur. Phys. J. C} \textbf{66} (2010), 585-686.



\bibitem{BESIII:2016bnd}
M.~Ablikim et al. [BESIII Collaboration],
\emph{Precise measurement of the $e^+e^-\to \pi^+\pi^-J/\psi$ cross section at center-of-mass energies from 3.77 to 4.60 GeV},
\emph{Phys. Rev. Lett.} \textbf{118} (2017) no.9, 092001.



\bibitem{ks-sys-err}
M.~Ablikim et al. [BESIII Collaboration],
\emph{Study of decay dynamics and $CP$ asymmetry in $D^+ \to K^0_L e^+ \nu_e$ decay},
\emph{Phys. Rev. D} \textbf{92} (2015) no.11, 112008.


\bibitem{c02pi0-bes3}
M.~Ablikim et al. [BESIII Collaboration],
\emph{Branching fraction measurements of $\chi_{c0}$ and $\chi_{c2}$ to $\pi^0\pi^0$ and $\eta\eta$},
\emph{Phys. Rev. D} \textbf{81} (2010), 052005.

\bibitem{KF}
M.~Ablikim et al. [BESIII Collaboration],
\emph{Search for hadronic transition $\chicJ\to \eta_c\pp$ and observation of $\chicJ\to K\overline{K}\pi\pi\pi$},
\emph{Phys. Rev. D} \textbf{87} (2013) no.1, 012002.



\bibitem{BESIII:2013fnz}
M.~Ablikim et al. [BESIII Collaboration],
\emph{Observation of $\EE\to\gamma X(3872)$ at BESIII},''
\emph{Phys. Rev. Lett.} \textbf{112} (2014) no.9, 092001.






\end{thebibliography}
\end{document}